\documentclass[twocolumn]{aastex631}
\usepackage{amsmath}
\usepackage{braket}
\usepackage{subfigure}
\usepackage{color}

\begin{document}

\title{Eliminating Primary Beam Effect in Foreground Subtraction of Neutral Hydrogen Intensity Mapping Survey with Deep Learning}

\author[0000-0002-5386-1627]{Shulei Ni}
\affiliation{Department of Physics, College of Sciences, Northeastern University, Shenyang 110819, China}

\author[0000-0003-1962-2013]{Yichao Li }
\affiliation{Department of Physics, College of Sciences, Northeastern University, Shenyang 110819, China}

\author[0000-0001-5469-5408]{Li-Yang Gao}
\affiliation{Department of Physics, College of Sciences, Northeastern University, Shenyang 110819, China}

\author[0000-0002-6029-1933]{Xin Zhang}
\affiliation{Department of Physics, College of Sciences, Northeastern University, Shenyang 110819, China}
\affiliation{Frontiers Science Center for Industrial Intelligence and Systems Optimizaiton, Northeastern University, Shenyang 110819, China}
\affiliation{Key Laboratory of Data Analytics and Optimization for Smart Industry (Ministry of Education), Northeastern University, China}

\correspondingauthor{Xin Zhang}
\email{zhangxin@mail.neu.edu.cn}




\begin{abstract}
In the neutral hydrogen (HI) intensity mapping (IM) survey, the foreground contamination on the cosmological signals is extremely severe, and the systematic effects caused by radio telescopes themselves further aggravate the difficulties in subtracting foreground. In this work, we investigate whether the deep learning method, concretely the 3D U-Net algorithm here, can play a crucial role in foreground subtraction when considering the systematic effect caused by the telescope's primary beam. We consider two beam models, i.e., the Gaussian beam model as a simple case and the Cosine beam model as a sophisticated case. 
The traditional principal component analysis (PCA) method is employed as a comparison and, more importantly, as the
preprocessing step for the U-Net method to reduce the sky map dynamic range.
We find that in the case of the Gaussian beam, the PCA method can effectively clean the foreground. However, the PCA method cannot handle the systematic effect induced by the Cosine beam, and the additional U-Net method can improve the result significantly. In order to show how well the PCA and U-Net methods can recover the HI signals, we also derive the HI angular power spectra, as well as the HI 2D power spectra, after performing the foreground subtractions. It is found that, in the case of Gaussian beam, the concordance with the original HI map using U-Net is better than that using PCA by $27.4\%$, and in the case of Cosine beam, the concordance using U-Net is better than that using PCA by $144.8\%$. Therefore, the U-Net based foreground subtraction can efficiently eliminate the telescope primary beam effect and shed new light on recovering the HI power spectrum for future HI IM experiments.


\end{abstract}




\section{Introduction}\label{sec:intro}

The cosmic large-scale structure (LSS) carries crucial information about the evolutionary history of the Universe. The neutral hydrogen (HI) intensity mapping (IM) has been proposed as a promising technique for the LSS survey~\citep[e.g.][]{Battye:2004re,McQuinn:2005hk,Pritchard:2008da}. HI IM relies on observing integrated HI 21-cm emission of many galaxies in large voxels and can be quickly carried out to cover a very large survey volume, which is ideal for cosmological studies~\citep{Chang:2007xk,Loeb:2008hg,Mao:2008ug,Wyithe:2007gz,Wyithe:2007rq,Bagla:2009jy,Seo:2009fq,Lidz:2011dx,Ansari:2011bv,Battye:2012tg}.


\citet{Chang:2010jp} for the first time explored the HI IM technique with Green Bank Telescope (GBT) by measuring the cross-correlation function between an HI IM survey and an optical galaxy survey. So far, several other explorations of HI IM have been done by measuring the cross-correlation power spectrum between the HI IM survey and the optical galaxy survey, e.g., using the GBT or Parkes telescope~\citep{Masui:2012zc,Anderson:2017ert,Wolz:2015lwa,Wolz:2021ofa}, but the HI IM auto-correlation power spectrum remains undetected~\citep{Switzer:2013ewa}. Besides, a number of specially designed radio telescopes or telescope arrays targeting at HI IM LSS survey, such as Tianlai~\citep{Chen:2012xu}, CHIME~\citep{Bandura:2014gwa}, BINGO~\citep{Pee:l2019BaryonAO}, FAST~\citep{Nan:2011um}, HIRAX~\citep{Newburgh:2016mwi}, are either under construction or collecting data~\citep{Wu:2020jwm,Li:2020ast,Wuensche:2021dcx,Abdalla:2021nyj,CHIME:2022kvg}. The HI IM is proposed as the major technique for cosmological exploration with the Square Kilometre Array (SKA)~\citep{Santos:2015gra,Bull:2014rha,SKA:2018ckk}. Moreover, with the newly built MeerKAT array, it is also proposed to have an HI IM survey in single-dish mode~\citep{MeerKLASS:2017vgf,Li:2020bcr,Wang:2020lkn}.

However, there are a couple of technical challenges for HI IM data analysis. The major challenge comes from extracting the weak HI brightness fluctuation from bright foreground radio emission in the same frequency range, i.e., the synchrotron or free-free emission of the Galaxy and nearby radio galaxy point sources. It is shown that the intensity of foreground emission, especially synchrotron emission from the Galaxy, is about $5$ orders of magnitude stronger than the HI signal. Besides, the free-free emission of the Galaxy, as well as the extragalactic point sources,  can be about $3$ orders of magnitude stronger than the HI signal.

Considerable work has been done to subtract foreground, proposing some methods for deducting HI foreground. There are advantages and disadvantages in each of these methods, which are mainly divided into two categories: non-blind and blind algorithms. Non-blind algorithms mainly rely on the assumption of a smooth foreground spectrum model by fitting a few of low-order polynomials to extract the smooth foreground components. However, the complex systematic effects of the actual data can seriously undermine the assumption of smooth foreground. The non-blind algorithm performs poorly in the actual data analysis, and it is difficult to eliminate the foreground interference. Blind algorithms are commonly used foreground subtraction methods in the analysis of HI IM real data, mainly including principal component analysis (PCA)~\citep{deOliveira-Costa:2008cxd}, fast independent component analysis~(fastICA)~\citep{Hothi:2020dgq}, correlated component analysis (CCA)~\citep{Bonaldi:2014zma}, Gaussian process~\citep{Mertens:2017gxw,Soares:2021ohm}, etc. The blind algorithms can cope with the simple systematic effects and improve the accuracy of foreground subtraction to a certain extent. But they usually cause serious signal loss and have some randomness and instability when dealing with complex beams.

These foreground subtraction methods all rely on the assumption of smoothness of the foreground emission spectrum. The instrument's systematic effects are entangled with the foreground, leading to unsatisfactory foreground subtraction by these algorithms. Therefore, the development of new algorithms for HI foreground subtraction is imminent. Deep learning forms more abstract high-level features through the underlying features to discover how the instrumental effects impact the signal. Therefore, we wish to use the deep learning approach to deal with this challenge.

Deep learning has unique advantages in feature recognition of complex systems~\citep{Makinen:2020gvh,Villanueva-Domingo:2020wpt}. In the work of \citet{Makinen:2020gvh}, the foreground subtraction with U-Net, one of the convlution network deep learning methods, has been generally discussed. The main objective of this study is to apply deep learning algorithms in foreground subtraction to eliminate instrumental effects and improve foreground subtraction. We plan to investigate the application of deep learning in eliminating instrumental effects in foreground subtraction through a series of simulations. The instrumental effects considered in this work mainly include the frequency dependence of the antenna beam and the sidelobe leakage. At present, the analysis is based primarily on the antenna beam model of MeerKAT~\citep{Matshawule:2020fjz}. We use simulation software to simulate the instrumental effects from the HI observed signal, foreground, and the introduction of beam models of different complexities. These simulated data are used as training set, test set, and validation set, respectively, to study applying the U-Net deep learning method to eliminate the instrumental effects and optimize the foreground reduction.

The rest of the paper is organized as follows: In Section~\ref{sec:Simulations} we introduce our simulation. In Section~\ref{sec:Beam_Model} we consider the frequency dependence of the antenna beam and the sidelobe leakage. In Section~\ref{sec:fg_rm} we explain how PCA and deep learning work to remove the foreground. In Section~\ref{sec:results} we show the results in our analysis and make some discussions. We conclude in Section~\ref{sec:con}.

\section{Sky simulation}\label{sec:Simulations}
In this section, we describe the simulated sky components used in our analysis, including the cosmological HI signal (Section~\ref{subsec:HI}), foreground emission (Section~\ref{subsec:fg}), and white Gaussian noise (Section~\ref{subsec:noise}). The simulated sky components are generated using the publicly available package, i.e., Cosmological Realizations for Intensity Mapping Experiments (CRIME; \url{http://intensitymapping.physics.ox.ac.uk/CRIME.html})~\citep{Alonso:2014sna}. In the following, we briefly introduce the simulation method. The simulated sky components are all given in full sky using the HEALPix pixelization strategy~\citep{Gorski:2004by}. Given the low angular resolution requirement, we use the map with $N_{\rm side}=256$ corresponding to an angular resolution of $13.73~{\rm arcmin}$, which is good enough for the HI IM experiments in the near future. In the meanwhile, we consider the frequency range spanning between $900~{\rm MHz}$ and $1050~{\rm MHz}$.

\subsection{Cosmological signal}\label{subsec:HI}
The mean brightness temperature of HI 21-cm emission from redshift $z$ is
\begin{equation}\label{equ:temp}
	\bar{T}_{\rm HI}(z)= 190.55~{\rm mK}
	\frac{\Omega_{\rm b}h(1+z)^2x_{\rm HI}(z)}{\sqrt{\Omega_{\rm m}(1+z)^3+\Omega_{\rm \Lambda}}},
\end{equation}
where $h=H_0/(100{\rm~km~s^{-1}~Mpc^{-1}})$ is the dimensionless Hubble constant,
$x_{\rm HI}$ is the fraction of the HI mass in the total baryons,
and $\Omega_{\rm b}$, $\Omega_{\rm m}$, and $\Omega_{\Lambda}$ are the baryon,
total matter, and dark energy density fractions, respectively.
The HI brightness temperature fluctuation is expressed as
\begin{equation}
	T_{\rm HI}(z, \hat{\bf n}) = \bar{T}_{\rm HI}(z) (1 + \delta_{\rm HI}(z, \hat{\bf n})),
\end{equation}
where $\delta_{\rm HI}(z, \hat{\bf n})$ is the HI overdensity field in redshift space.
The final simulated maps are generated using uniform frequency slices according to
$z + 1 = \nu_0 / \nu$, where $\nu_0=1420.406~{\rm MHz}$ is the HI 21-cm emission rest frame frequency.
The HI fluctuation signal is simulated using the lognormal realization with
power spectrum of~\citep{Alonso:2014sna,Olivari:2017bfv}
\begin{equation}
	C_{\rm \ell}^{\rm HI}(\nu_1, \nu_2) = \frac{2}{\pi} \int {\rm d}k~k^2
	P_{\rm DM}(k) W_{\ell, \nu_1}(k) W_{\ell, \nu_2}(k),
\end{equation}
where $P_{\rm DM}(k)$ is the underlying dark matter power spectrum at redshift $z=0$, and
$W_{\ell, \nu}$ is the window function defined as
\begin{equation}
	\begin{split}
		W_{\ell, \nu_i}(k) = \int& {\rm d}z~\bar{T}_{\rm HI}(z)\phi(\nu_i, z) D(z) \\
		&\times \Big( b(z) j_\ell(k\chi) - f(z) j''_\ell(k\chi) \Big),
	\end{split}
\end{equation}
where $j_\ell$ is the $\ell$-th spherical Bessel function,
$\phi(\nu_i, z)$ is the frequency selection function centering
at $\nu_i$, $D(z)$ is the growth factor, $f(z)$ is the growth rate
and $b(z)$ is the linear bias of the HI density with respect to the
underlying dark matter density, i.e. $\delta_{\rm HI} = b(z) \delta_{\rm DM}$.

%
%

In our simulation, we use the best-fit cosmological parameters from the {\it Placnk} 2018 $\Lambda$CDM results~\citep{Planck:2018vyg}, i.e.,
$H_0=67.7$ km s$^{-1}$ Mpc$^{-1}$, $\Omega_{\rm b}=0.049$, $\Omega_{\rm m}=0.311$, $\Omega_\Lambda=0.689$, and $\sigma_8=0.81$.
As an example, one simulated HI map of frequency slice at $\nu=1000~{\rm MHz}$ is shown in
Figure~\ref{fig:HI}.

\subsection{Foreground components}\label{subsec:fg}

The foreground consists of many distinct components that emit via different emission mechanisms.
Following the analysis in the literature, we consider four different foreground components
in our simulation, i.e., the Galactic synchrotron emission, the Galactic free-free emission,
the extragalactic free-free emission, and the extragalactic point sources,
which are known as the major contamination of the HI signal.

The foreground is simulated using two different methods according to two different angular distributions, i.e., the anisotropic and isotropic ones. The Galactic synchrotron is produced by the accelerated motion of energetic charged particles dispersed in the Galactic magnetic field. Its angular structure is highly anisotropic, where the brightness temperature grows steeply towards the Galactic plane.

In the simulations, the Galactic synchrotron amplitude template is based on the Haslam map at $408$~MHz~\citep{Haslam:1982zz}. The Galactic synchrotron emission is simulated by extrapolating the Haslam map's brightness temperature $T_{\rm Haslam}(\hat{\bf n})$ to other frequencies with direction dependent spectral index. The spectral index, $\gamma(\hat{\bf n})$, is generated with the Planck Sky Model (PSM)~\citep{Delabrouille:2012ye}. The anisotropic Galactic synchrotron emission is simulated via
\begin{equation}
	T_{{\rm syn}, 0}(\nu, \hat{\bf n}) = T_{\rm Haslam}(\hat{\bf n})
	\left(\frac{408~{\rm MHz}}{\nu}\right)^{\gamma(\hat{\bf n})}.
\end{equation}

\begin{figure*}
	\centering
	\subfigure[]{%
		\includegraphics[width=0.45\textwidth]{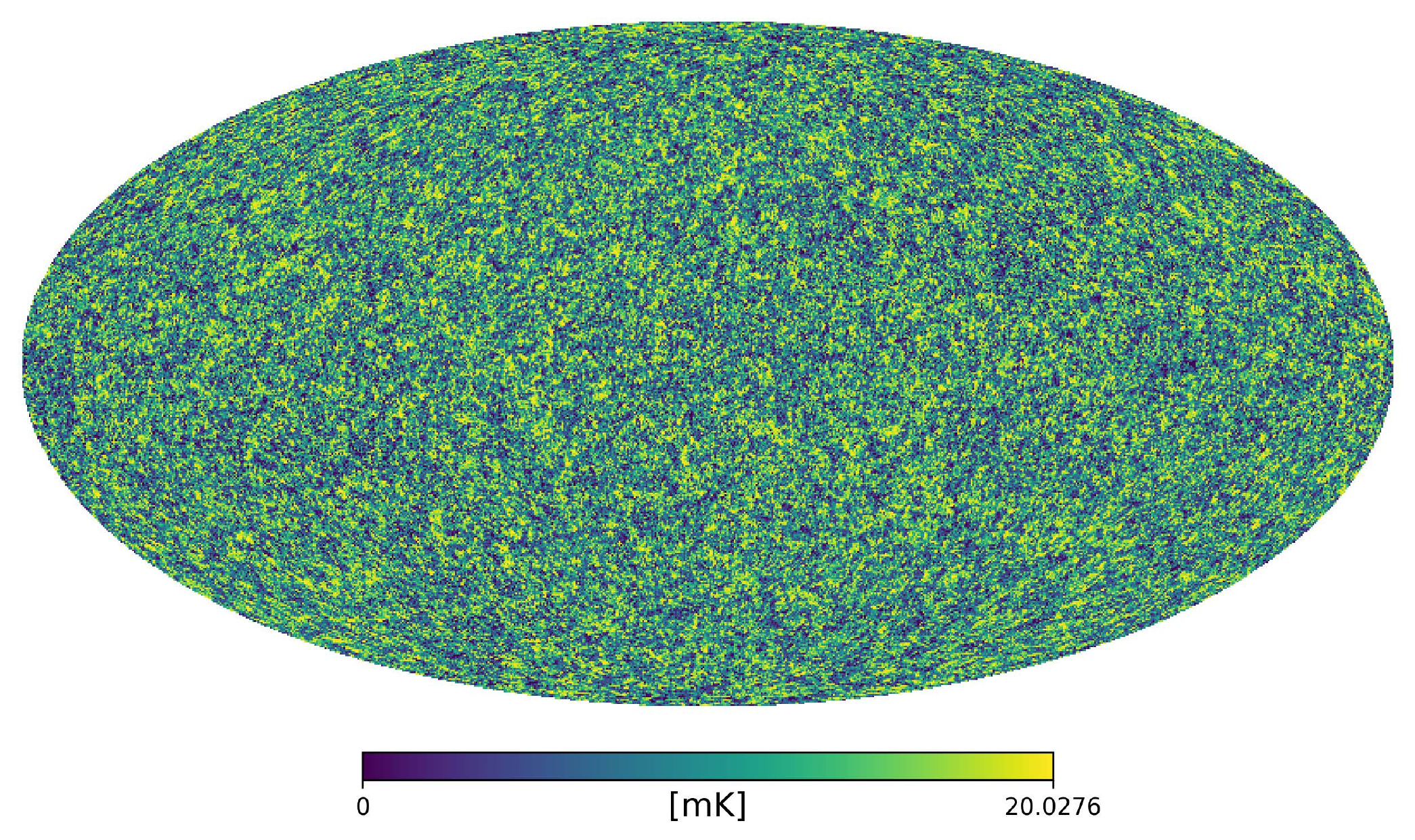}%
		\label{fig:HI}%
	}\
	\subfigure[][]{%
		\includegraphics[width=0.45\textwidth]{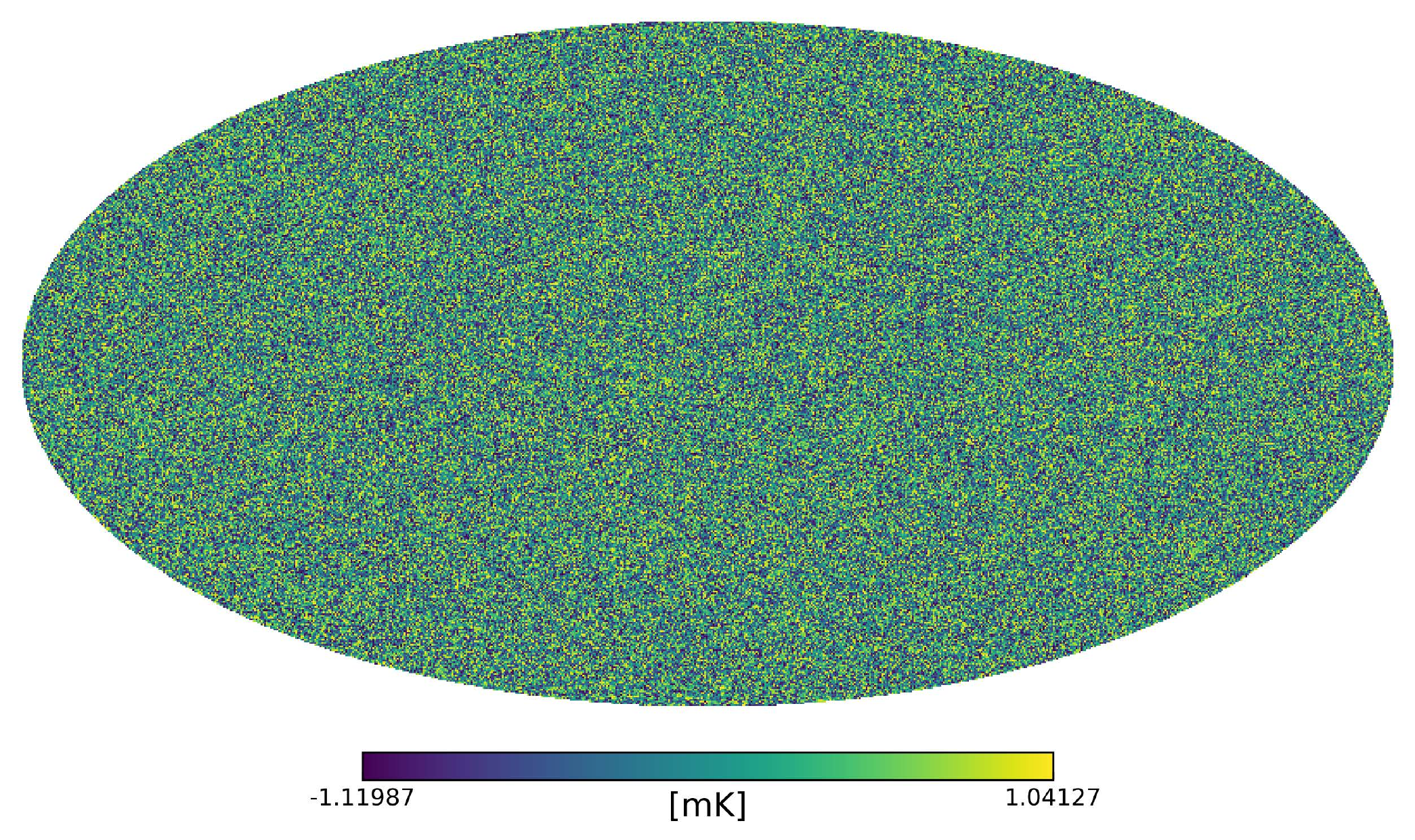}%
		\label{fig:noise}%
	}\\
	\subfigure[][]{%
		\includegraphics[width=0.45\textwidth]{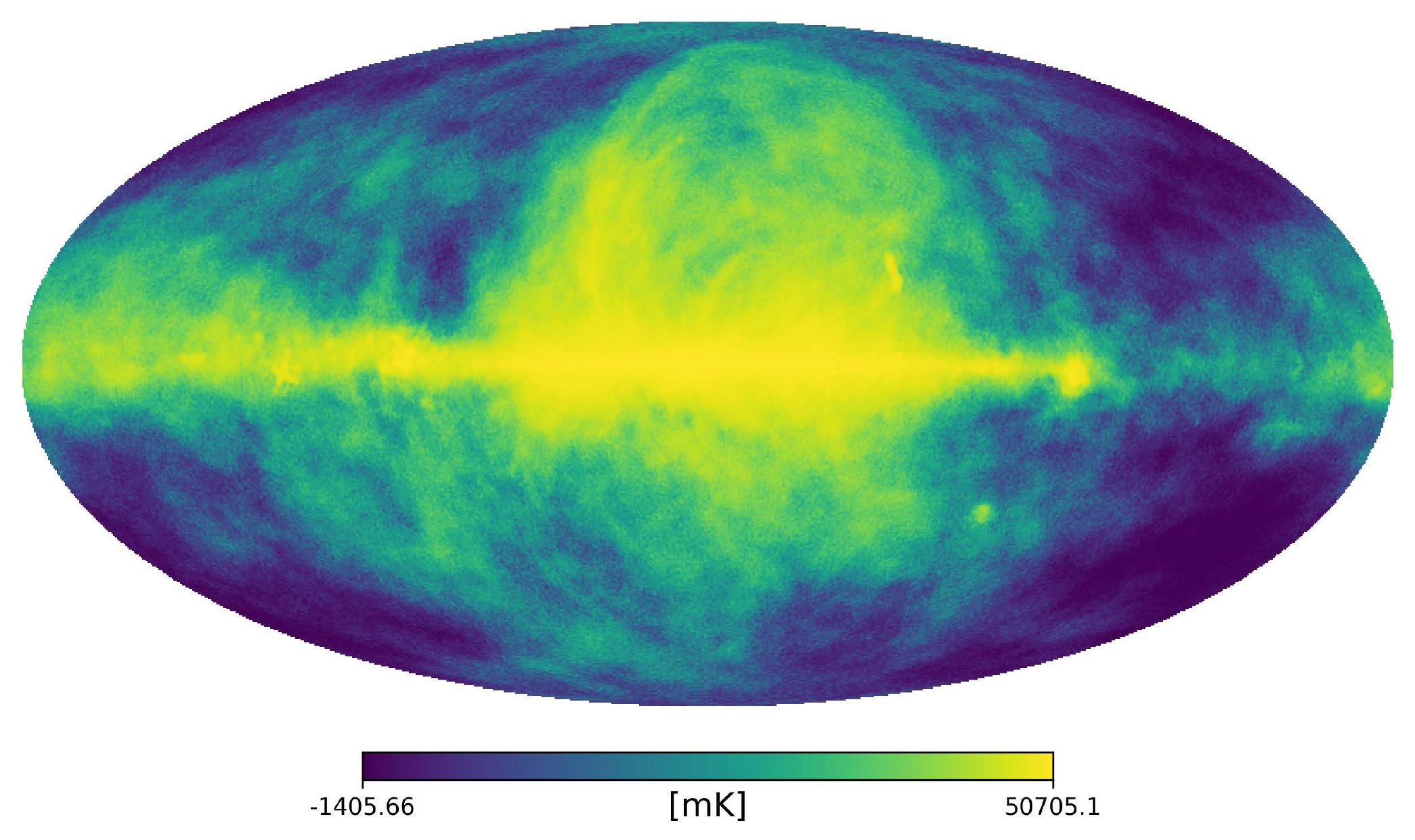}%
		\label{fig:sy}%
	}
	\subfigure[][]{%
		\includegraphics[width=0.45\textwidth]{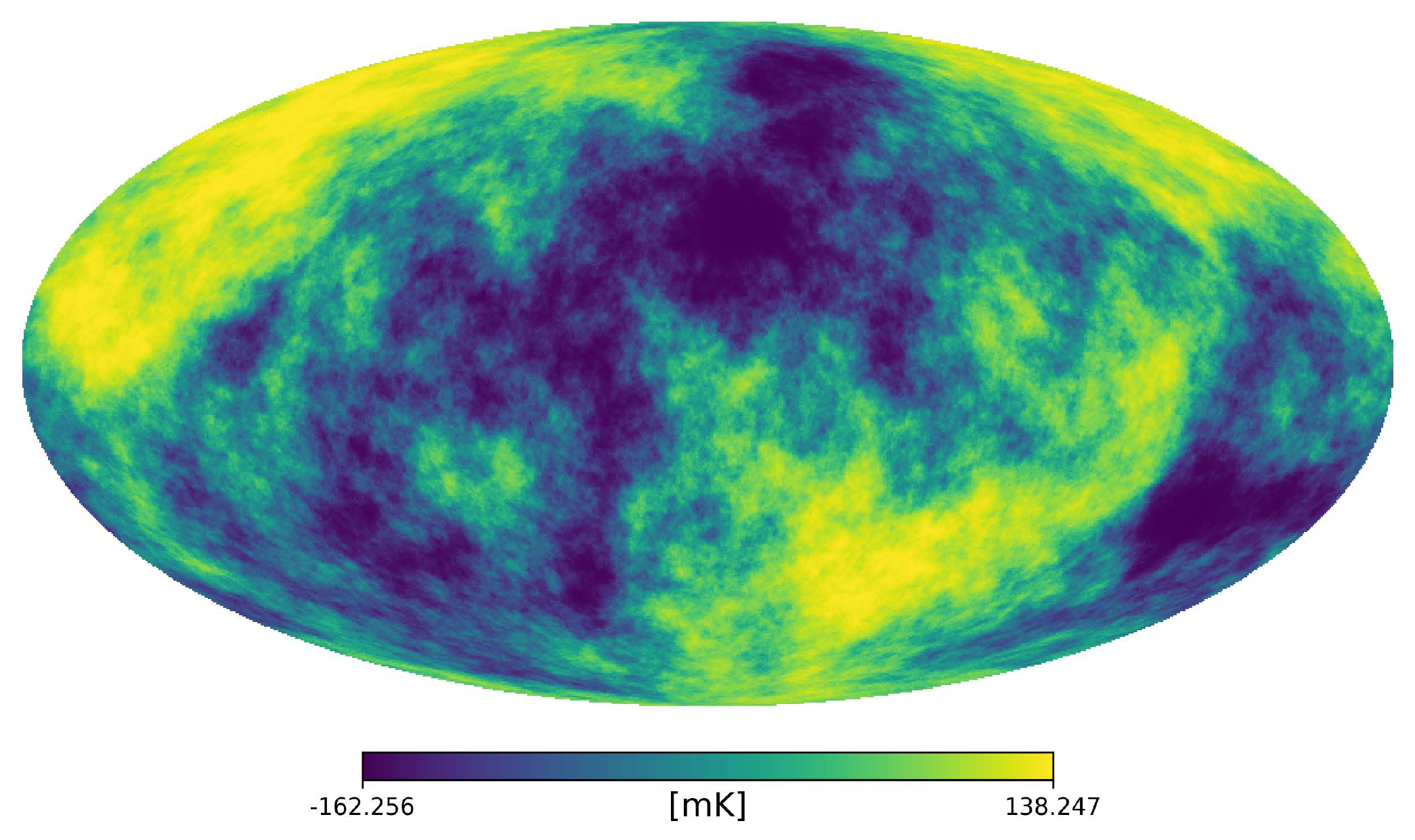}%
		\label{fig:gf}%
	}\\
	\subfigure[][]{%
		\includegraphics[width=0.45\textwidth]{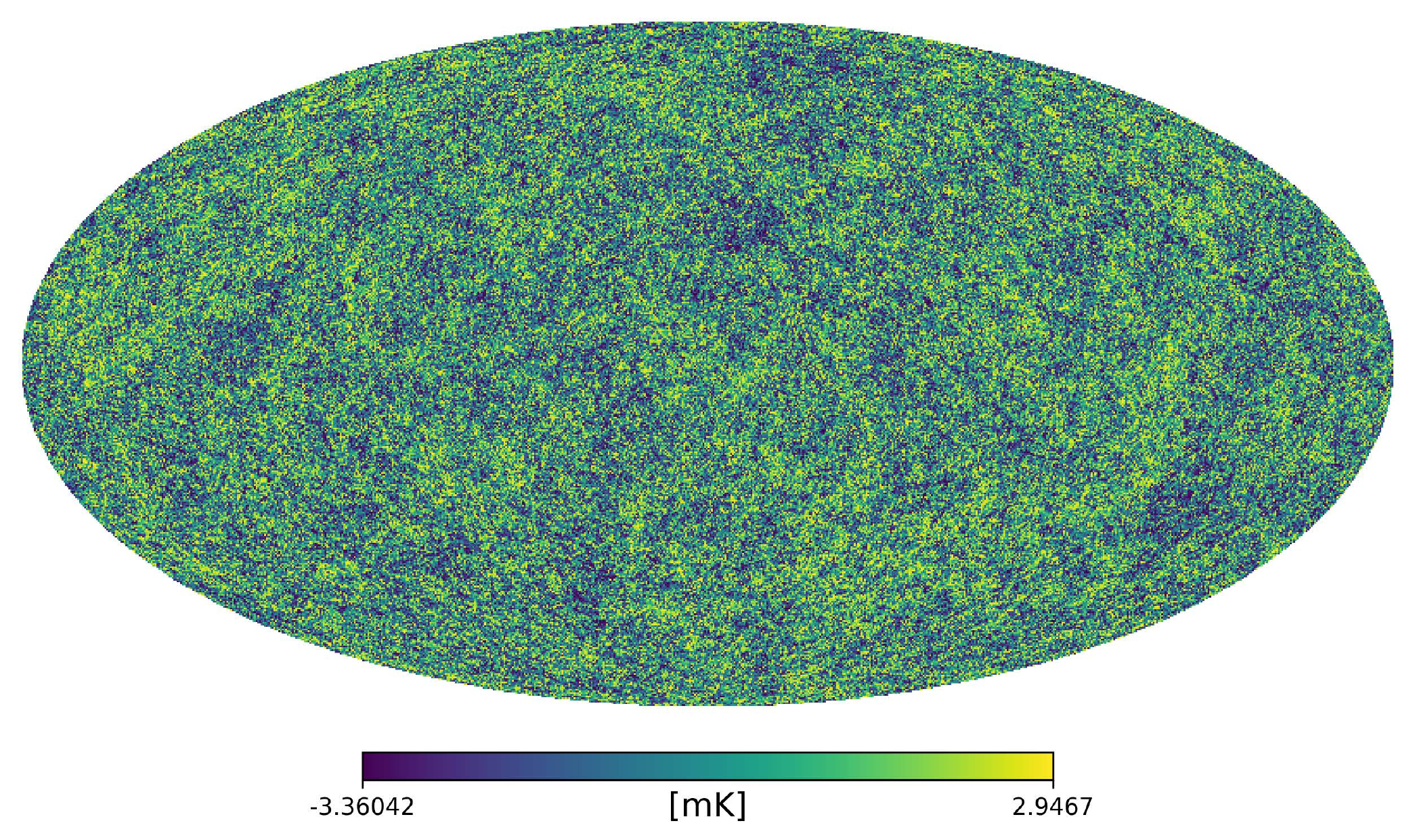}%
		\label{fig:egf}%
	}
	\subfigure[][]{%
		\includegraphics[width=0.45\textwidth]{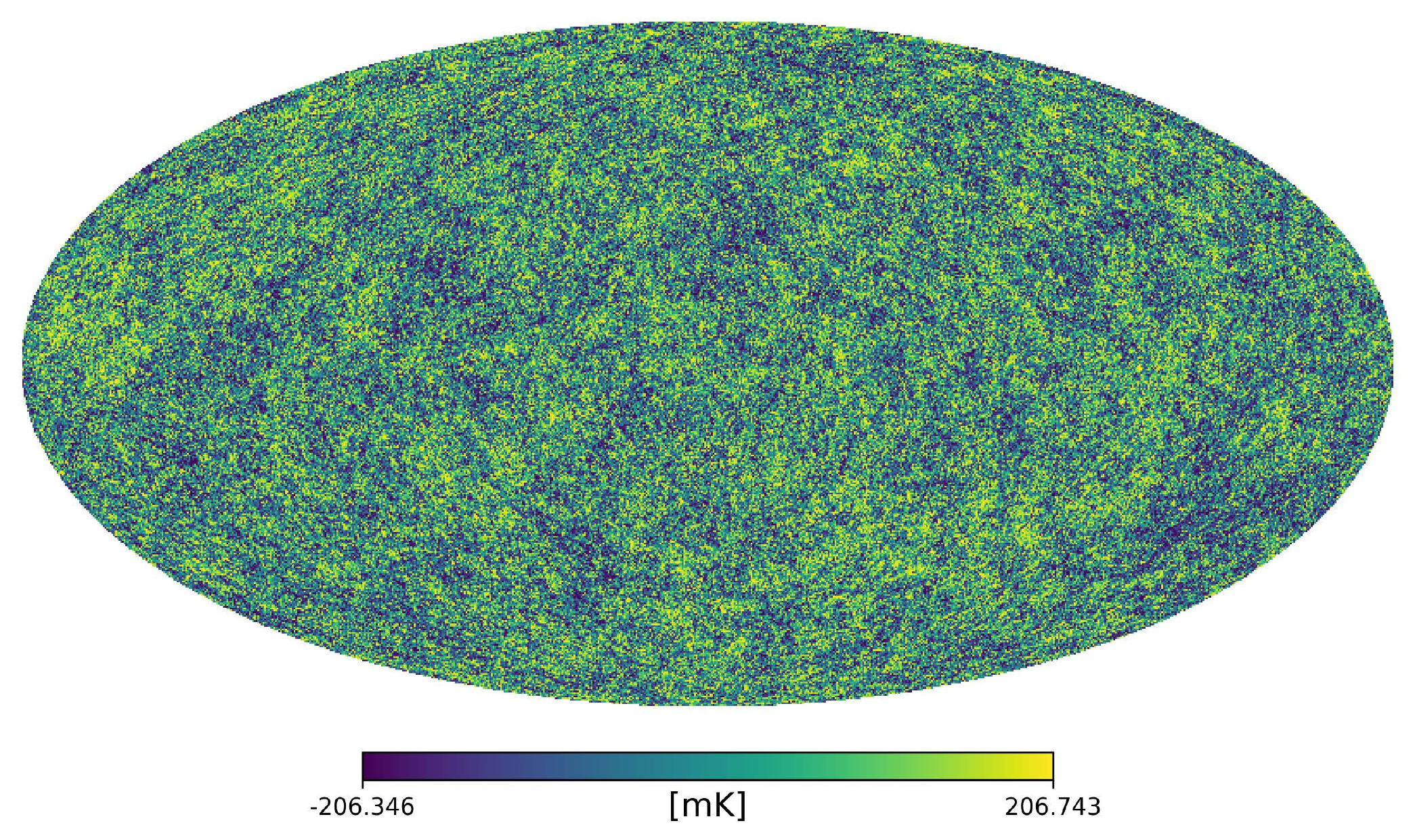}%
		\label{fig:ps}%
	}
  \caption{
  Simulated sky maps from different components. Panels (a)--(f) are the maps of HI, white noise, Galactic synchrotron, Galactic free-free, extragalactic free-free, and extragalactic point source, respectively.
  The maps are shown using the histogram equalized color mapping to amplify the contrast of the image.
  }
  \label{fig:fg_maps}
\end{figure*}

Due to the low resolution of the Haslam sky map, the angular structure information is lost
at small scales.
We assume that the anisotropy of Galactic synchrotron emission can be neglected at small scales. Therefore, based on the following form of spectrum,
\begin{equation}\label{equ:fg_equation}
	C^{\rm fg}_\ell(\nu_1, \nu_2) = A \left( \frac{\ell_{\rm ref}}{\ell} \right) ^\beta
	\left( \frac{\nu_{\rm ref}^2}{\nu_1\nu_2} \right)^\alpha
	\exp{\left(-\frac{\log^2(\nu_1/\nu_2)}{2\xi^2}\right)},
\end{equation}
we generate the isotropic structure of synchrotron emission using the method implemented by Santos, Cooray and Knox (SCK)~\citep{Santos:2004ju} with random realizations, $\delta T_{\rm syn}(\nu, \hat{\bf n}) $.
In Equation (\ref{equ:fg_equation}), $A$ is the amplitude of power spectrum,
$\xi$ is the frequency-space correlation length of the emission.
We use the reference scale $\ell_{\rm ref}=1000$ and the reference frequency
$\nu_{\rm ref}=130~{\rm MHz}$.
The rest parameters are listed in Table~\ref{tab:fg}.
In order to restrict the synchrotron simulation to only the small scales, the random realizations are rescaled to $0$ on the scales 
overlapping with the Haslam map's resolution scale before adding to the 
final Galactic synchrotron map,
\begin{equation}
	T_{{\rm syn}}(\nu, \hat{\bf n})  = T_{{\rm syn}, 0}(\nu, \hat{\bf n})
	+ \delta T_{\rm syn}(\nu, \hat{\bf n}).
\end{equation}
Figure~\ref{fig:sy} shows an example of the simulated Galactic synchrotron map at the
frequency of $1000~{\rm MHz}$.

\begin{table}
	\caption{
		The parameters of isotropic foreground power spectrum model
		[see Equation~(\ref{equ:fg_equation})] for different components.
		The pivot values are $\ell_{\rm ref}=1000$ and $\nu_{\rm ref}=130$~MHz.
	}\label{tab:fg}
	\begin{center}
		\begin{tabular}{lcccc}
			\hline\hline
			Foreground Component     & $A~[{\rm mK}^2]$ & $\beta$ & $\alpha$ & $\xi$ \\
			\hline
			Iso. Gal. Syn.           & $700   $    & $2.4 $    &   $2.80 $  & $4.0$   \\
			Point sources            & $57.0  $    & $1.1 $    &   $2.07 $  & $1.0$   \\				
			Galactic free-free       & $0.088 $    & $3.0 $    &   $2.15 $  & $35 $ \\
			Extragalactic free-free  & $0.014 $    & $1.0 $    &   $2.10 $  & $35 $ \\
			\hline
		\end{tabular}
	\end{center}
\end{table}

Besides, the SCK model is also used to simulate free-free emission and extragalactic point sources.

Extragalactic point sources are radio galaxies radiating from outside the Milky Way.
Both neighboring radio galaxies and high-redshift radio galaxies are included.
For high-redshift galaxies, their association with cosmological signals needs to be considered.
We use the SCK model to simulate radio galaxies with certain clustering properties by
introducing an LSS power spectrum model of radio galaxies
through Equation~(\ref{equ:fg_equation}), and the specific simulation parameters are shown in Table~\ref{tab:fg}.
Strictly speaking, high-redshift radio galaxies are the background components of the HI signal.
However, their interference effect on the HI is the same as the foreground, and they are
also used as foreground components in the simulation.
Figure~\ref{fig:ps} shows an example of the simulated extragalactic point source map
at the frequency of $1000~{\rm MHz}$.

Free-free emission is the electromagnetic radiation produced by the deceleration of high-speed electrons when deflected by an interstellar ion, i.e., bremsstrahlung. Free-free emission is not homogeneous, and considering that it is exceptionally smooth frequency dependence and its subdominant amplitude, to simplify the simulation, we approximate the free-free emission as isotropic and introduce its distribution power spectrum model through Equation~(\ref{equ:fg_equation}) to simulate the free-free emission sky map. An example of Galactic and extragalactic free-free emission maps at $1000~{\rm MHz}$
are shown in Figures~\ref{fig:gf} and \ref{fig:egf}.

The total foreground map is generated by combining all different components,
\begin{equation}
	\begin{split}
		T_{\rm fg}(\nu, \hat{\bf n}) &= T_{\rm syn}(\nu, \hat{\bf n}) + T_{\rm PS}(\nu, \hat{\bf n}) \\
		&+ T_{\rm GFF}(\nu, \hat{\bf n})  + T_{\rm EGFF}(\nu, \hat{\bf n}),
	\end{split}
\end{equation}
in terms of Galactic synchrotron map $T_{\rm syn}(\nu, \hat{\bf n})$,
extragalactic point source map $T_{\rm PS}(\nu, \hat{\bf n})$,
Galactic free-free emission map $T_{\rm GFF}(\nu, \hat{\bf n})$,
and extragalactic free-free emission map $T_{\rm EGFF}(\nu, \hat{\bf n})$.

\subsection{White noise}\label{subsec:noise}
We also consider the instrumental noise and its impact on the observed HI signal. The instrument generates both the uncorrelated white noise and the correlated $1/f$ noise~\citep{Li:2020bcr}. The effect of $1/f$ noise is related to the detailed scanning strategy. To simplify the simulation, we only consider the white noise and add onto our simulated maps a Gaussian random realization, $T_{\rm noise}(\nu, \hat{\bf n})$, with root mean square~(RMS)
\begin{equation}\label{equ:noise}
\sigma_{\rm noise}=T_{\rm sys}\sqrt{\frac{4\pi f_{\rm sky}}
{\Omega_{\rm beam} N_{\rm dish}t_{\rm obs}\Delta\nu}},
\end{equation}
where $N_{\rm dish}$ is the number of telescope dishes or feeds,
$\Omega_{\rm beam}\approx{(\lambda}/{\rm D})^2$ is the solid angle of the telescope main beam,
$t_{\rm obs}$ is the total observation time, and $\Delta\nu$ is the frequency resolution.

Since there is no full-sky experiment for HI, we refer to the observational parameters of a currently available experiment, the MeerKAT telescope, for a simple simulation of the noise. The analysis reveals that the instrumental noise is much smaller relative to the foreground, so we only briefly consider the instrumental noise. We fix the sensitivity of telescope $T_{\rm sys}=16~{\rm K}$, the number of dishes $N_{\rm dish}=64$, the observation time $t_{\rm obs}=8000~{\rm h}$, the sky coverage $f_{\rm sky}=1$, the frequency resolution $\Delta\nu = 2.34375~{\rm MHz}$, and the dish diameter $D=13.5{\rm m}$. So, the uncorrelated Gaussian white noise with an RMS $\sigma_{\rm noise}=0.25~{\rm mK}$ . As an example, one simulated noise map of frequency slice at $\nu=1000~{\rm MHz}$ is shown in Figure~\ref{fig:noise}.

{The foreground parameters listed in Table~\ref{tab:fg} are fixed in our simulations.
In practice, these parameters may differ, resulting in different foreground features.
However, such variation of parameters does not increase the degrees of freedom of the
foreground components. A preprocessing with blind foreground subtraction, such as 
PCA, can remove the major foreground components without accurate measurements of
the foreground parameters. 
The residuals have subdominant effects when the foreground parameters 
are changed. To make the analysis easier, we skip generating the robust training 
set simulations by varying the foreground parameters.}

\section{Primary beam model}\label{sec:Beam_Model}

The final simulated sky map is the linear combination of the different components,
\begin{equation}\label{equ:T_tot}
    T_{\rm sky}(\nu, \hat{\bf n}) = T_{\rm HI}(\nu, \hat{\bf n}) + T_{\rm fg}(\nu, \hat{\bf n})
    + T_{\rm noise}(\nu, \hat{\bf n}).
\end{equation}
The observed sky map is the observational sky map convolved with the telescope's primary beam.

\begin{figure*}
    \centering
	\subfigure[]{%
		\includegraphics[width=0.495\textwidth]{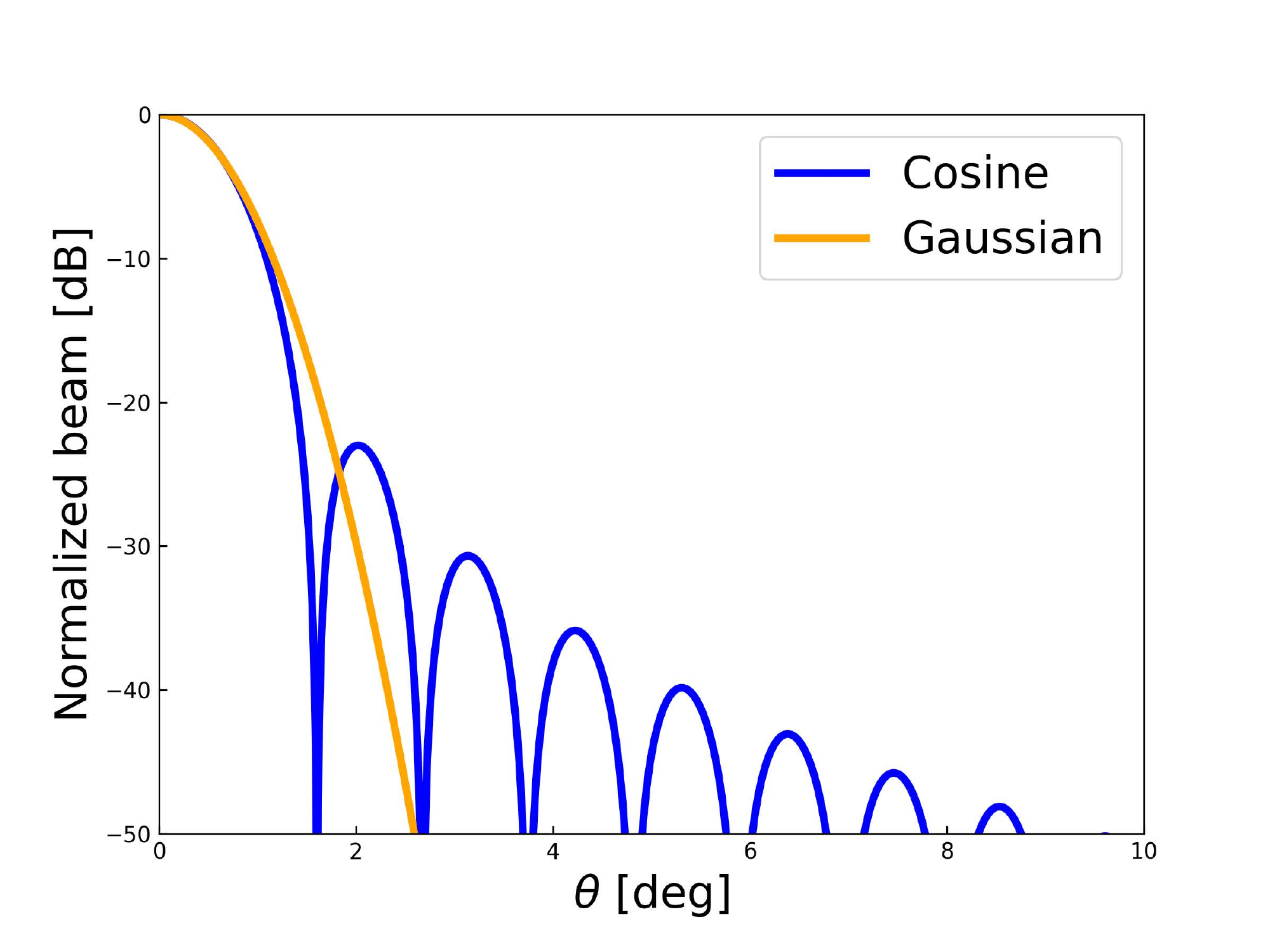}%
		\label{fig:btf}%
	}
	\subfigure[]{%
		\includegraphics[width=0.495\textwidth]{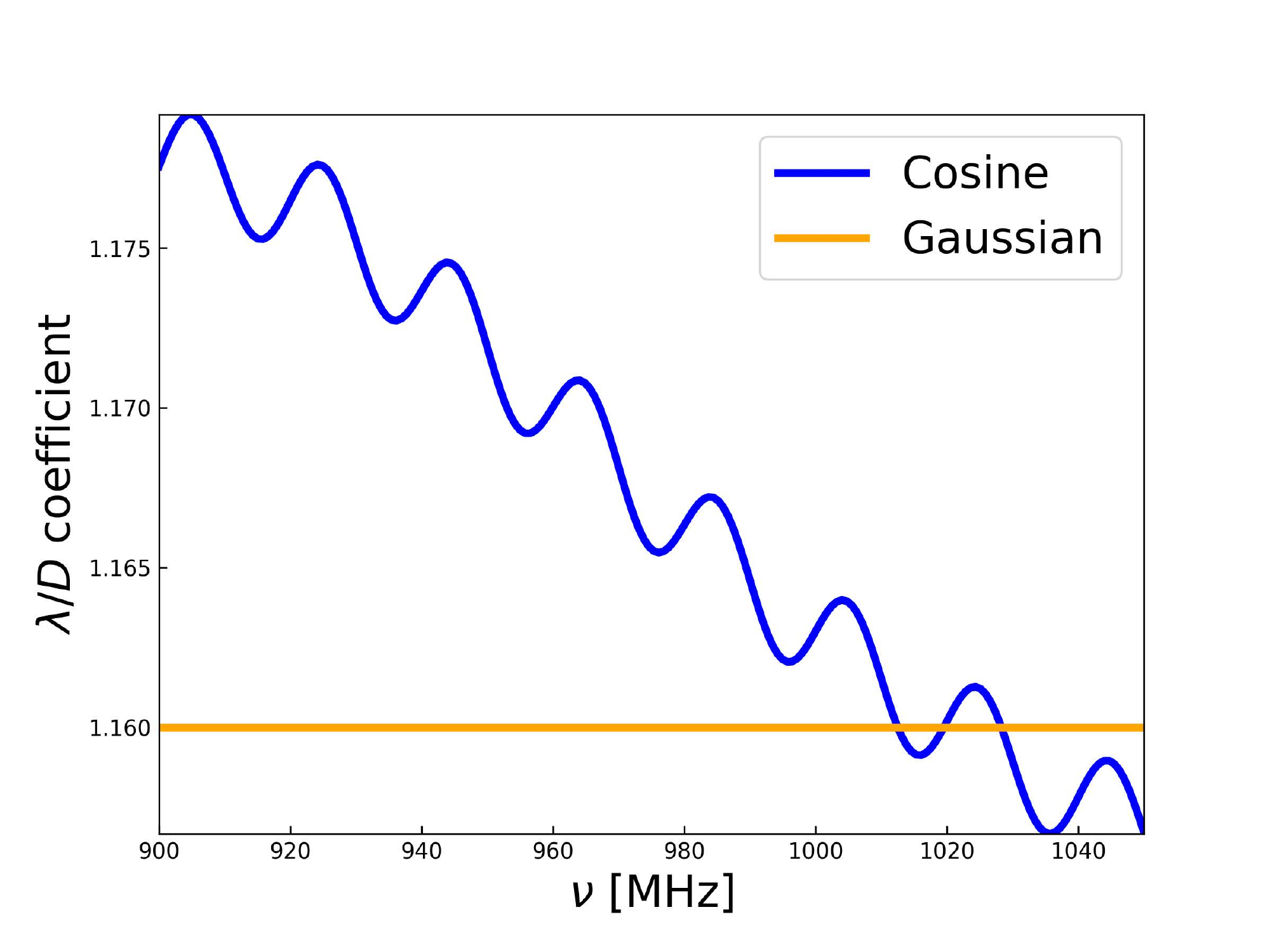}%
		\label{fig:coeff}%
	}\caption{MeerKAT beam model~\citep{Matshawule:2020fjz}. Panel (a): Comparison of primary beam models, the standard Gaussian model (orange) and the Cosine model (blue), at 1000 MHz. Panel (b): The $\lambda/D$ normalized beam size as a function of frequency for the Gaussian model [orange, see Equation~(\ref{equ:gaussian_fwhm})] and the Cosine model [blue, see Equation~(\ref{equ:cosine_fwhm})].}\label{fig:MeerKAT_beam}
\end{figure*}

The frequency dependence of the primary beam, such as the frequency evolution of the beam size, can complicate the foreground emission. Such a feature has been noticed in most of the HI IM experiments and can be overcome by degrading the beam size to the largest one~\citep{Switzer:2013ewa}. To simulate such an effect, we use a simple Gaussian beam model with frequency-dependent beam size,
\begin{equation}\label{equ:gaussian_beqm}
B_{\rm G}(\nu, \theta)=\exp \left[-4 \ln 2 \left(
\frac{\theta}{\Delta\theta_{\rm G}(\nu)}\right)^2\right],
\end{equation}
where $\Delta\theta_{\rm G}(\nu)$ is the full width at half maximum (FWHM) of the primary beam,
i.e., frequency-dependent beam size. Such a frequency dependence is modeled as \citep{Matshawule:2020fjz}
\begin{equation}\label{equ:gaussian_fwhm}
\Delta\theta_{\rm G}(\nu) = 1.16\frac{\lambda(\nu)}{D},
\end{equation}
where $\lambda(\nu) = c/\nu$ is wavelength at frequency $\nu$ and $c$ is the speed of light. The factor of $1.16$ is a good approximation to the MeerKAT primary beam. We take the MeerKAT beam as the example throughout this work and use $D=13.5~{\rm m}$.

The Gaussian beam model only includes the frequency variation of the primary beam size and is not accurate enough to recover the actual primary beam feature. Using the MeerKAT L-band full-polarization `astro-holographic' observation, \citet{Asad:2021mnras} reconstructed a Zernike-based beam model and provided a software tool named EIDOS (\url{https://github.com/ratt-ru/eidos}), which is the most accurate beam model for MeerKAT dishes. With such a model, two important features can be seen clearly. One is the primary beam sidelobes. The EIDOS software provides the accurate primary beam sidelobe feature within $5^{\circ}$ radius. To extend the sidelobes to a large separation angle, following \citet{Matshawule:2020fjz}, we use a cosine-tapered field illumination function to produce the sidelobe pattern,
\begin{equation}\label{equ:cosine_beam}
B_{\rm C}(\nu, \theta) = \left[\frac{\cos(1.189\pi/\Delta\theta_{\rm C}(\nu))}
{1-4(1.189\theta/\Delta\theta_{\rm C}^2(\nu))}\right]^2,
\end{equation}
which is called the Cosine beam model in this work.
The primary beam patterns of the Gaussian and Cosine beam models at the frequency of $1000$ MHz
are shown in Figure~\ref{fig:btf}.

The Cosine beam model also has the frequency-dependent beam size, $\Delta\theta_{\rm C}(\nu)$. Another important primary beam feature produced with the EIDOS software is that the beam size follows the frequency evolution as expressed in Equation~(\ref{equ:gaussian_fwhm}) but also exhibits a low-level frequency-dependent ripple. Such a ripple can be well modeled with a sinusoidal oscillation on top of a high order polynomial function~\citep{Matshawule:2020fjz},
\begin{equation}\label{equ:cosine_fwhm}
\Delta \theta_{\rm C}(\nu) = \frac{\lambda(\nu)}{D}\left[\sum_{d=0}^8a_d\hat{\nu}^d+A\sin\left(\frac{2\pi\hat{\nu}}{T}\right)\right],
\end{equation}
where $\hat{\nu}=\nu/{\rm MHz}$, $A = 0.1$, $T = 20$,
and $a_d = \{ 3.4 \times 10^{-21}, \ -3.0 \times 10^{-17}, \ -1.2 \times 10^{-11},
\ -2.6 \times 10^{-10}, \ 3.5 \times 10^{-7},\ -3.0 \times 10^{-4},
\ 0.16449, \ -50.37020, \ 6704.28133\}$.
The $\lambda/D$ normalized beam sizes for both Gaussian and Cosine beam models are shown in Figure~\ref{fig:coeff}.

In this work, we use the Gaussian beam model as a simple case and consider the Cosine beam model as a realistic case. Although the model is based on the MeerKAT primary beam measurements, the frequency-dependent effects addressed here exist in most of HI IM experiments, as most of the radio telescope dishes have similar sidelobe and frequency-dependent ripple features.

\begin{figure*}
    \centering
	\subfigure[]{%
		\includegraphics[width=0.495\textwidth]{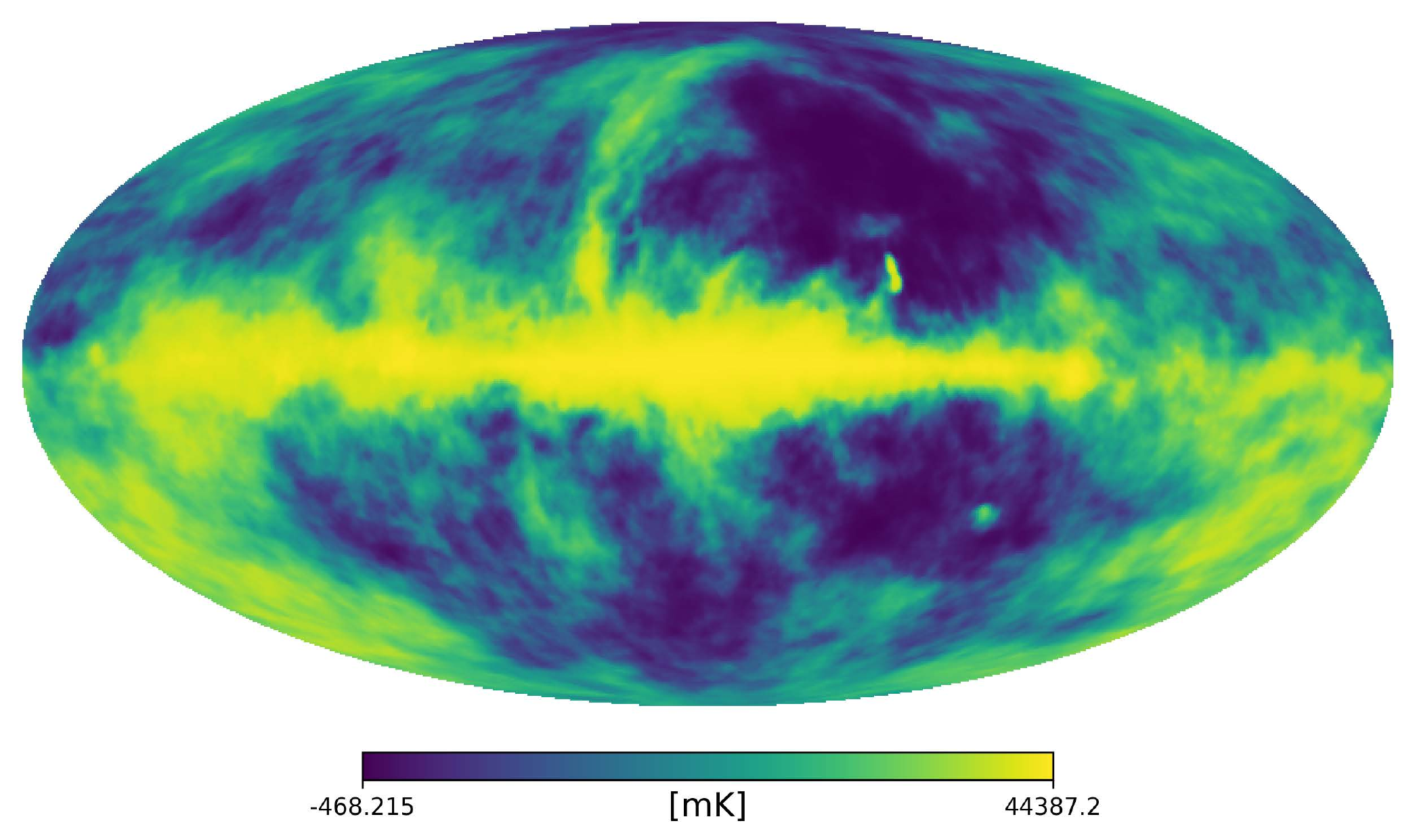}%
		\label{fig:G_smoothed}%
	}
	\subfigure[]{%
		\includegraphics[width=0.495\textwidth]{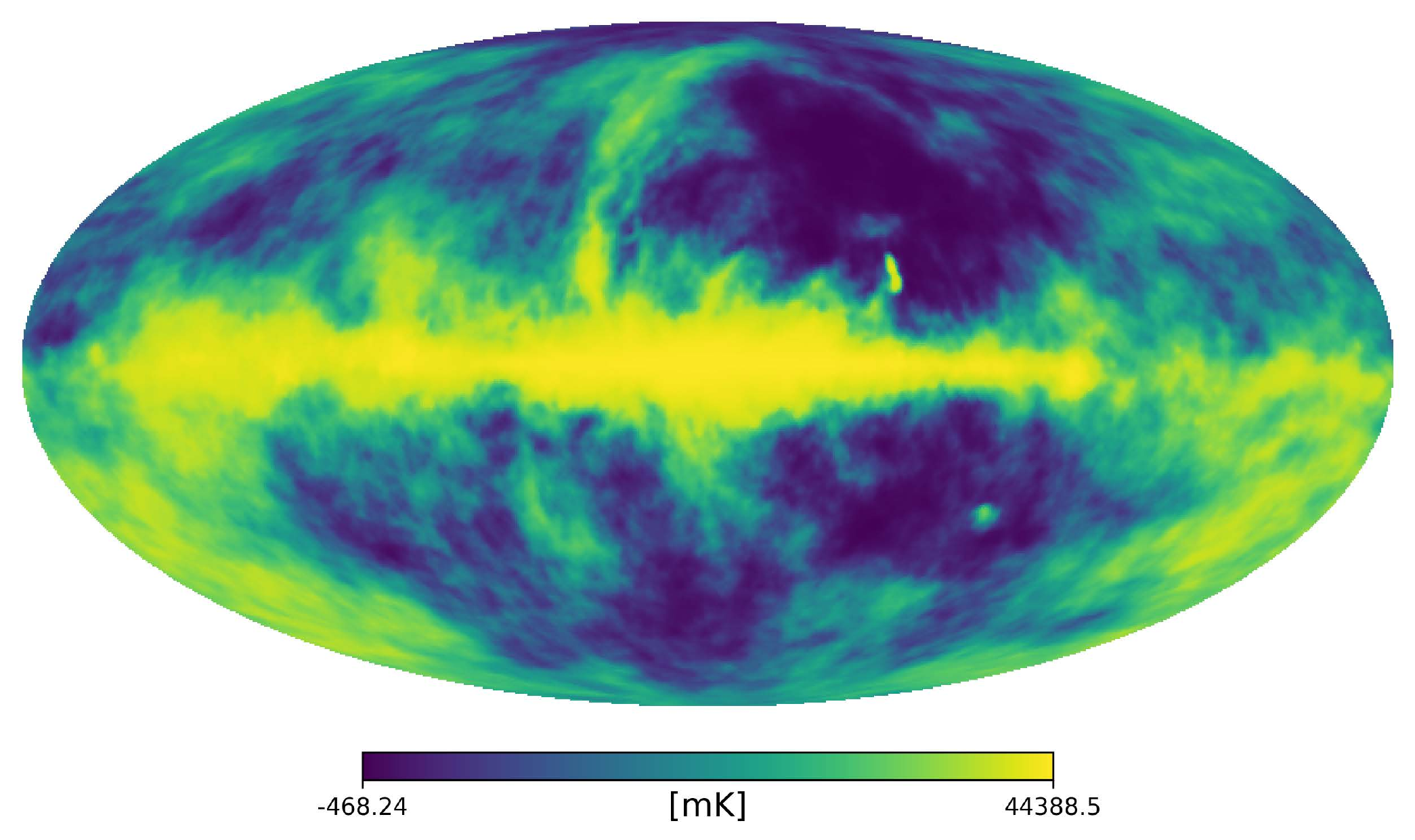}%
		\label{fig:C_smoothed}%
	}
    \caption{Different beam convolved full-sky maps at 1000 MHz: Gaussian beam convolved (a) and Cosine beam convolved (b) maps. The maps are shown using the histogram equalized color mapping to amplify the image's contrast.}
  \label{fig:smoothed_maps}
\end{figure*}

The measured sky brightness temperature map is the convolution of the original sky map with one of the beam models (see Figure~\ref{fig:smoothed_maps}).

\section{U-Net network}\label{sec:fg_rm}
\begin{figure*}
  \centering
    \includegraphics[width=\textwidth]{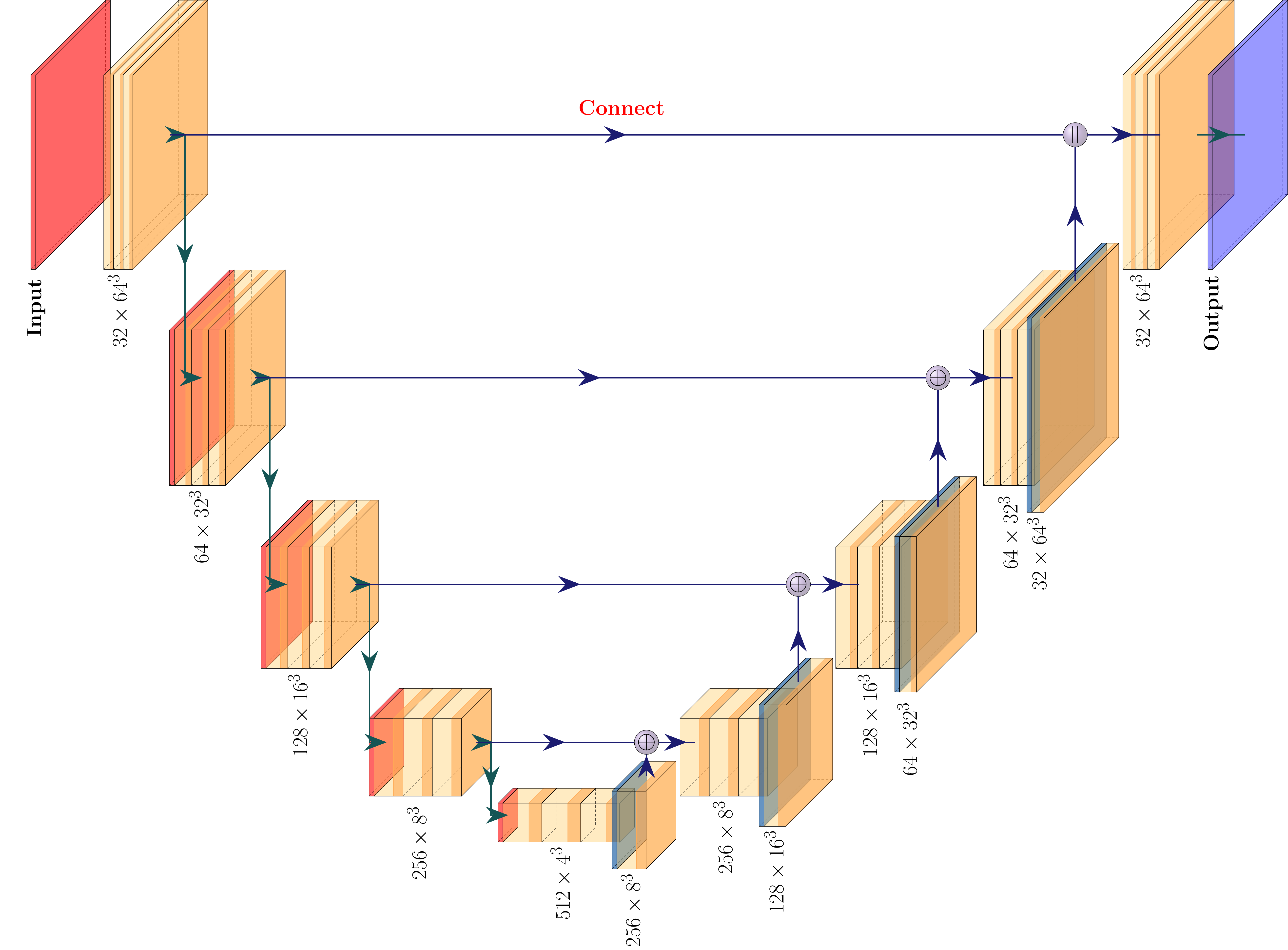}
  \caption{The U-Net architecture. Each orange box corresponds to a multi-channel feature map and a series of convolutions. The bottom of each box illustrates the number of channels and the size of the output. The dark red box and the dark blue box indicate the input and output. The light blue box shows the connected portion of the box during transpose convolutions. The down, right, and up arrows indicate maximum pooling, skip connection, and transpose convolutions, respectively. The last arrow operation denotes a $1\times1\times1$ convolution for mapping 64 features to the final 3D data. This visualization is made with the PlotNeuralNet library (\url{https://github.com/HarisIqbal88/PlotNeuralNet}).}
  \label{fig:unet}
\end{figure*}

This section introduces the deep neural network architecture (3D U-Net) used in our foreground subtraction analysis~\citep{Cicek:2016abs,Isensee:2019AnAA,Makinen:2020gvh}. 
The U-Net network is a convolutional neural network (CNN) originally developed for  
biomedical image segmentation~\citep{Ronneberger:2015UNetCN}.
It is based on the CNN, but with significant structural modifications. 
The basic U-Net architecture is shown in Figure~\ref{fig:unet}.
In addition to the standard CNN architecture, U-Net has many feature channels in the upsampling part, 
allowing the network to propagate contextual information to higher resolution layers through a series 
of transpose convolutions.
The main idea is to add successive layers to the traditional contractual network, 
with an upsampling operation replacing the convergence operation.
As a result, these layers increase the output resolution. 
The extended path is more or less symmetric with the contracted part and produces an U-shaped structure~\citep{Ronneberger:2015UNetCN}

The U-Net network is mainly used to process two-dimensional images. In contrast, HI observations can be obtained for multiple frequency channels in a specific frequency range. The correlation between frequencies is the main feature distinguishing the HI signal from the foreground. It is shown that the foreground signal has a stronger frequency correlation than the HI signal. {In addition, the complex variation of the primary beam shape of the telescope with frequency is the main difficulty of foreground subtraction.} Therefore, introducing frequency information in deep learning is necessary to eliminate systematic effects and extract foreground information accurately. For the HI foreground subtraction problem, we adopt the generalized model of the U-Net network, i.e., 3D U-Net, which can include both angular and frequency information~\citep{Ronneberger:2015UNetCN, Cicek:2016abs, Villanueva-Domingo:2020wpt, Isensee:2019AnAA, Makinen:2020gvh}.

\subsection{Data preprocessing}

\begin{figure*}
    \centering
    \subfigure[]
        {\includegraphics[width=0.33\textwidth]{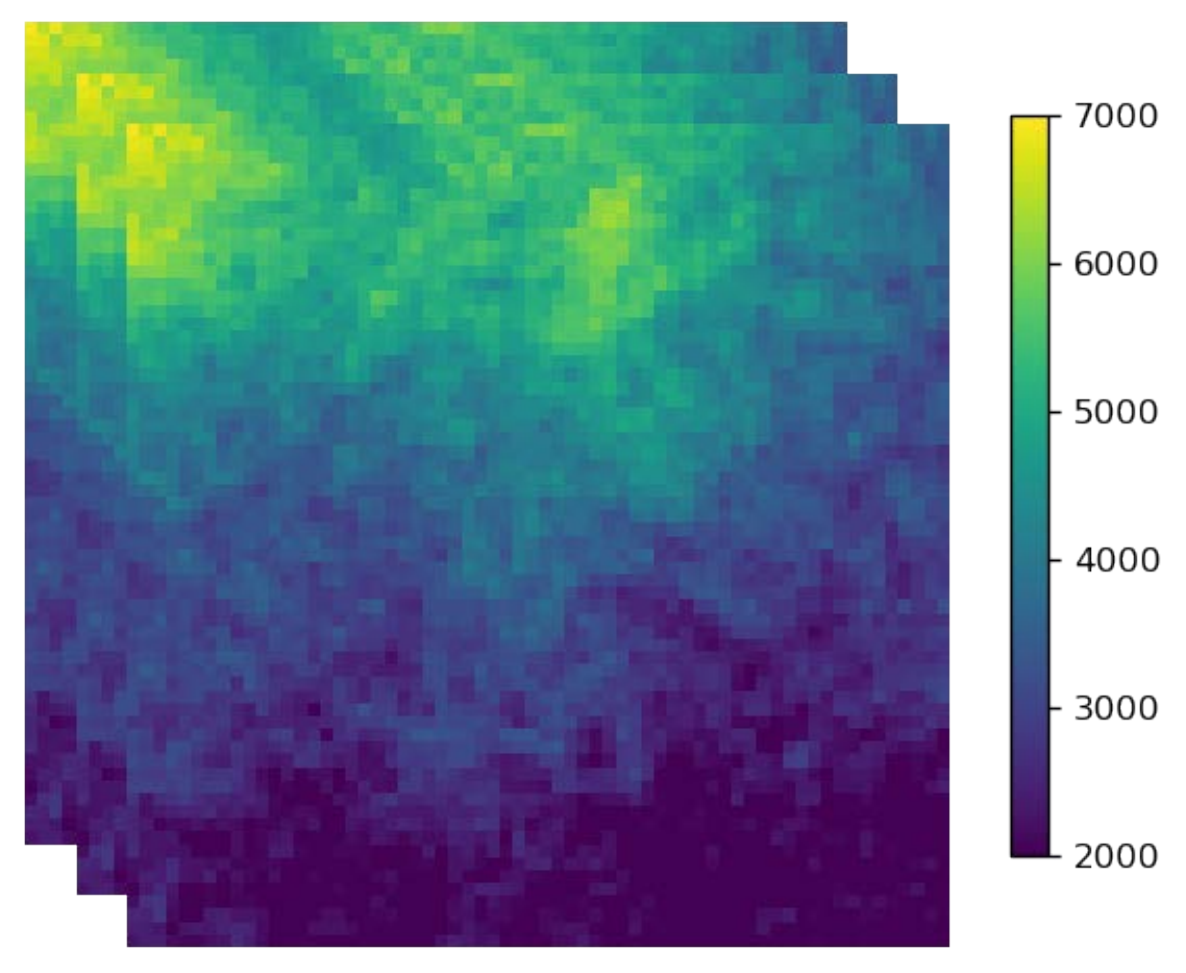}
        \includegraphics[width=0.33\textwidth]{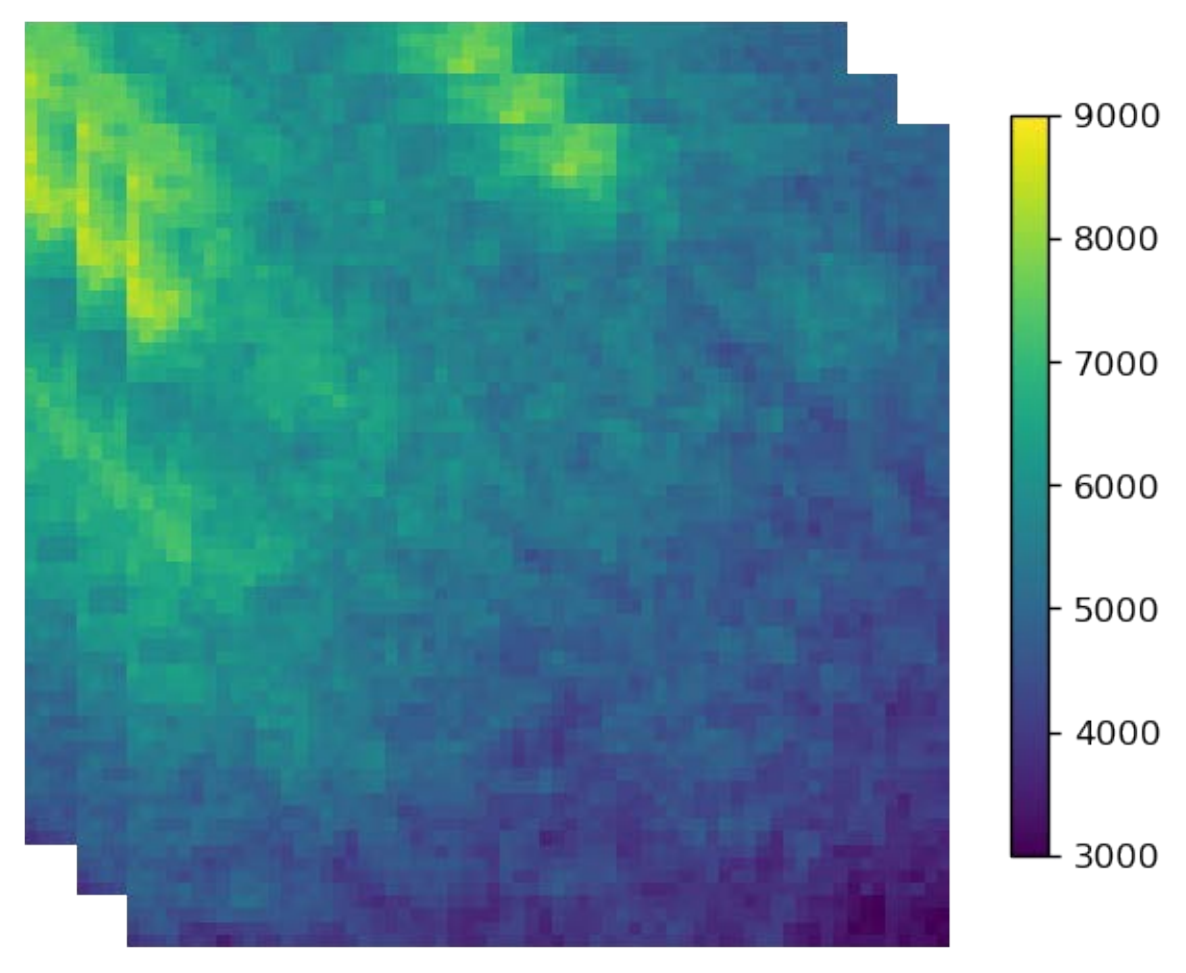}
        \includegraphics[width=0.33\textwidth]{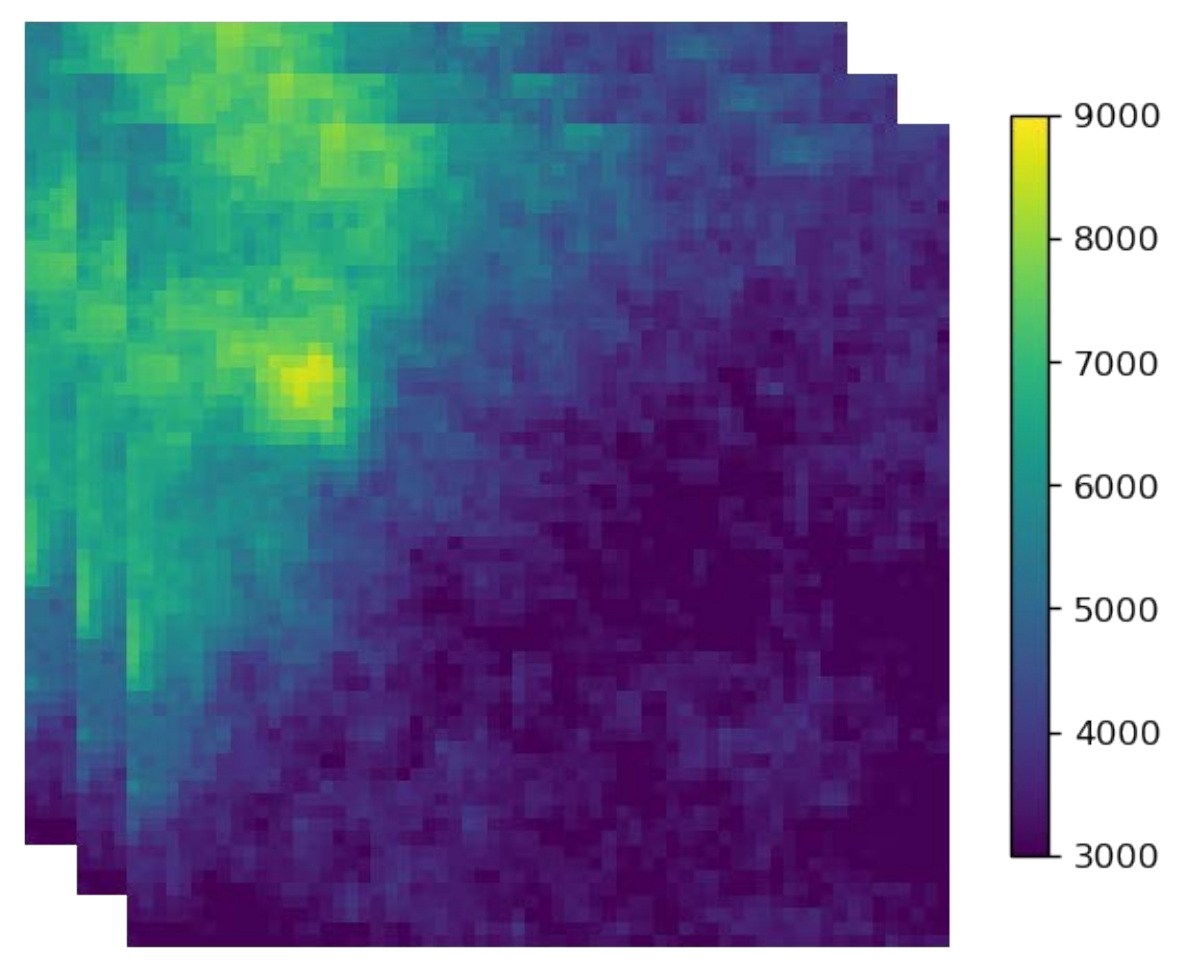}\label{sim_sky_map}}\\
    \subfigure[]
        {\includegraphics[width=0.33\textwidth]{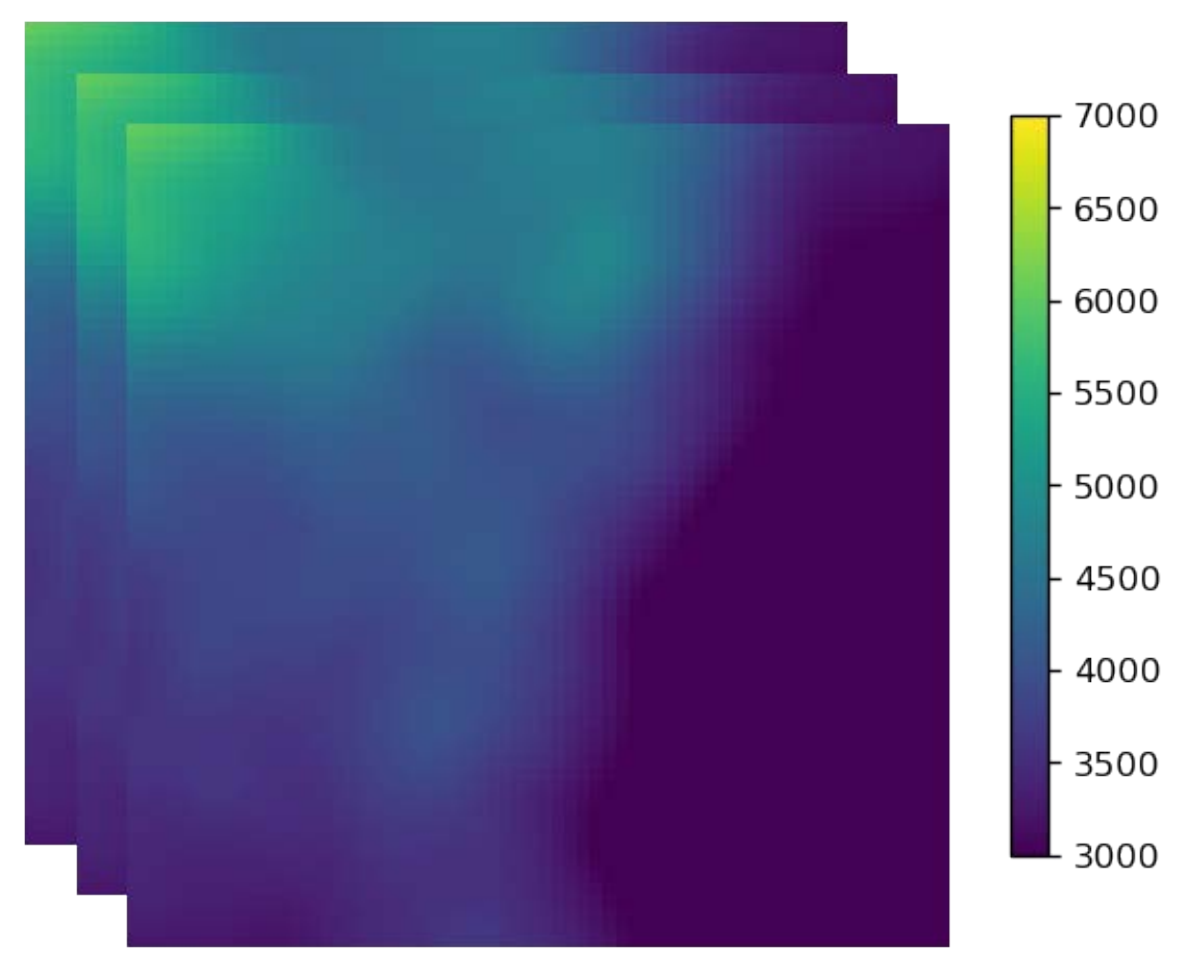}
        \includegraphics[width=0.33\textwidth]{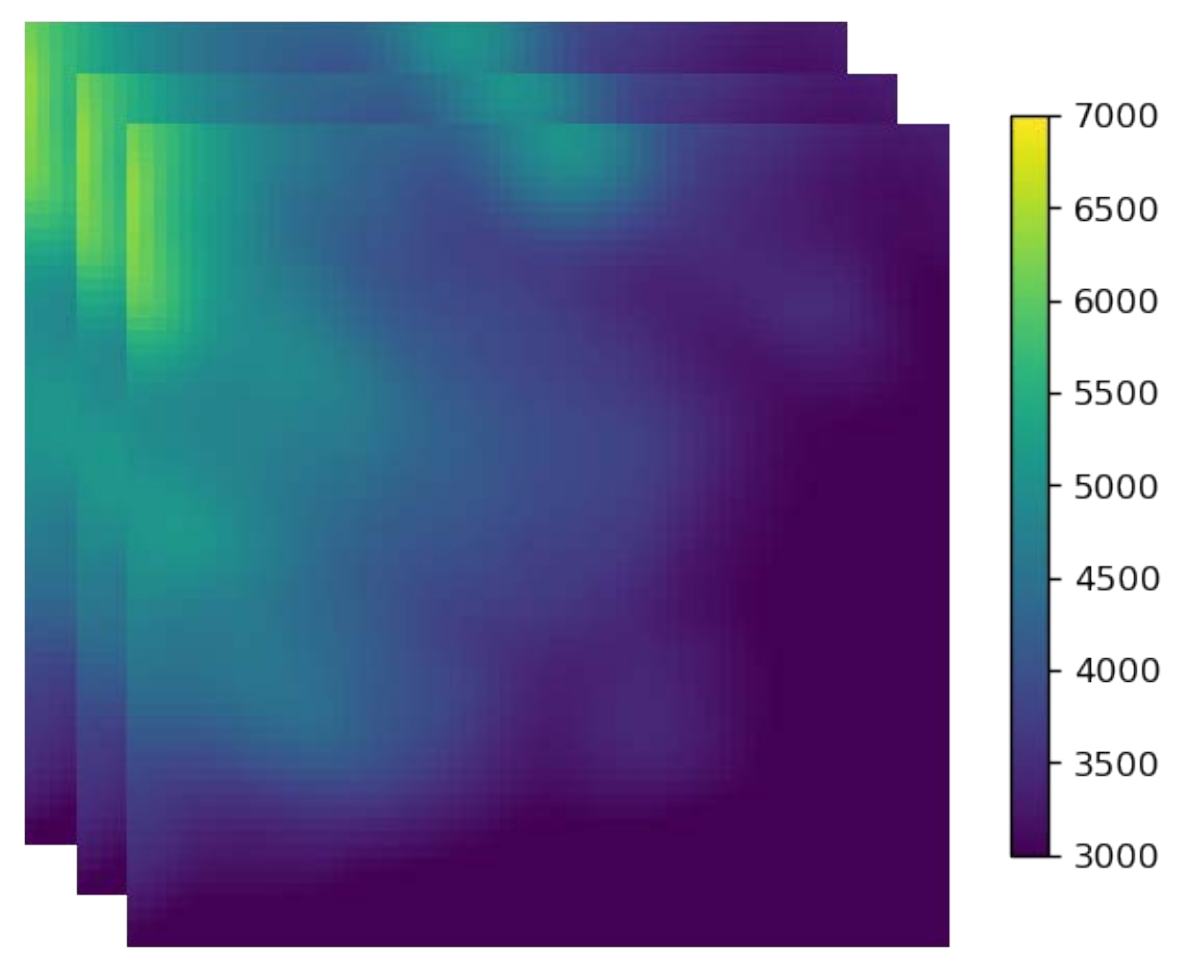}
        \includegraphics[width=0.33\textwidth]{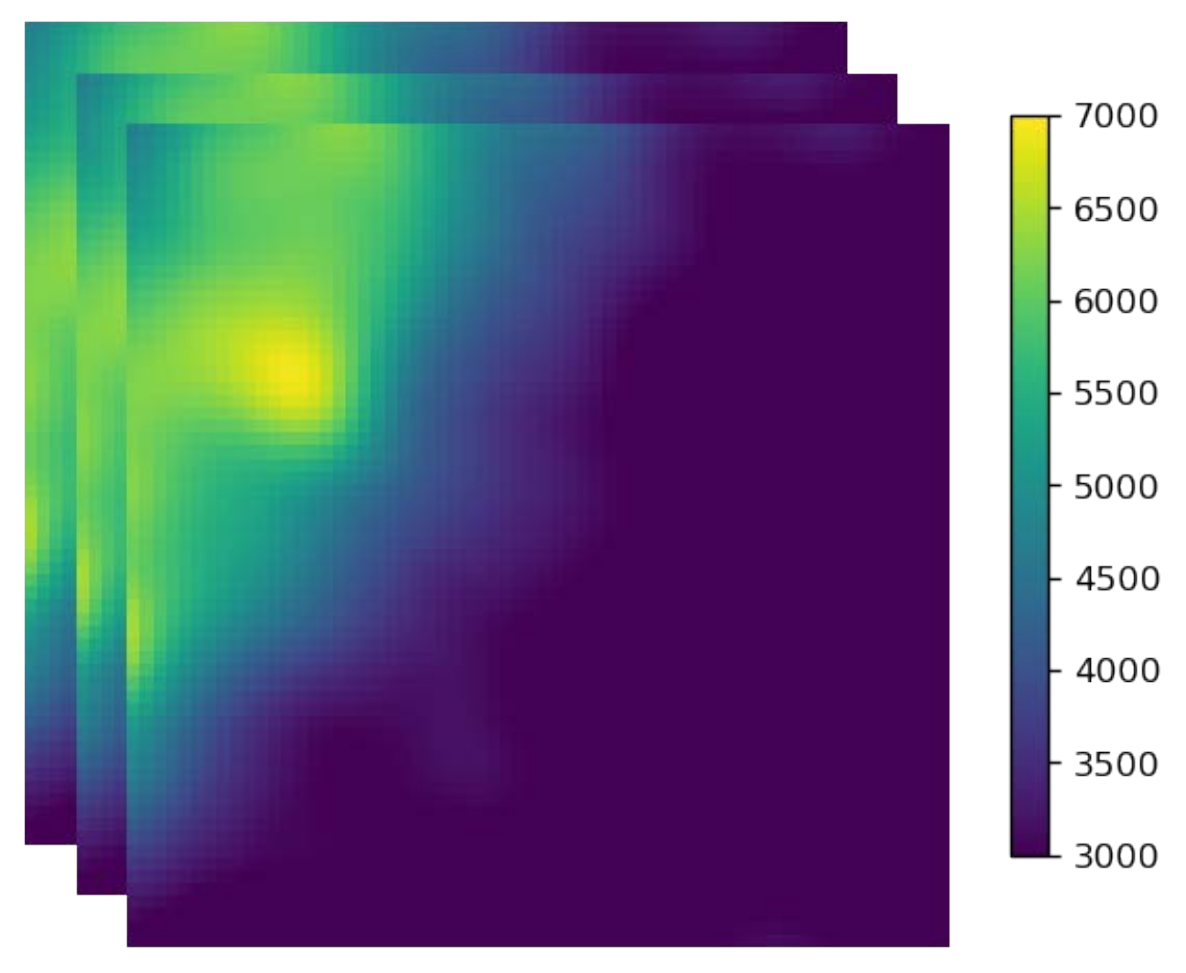}\label{sim_sky_map_G}}\\
    \subfigure[]
        {\includegraphics[width=0.33\textwidth]{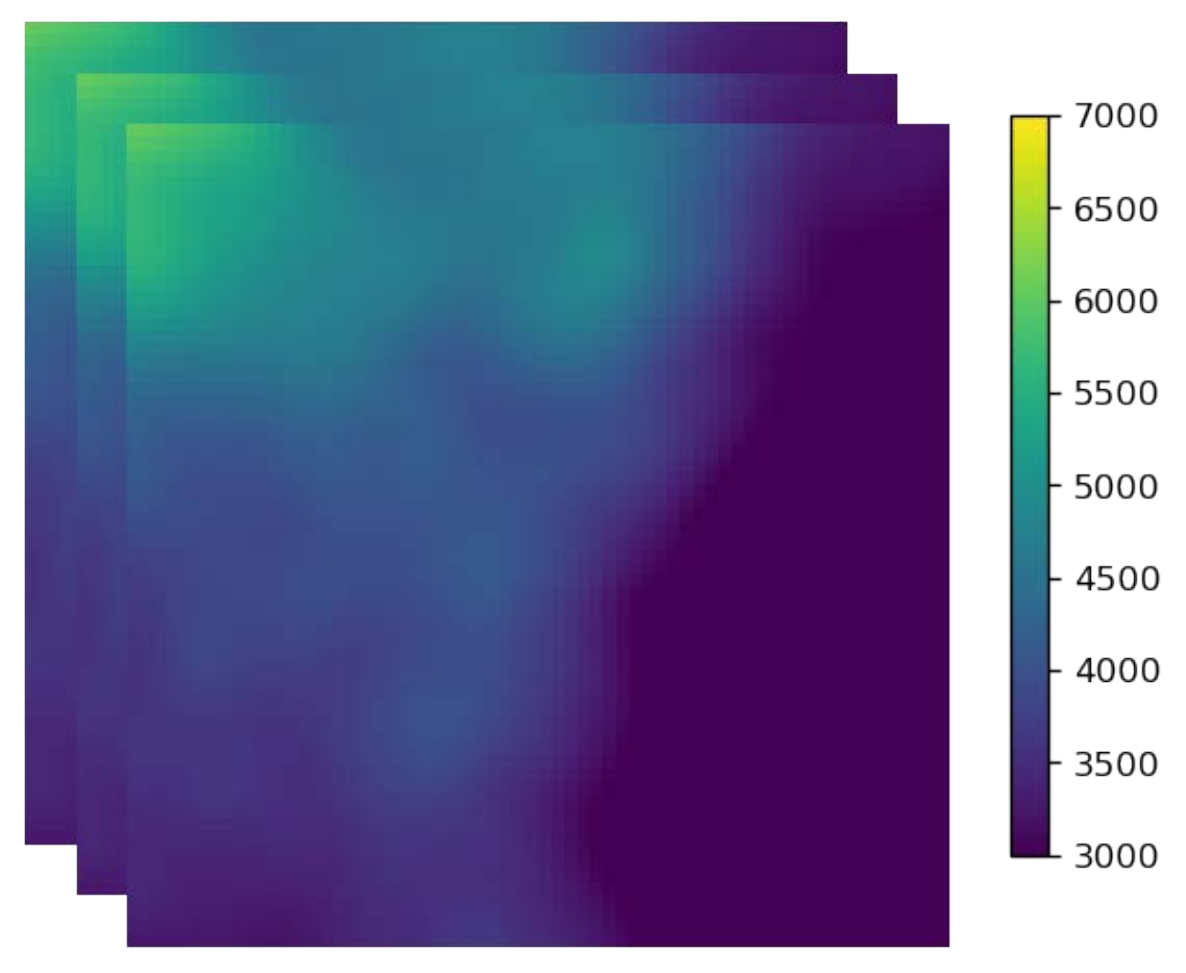}
        \includegraphics[width=0.33\textwidth]{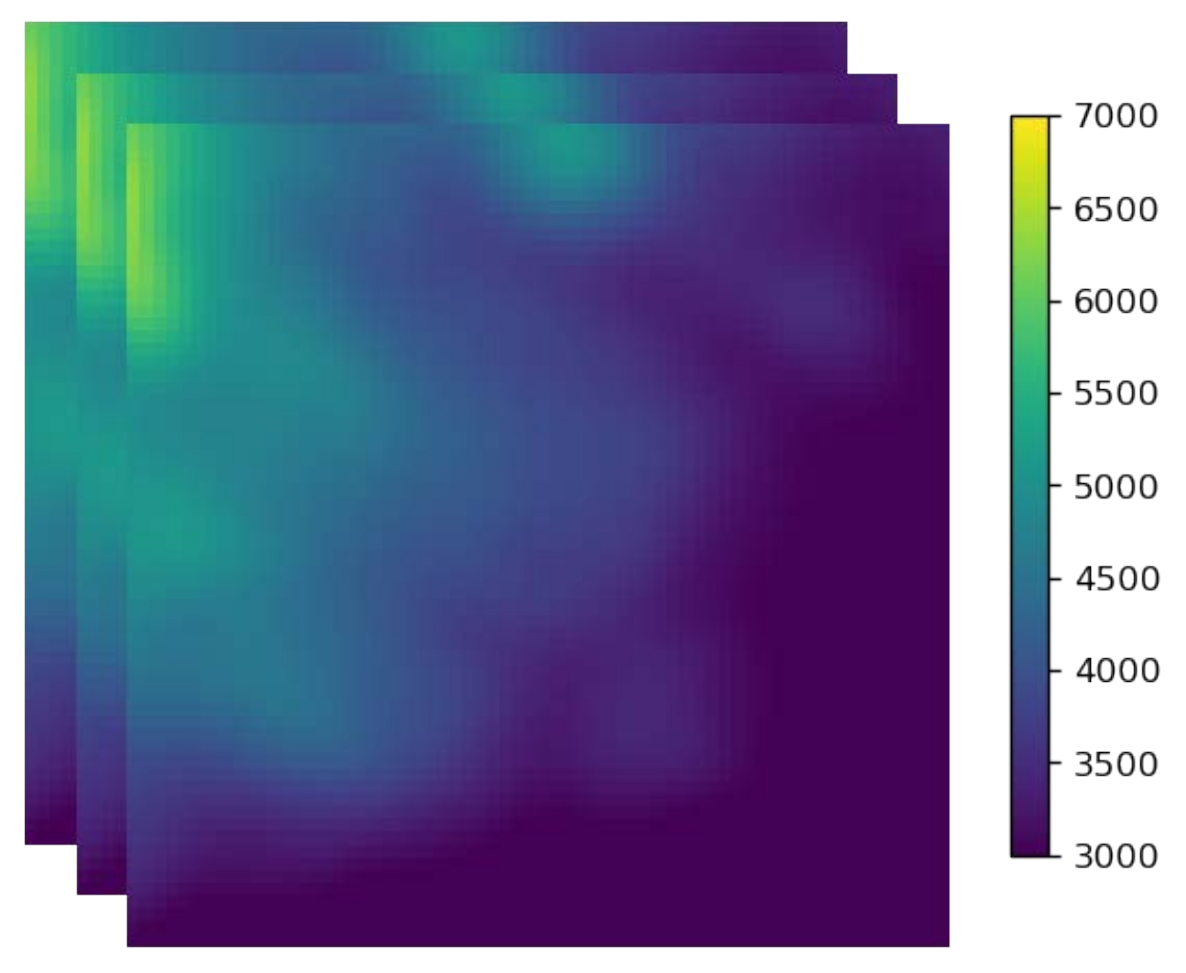}
        \includegraphics[width=0.33\textwidth]{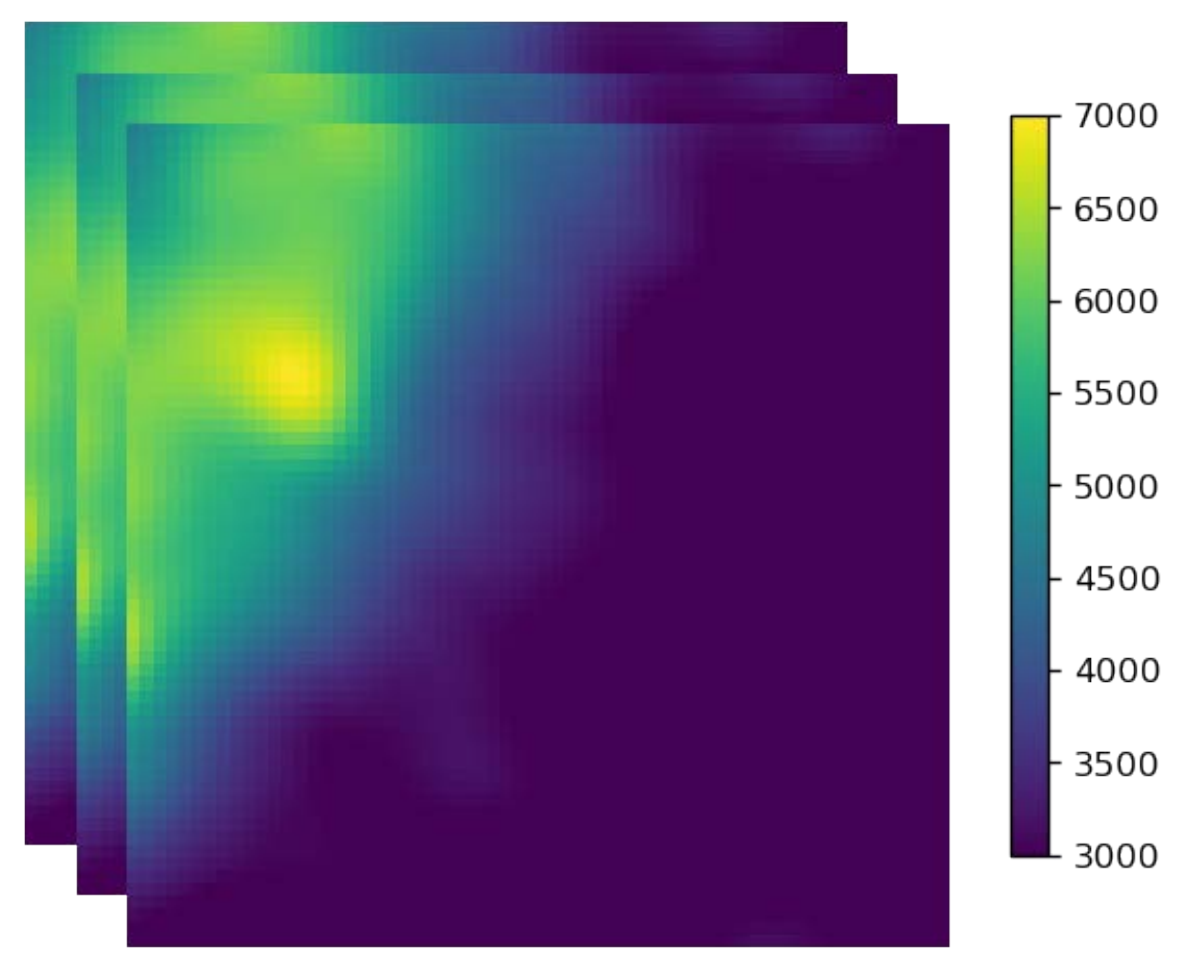}\label{sim_sky_map_C}}

    \caption{Data cubes of $214.86~{\rm{deg}}^2$ sky patches.
    From left to right, the center of the sky patches are
    $(l, b) = (179.648^\circ, 33.331^\circ)$, $(l, b) = (224.824^\circ, 9.594^\circ)$
    and $(l, b) = (303.398^\circ, 2.836^\circ)$, where $l$ and $b$ are the
    Galactic longitude and Galactic latitude, respectivly. The redshift range of the sky patches
    is $z\in[0.353, 0.578]$. The unit of the sky patches is mK.
     Panel (a) shows the raw simulated sky patch without beam convolved,
    Panel (b) shows the Gaussian beam convolved sky patch, and Panel (c) shows the Cosine beam convolved sky patch.}\label{fig:patch}
\end{figure*}

The core algorithm of U-Net is semantic segmentation, which is an approach detecting the belonging class of the object for every pixel~\citep{Ronneberger:2015UNetCN,Makinen:2020gvh,Villanueva-Domingo:2020wpt}. It can only handle 2D flat images or 3D sky patches, but not curved images or sky patches directly. 
As shown in the literature~\citep{Makinen:2020gvh}, the smooth components of the foreground contamination can be removed efficiently.
However, the non-smooth foreground residual, which is mainly caused due to the systematic effect,
challenges the detection of HI LSS signal.
The foreground residual can be further reduced with model independent foreground subtraction method, such as PCA. Since the smooth components of the foreground are not the major problem and can be efficiently removed with the
current foreground cleaning method, we first perform PCA on the simulated data before feeding to the U-Net architecture.

{On the other hand, the preprocessing with PCA is necessary. Without the preprocessing step, 
the U-Net performs poorly in recovering the HI fluctuation signal.
It is primarily due to the large dynamic range of the map amplitude \citep{Liu:2014bba,deOliveira-Costa:2008cxd,DiMatteo:2001gg}, 
which makes extracting correct information with the U-Net network difficult.
However, an aggressive PCA mode subtraction can result in HI signal loss~\citep{Switzer:2015ria}.
According to the simulation in the literature, the smooth foreground contamination 
can be significantly removed with a few PCA modes \citep{2022MNRAS.509.2048S}.
To avoid any HI signal loss, we subtract only the first three PCA modes. 
Subtracting a few more PCA modes, such as the first six PCA modes 
in \citet{Makinen:2020gvh}, can potentially improve the benchmark. 
As long as the U-Net can handle the residuals, we wish to keep the 
preprocessing step to a minimum of mode subtraction.
However, in practice, the preprocessing needs to be further improved because the actual foreground contamination is more complex than the simulated one.}


The 3D U-Net architecture requires equal-area square image data with HEALPix pixelization.
The full-sky map needs to be divided into several small sky patches.
We split the full sky map into $192$ equal-area patches corresponding to the pixel area of HEALPix Nside of $4$.
Each patch contains $64\times64$ pixels.
The full frequency range, i.e. $900\sim1050$ MHz, is binned into $64$ frequency bands.
Finally, the simulated full sky is split into $192$ sky patches with shape of $64\times64\times64$.
Three sky patches, as the example, are illustrated in Figure~\ref{fig:patch}.
The sky patches from the raw simulated sky map, Gaussian beam convolved sky map and Cosine beam convolved
sky map are shown in the top, middle and bottom panels, respectively.
We use the same sky segmentation method for the training, test, and validation datasets.

{The direction-dependent spectral index is used in our simulation. The PCA is performed by taking the Singular Value Decomposition (SVD) of the input data cube. The data cube has already been reshaped into $N_\nu \times N_\theta$, where $N_\nu$ is the number of frequency channels and $N_\theta$ is the number of pointing directions. The singular modes are used as the basis functions as expressed in Equation (3.2) of \citet{Makinen:2020gvh} and the first few modes represent the foreground contamination. It is not necessary to assume each of the foreground components is separable in frequency and angular direction. The foreground modes estimated with PCA is not exactly the same sets of foreground components as the simulation input. Instead, it is the linear combination of the input components. As long as the foreground components can be expressed in the form of power law spectrum, they can be significantly subtracted.}

\subsection{Loss function}
Although the dynamic amplitude range is significantly reduced with the PCA mode subtraction,
there is still large variability between the residual foreground and HI signal.
Following \citet{Makinen:2020gvh}, the Logcosh loss function is used in our analysis,
\begin{equation}\label{equ:loss_fun}
    L(p, t)=\sum_i\log\cosh(p_i-t_i),
\end{equation}
where $p_i\in p$ denotes prediction and $t_i\in t$ denotes simulation target.
The Logcosh function behaves much like the L1 norm in that it is more robust and
less sensitive to outliers, which is the main reason we chose it.

\subsection{Training and testing}
\begin{table*}
\caption{Description of the hyperparameters in the U-Net architecture design.}
\centering
	\begin{tabular}{llcc}
		\hline\hline
		Hyperparameter    & Description                            & Prior Value                     & Optimum Value\\
		\hline
		$\eta$            & learning rate for optimizer            & $10^{-2}, 10^{-3}, 10^{-4}, 10^{-5}$ & $10^{-4}$  \\
		$\omega$          & weight decay for optimizer             & $10^{-4}, 10^{-5}, 10^{-6}, 10^{-7}$ &$10^{-5}$   \\
		$n_{\rm filter}$ & initial number of convolution filters  & 8, 16, 32                            & 32   \\
		$b$               & batch size, i.e. number of samples per gradient descent step  & 8, 16  & 16   \\
		$\Omega$          & optimizer for training & Adam, NAdam     & NAdam   \\
		\hline
	\end{tabular}
	\label{tab:unet}
\end{table*}

\begin{figure*}
    \centering
    \subfigure[]
    {\includegraphics[width=0.495\textwidth]{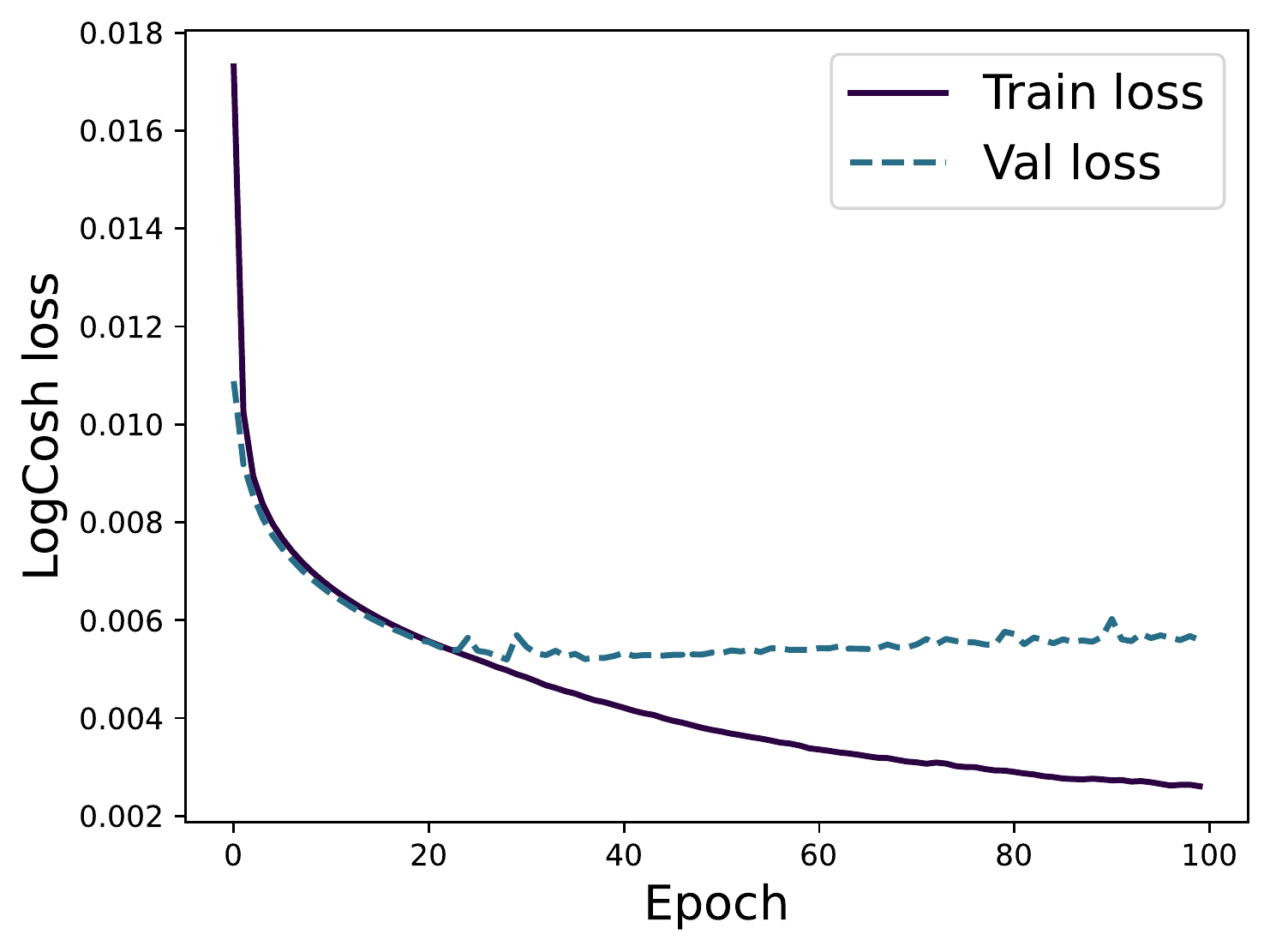}\label{fig:rate_gaussian}}
    \subfigure[]
    {\includegraphics[width=0.495\textwidth]{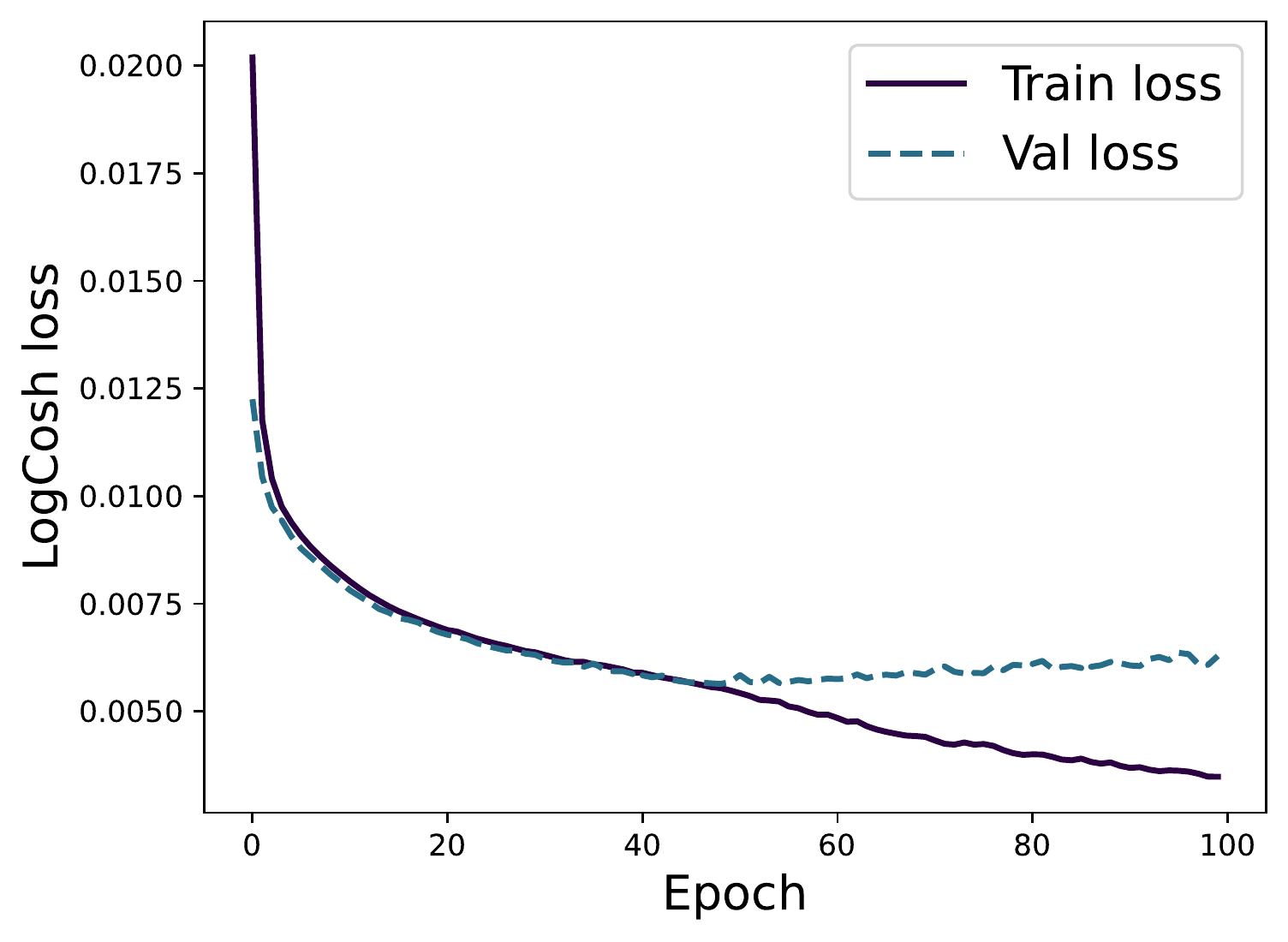}\label{fig:rate_cosine}}
  \caption{Loss function evolution for each network as a function of the number of epochs. The dark blue solid line indicates the training set loss function evolution and the light blue dashed line indicates the validation set loss function evolution. Panel (a) is U-Net foreground deduction for Gaussian beam convolved maps, and Panel (b) is U-Net foreground deduction for Cosine beam convolved maps.}
  \label{fig:rate_down}
\end{figure*}

\begin{figure*}
    \centering
    \subfigure[]
    {\includegraphics[width=0.495\textwidth]{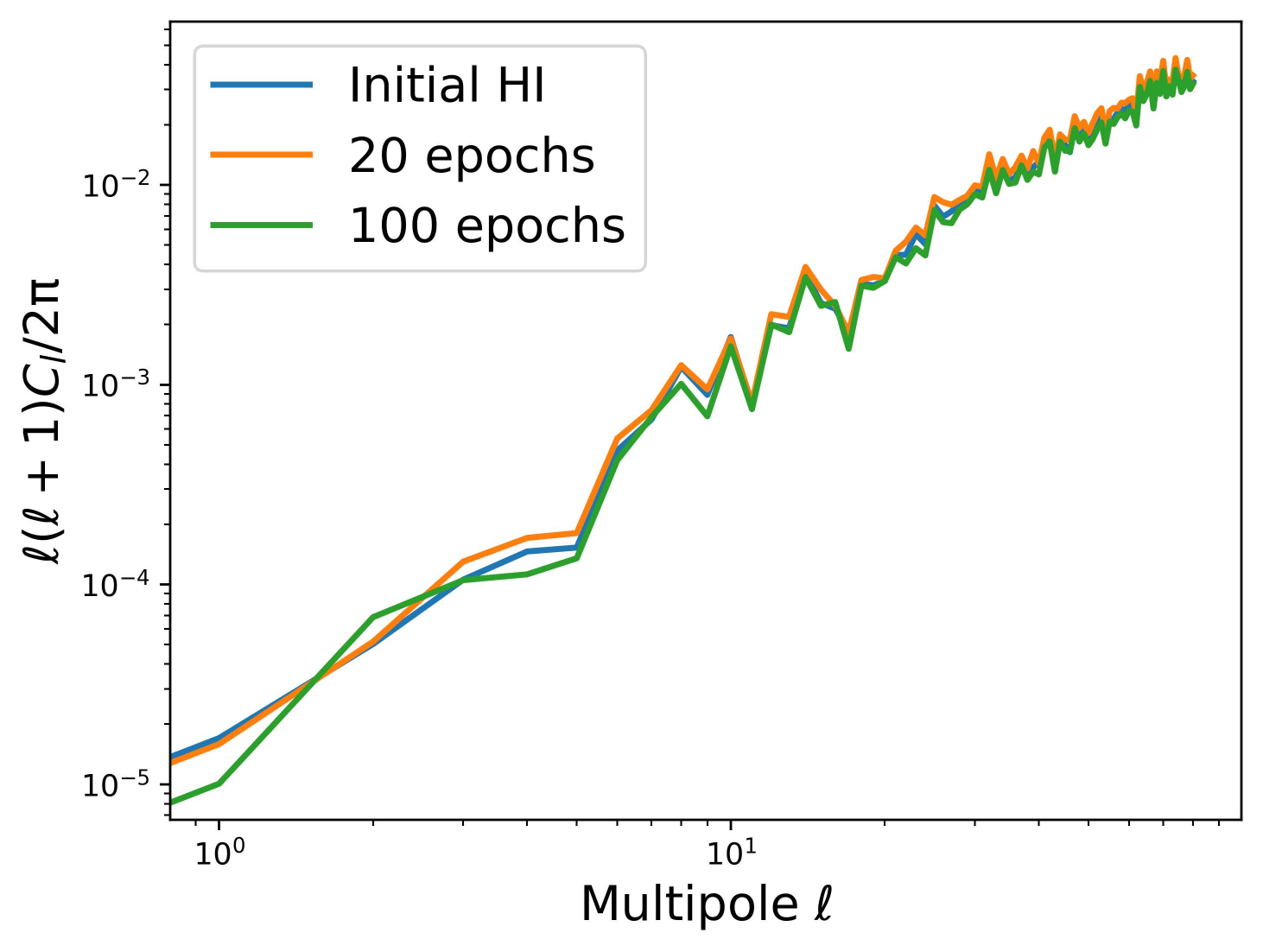}\label{fig:result_100}}
    \subfigure[]
    {\includegraphics[width=0.495\textwidth]{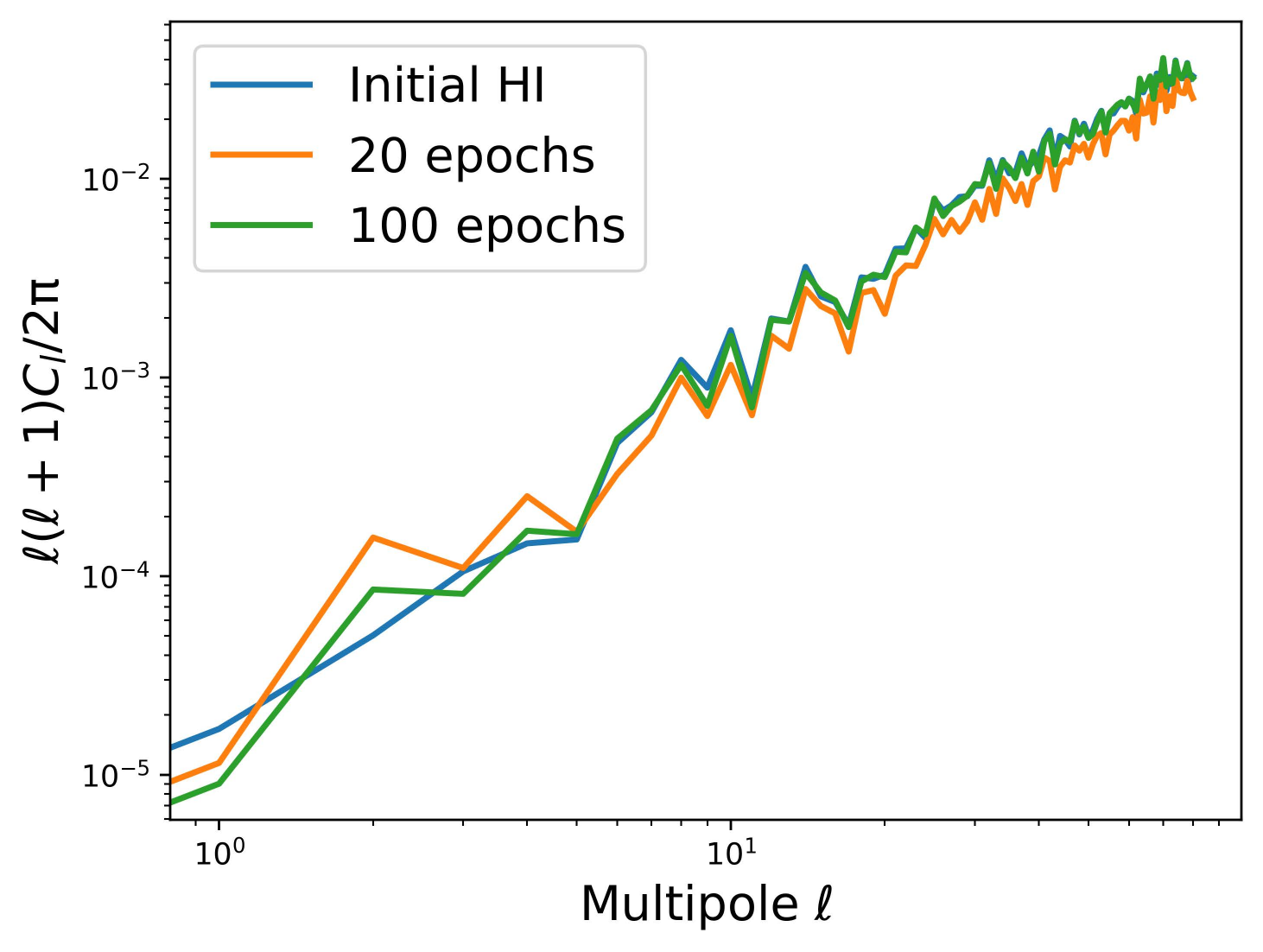}\label{fig:result_100}}
    \caption{The angular power spectra of the simulated sky maps after PCA+U-Net foreground subtraction. Panel (a) shows the angular power spectra from the Gaussian beam convolved sky map, and Panel (b) shows the angular power spectra from the Cosine beam convolved sky map.}
    \label{fig:result_gc_100}
\end{figure*}


{The core of any CNN is the convolutional layers which are defined by the convolution of a number of filters on the input.
The network is basically a stack of layers.
Following~\citet{Makinen:2020gvh}, we fixed the number of convolution kernels to be $32$. 
The kernel size defines the field of view of the convolution, which is fixed to $3\times3\times3$. 
To obtain the required output dimensionality, we used ``same" padding to handle the boundaries of the samples in the convolutions and transpose convolutions. 
The stride defines the step size of the kernel traversing the images. 
We use the default settings of stride equal to 1 in the convolutions and stride equal to 2 in the transpose convolutions.}


{Our selected U-Net architecture is trained end-to-end to the foreground removal using a set of HI foreground simulations. 
Table~\ref{tab:unet} presents the details of the hyperparameters used in this network. 
The NAdam optimizer is used in this analysis with the default TensorFlow parameters~\citep{Sashank:2019abs}.
The hyperparameters are carefully fine-tunned to optimize the network.  
The batch size is optimized to $16$ and the number of initial convolution filters is optimized to $32$, 
both of which are limited by the GPU memory. 
The initial learning rate is set to $10^{-2}$ and decayed by a factor of $10$, 
if the validation loss function is not improved after 20 epochs.
In the meanwhile, the weight decay is also test with a set of prior values as list in Table~\ref{tab:unet}.
Finally, we find that optimized values for learning rate and weight decay are $10^{-4}$ and $10^{-5}$, respectively.
The total number of trainable parameters is $7.4\times10^{7}$. 
Following~\citet{Makinen:2020gvh}, we apply a rectified linear unit (ReLU) activation in every convolution. 
In Figure~\ref{fig:rate_down}, we show the evolution of the loss function.The dark blue solid line indicates the training set loss function evolution and the light blue dashed line indicates the validation set loss function evolution. 
The results for Gaussian beam convolved maps and Cosine beam convolved maps are shown in the left and right panels, respectively.} 

{The loss function of the validation set is converged after about $20$ epochs and $50$ epochs for Gaussian and Cosine beam convolved maps, respectively. The recovered power spectra in the cases with $20$ and $100$ epochs are shown in Figure~\ref{fig:result_gc_100}, where the results for Gaussian maps are shown in the left panel and the results for Cosine maps are shown in the right panel. Significantly, the recovered angular power spectra from Gaussian beam convolved maps are consistent between the cases with $20$ and $100$ epochs, which is because the loss function is converged after $20$ epochs. Although the loss function in the Cosine case converges after about $50$ epochs, the improvement in recovering the power spectrum is not obvious after $20$ epochs. Therefore, in our analysis, we adopt a 20-epoch calculation scheme to make our computations faster.}

After completed the training, we investigated the accuracy of the network's prediction. 
In Figure~\ref{fig:result_gc}, we show the foreground subtraction results for PCA and PCA+U-Net in the presence of Gaussian and Cosine instrumental effects.

\section{Results and discussion}\label{sec:results}

\begin{figure*}
    \centering
    \subfigure[]
    {\includegraphics[width=0.495\textwidth]{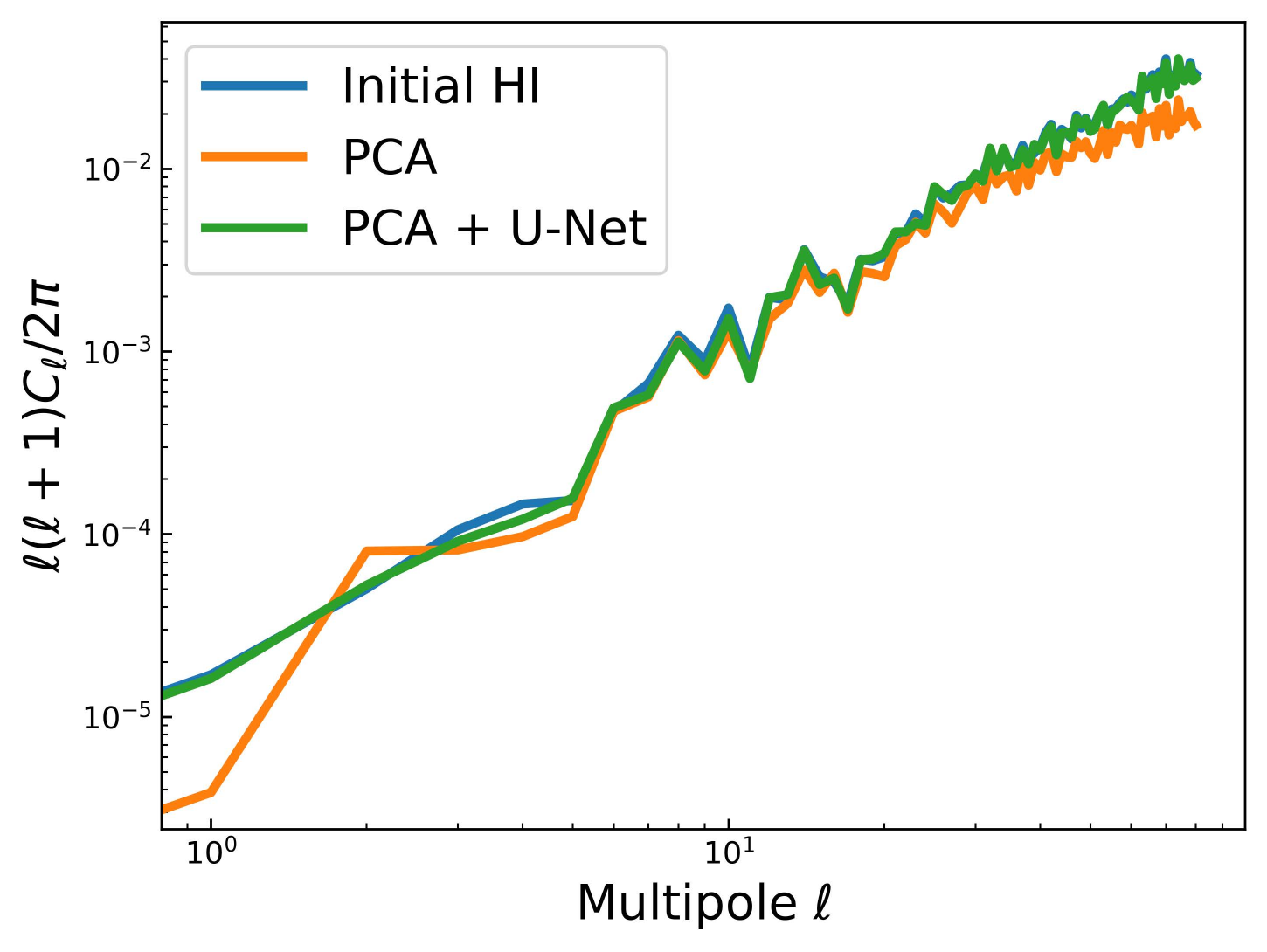}\label{fig:result_g}}
    \subfigure[]
    {\includegraphics[width=0.495\textwidth]{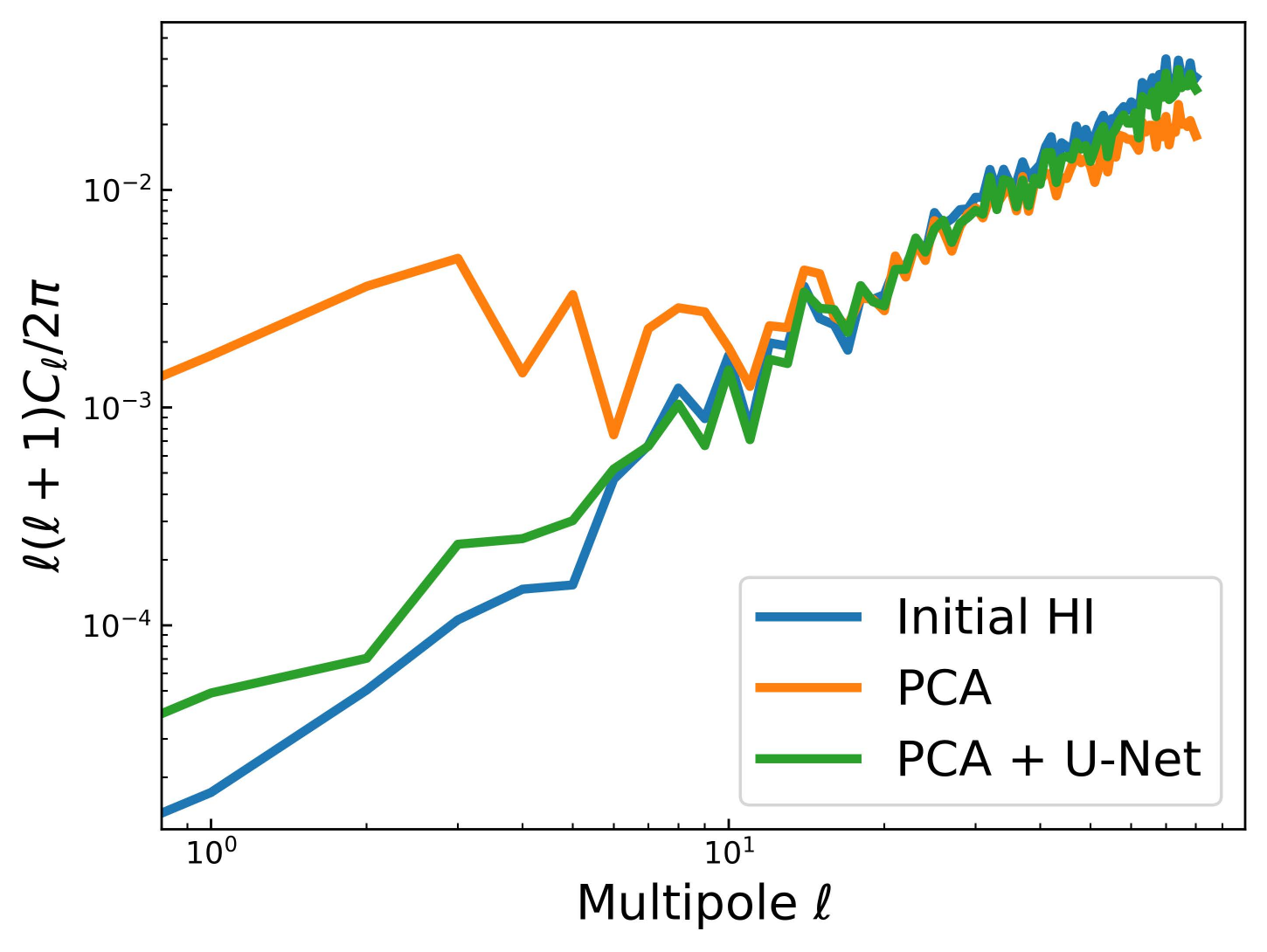}\label{fig:result_c}}
  \caption{
  The angular power spectra of the simulated sky maps after foreground subtraction.
  The blue, orange, and green lines show the results of the initial HI model, the foreground subtraction with the first 3 PCA modes, and the foreground subtraction with the combination of PCA and U-Net, respectively.
  Panel (a) shows the angular power spectra from the Gaussian beam convolved sky map, and Panel (b) shows the angular power spectra from the Cosine beam convolved sky map.
  } \label{fig:result_gc}
\end{figure*}

\begin{figure*}
    \centering
    \subfigure[]
    {\includegraphics[width=0.25\textwidth]{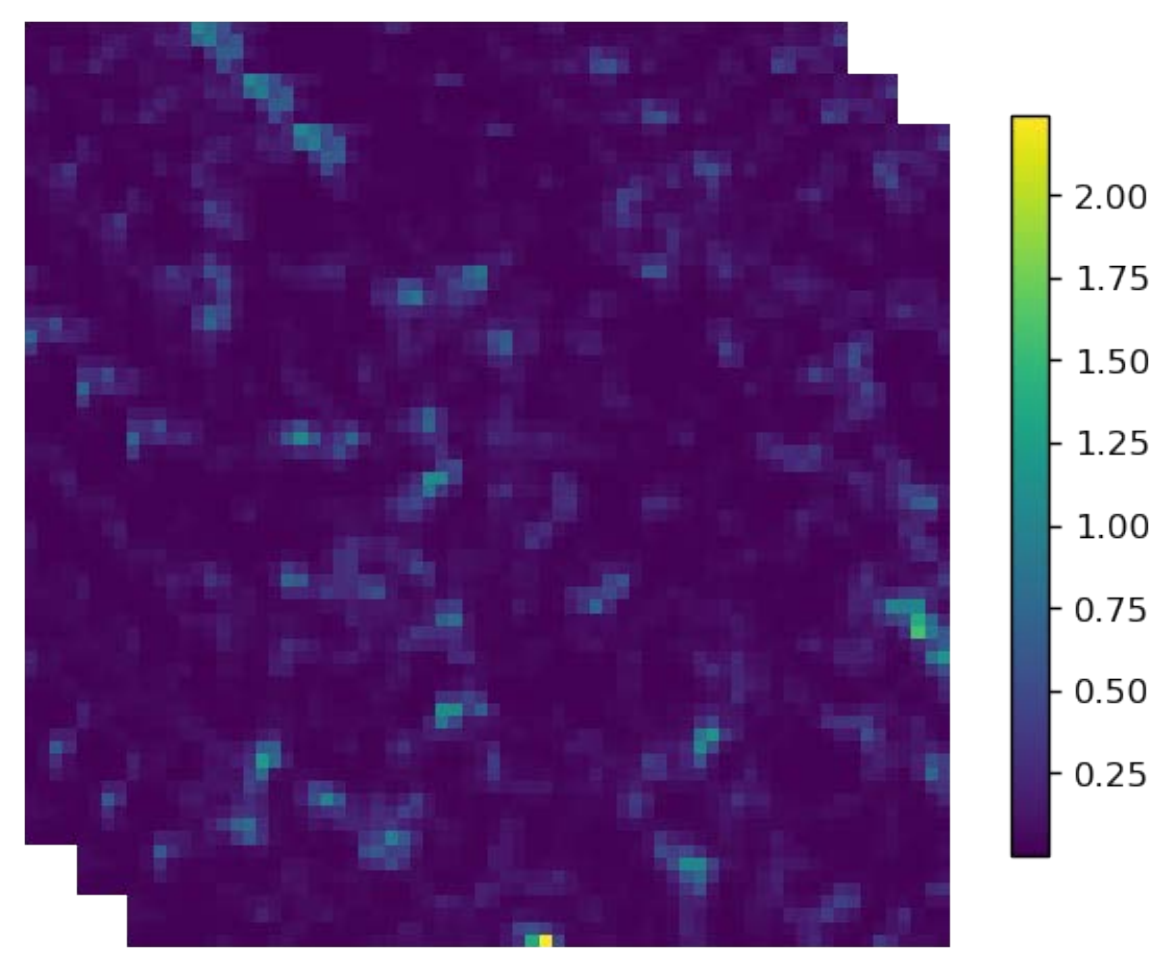}
    \includegraphics[width=0.25\textwidth]{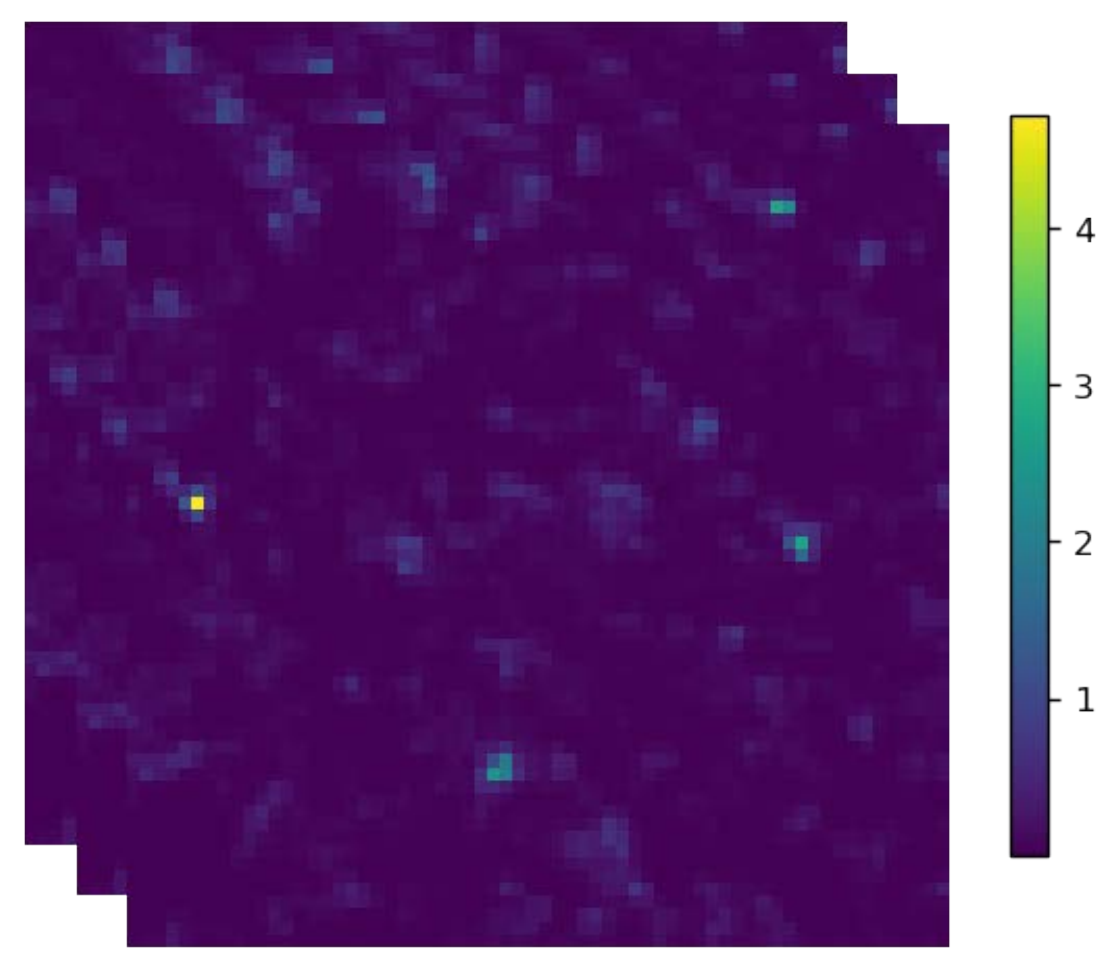}
    \includegraphics[width=0.25\textwidth]{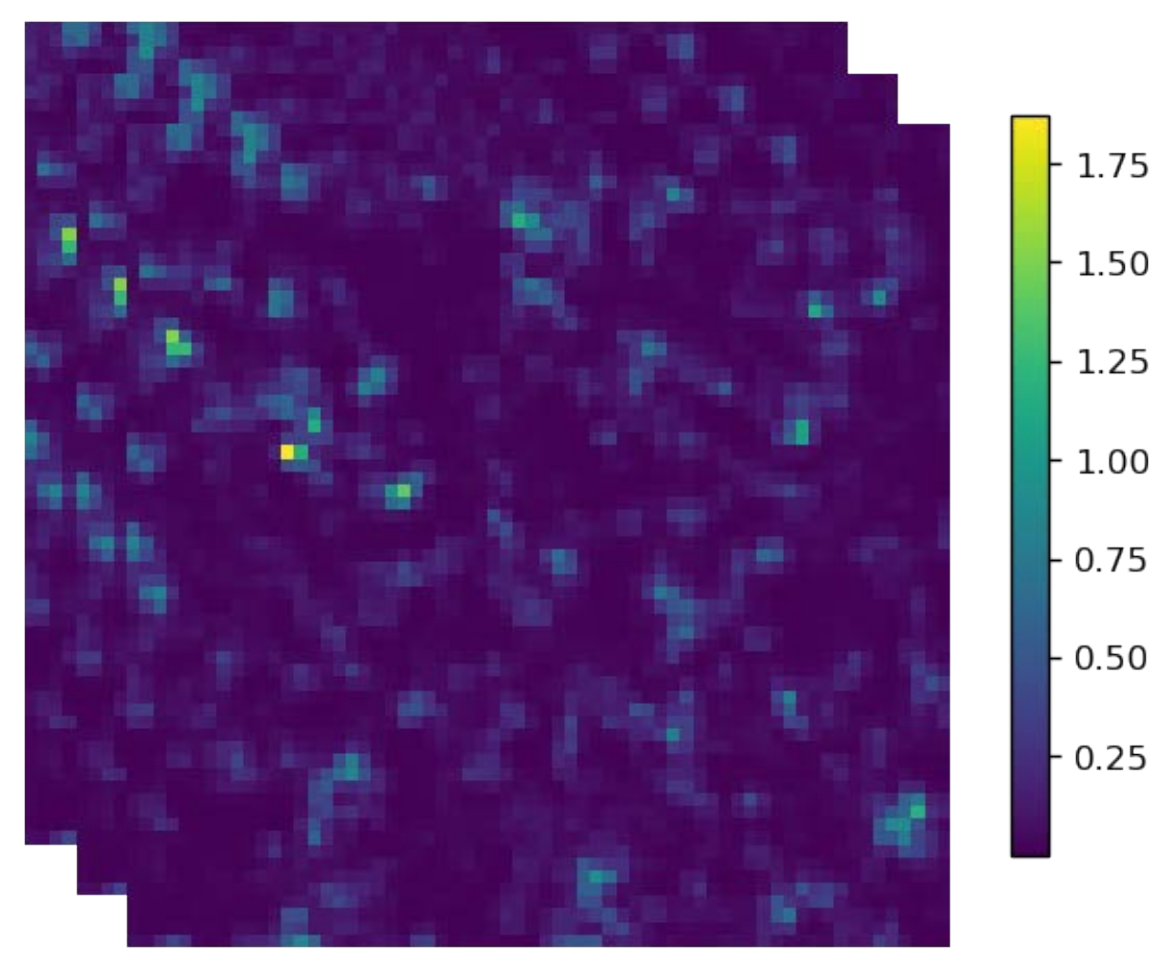}\label{fig:HImap}}\\
    \subfigure[]
    {\includegraphics[width=0.25\textwidth]{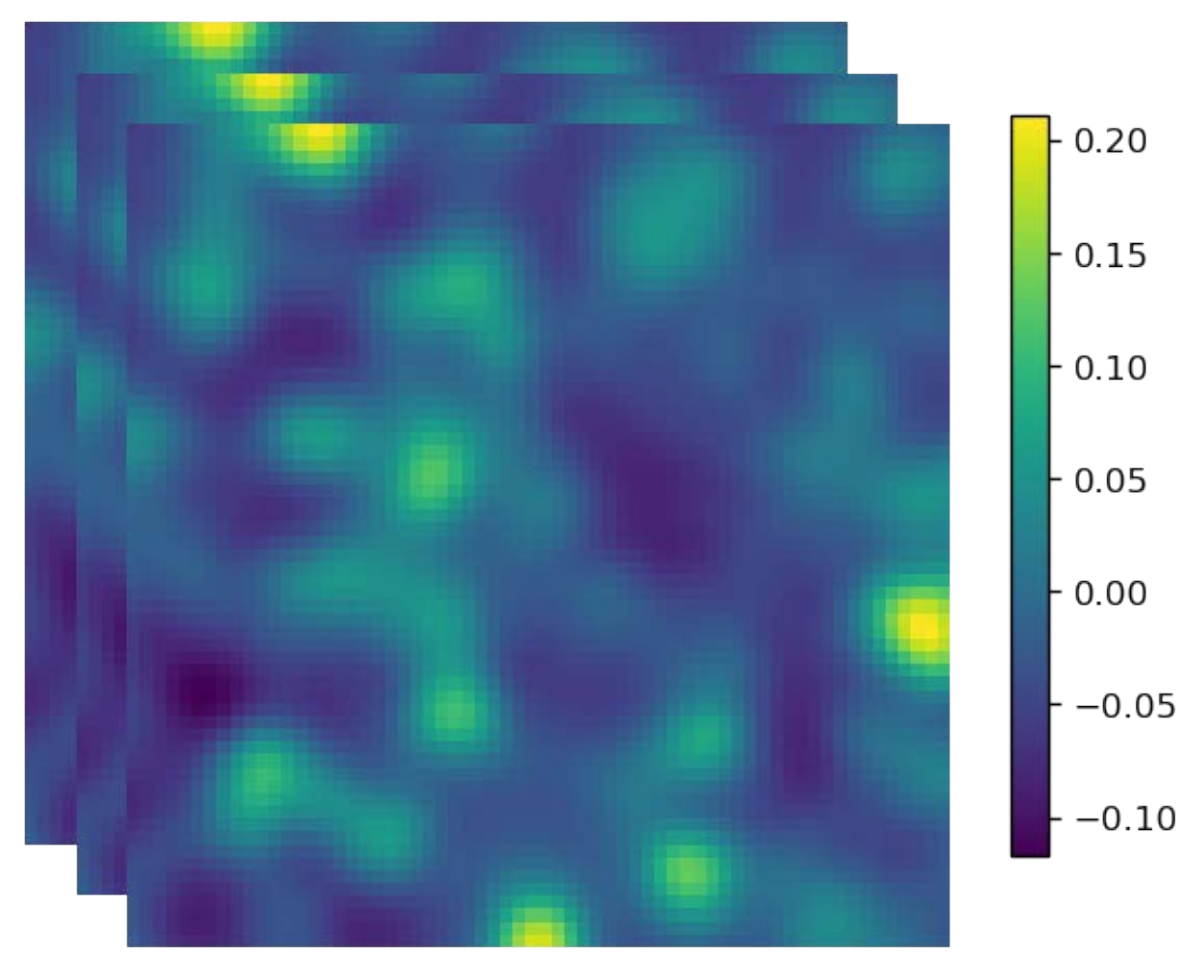}
    \includegraphics[width=0.25\textwidth]{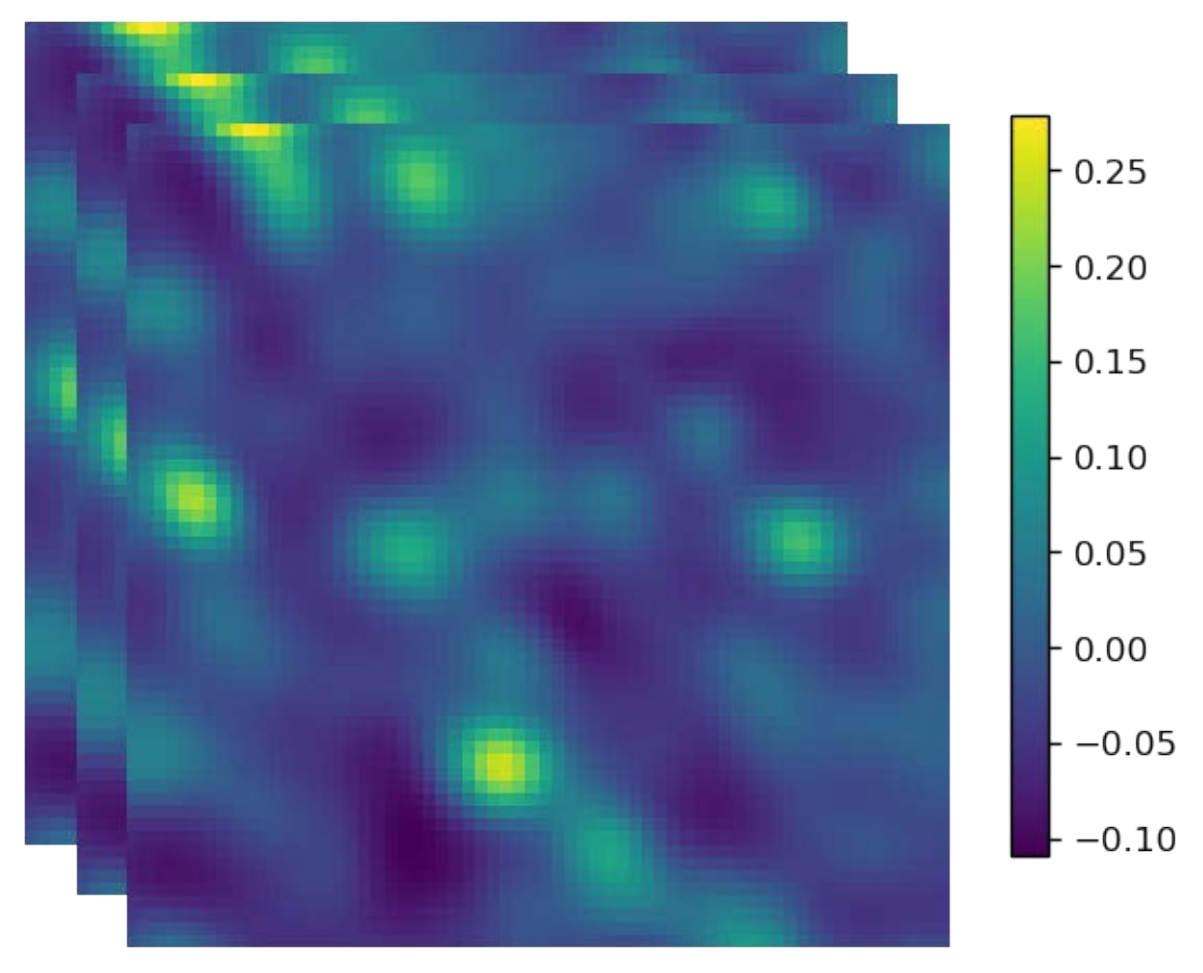}
    \includegraphics[width=0.25\textwidth]{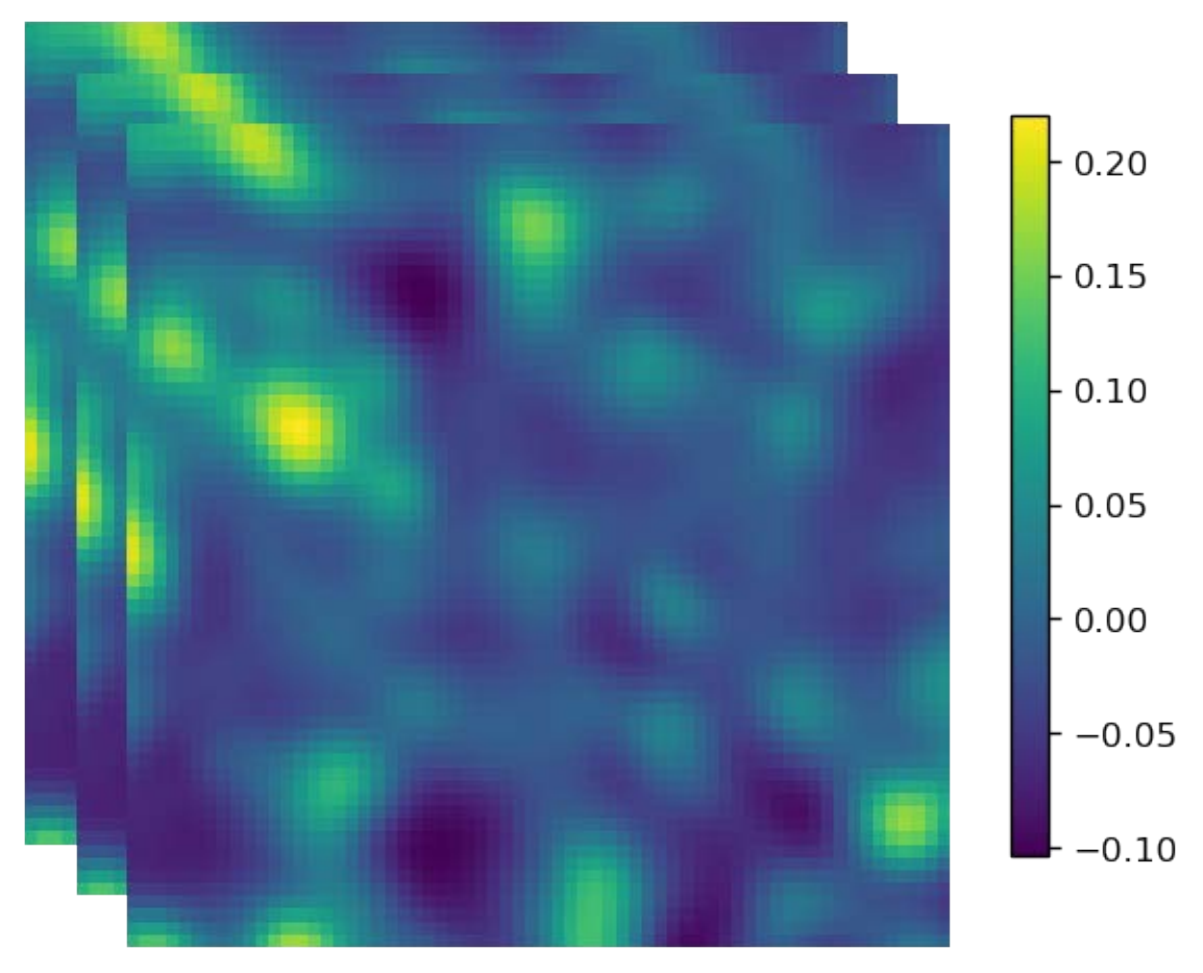}\label{fig:mapGpca}}\\
    \subfigure[]
    {\includegraphics[width=0.25\textwidth]{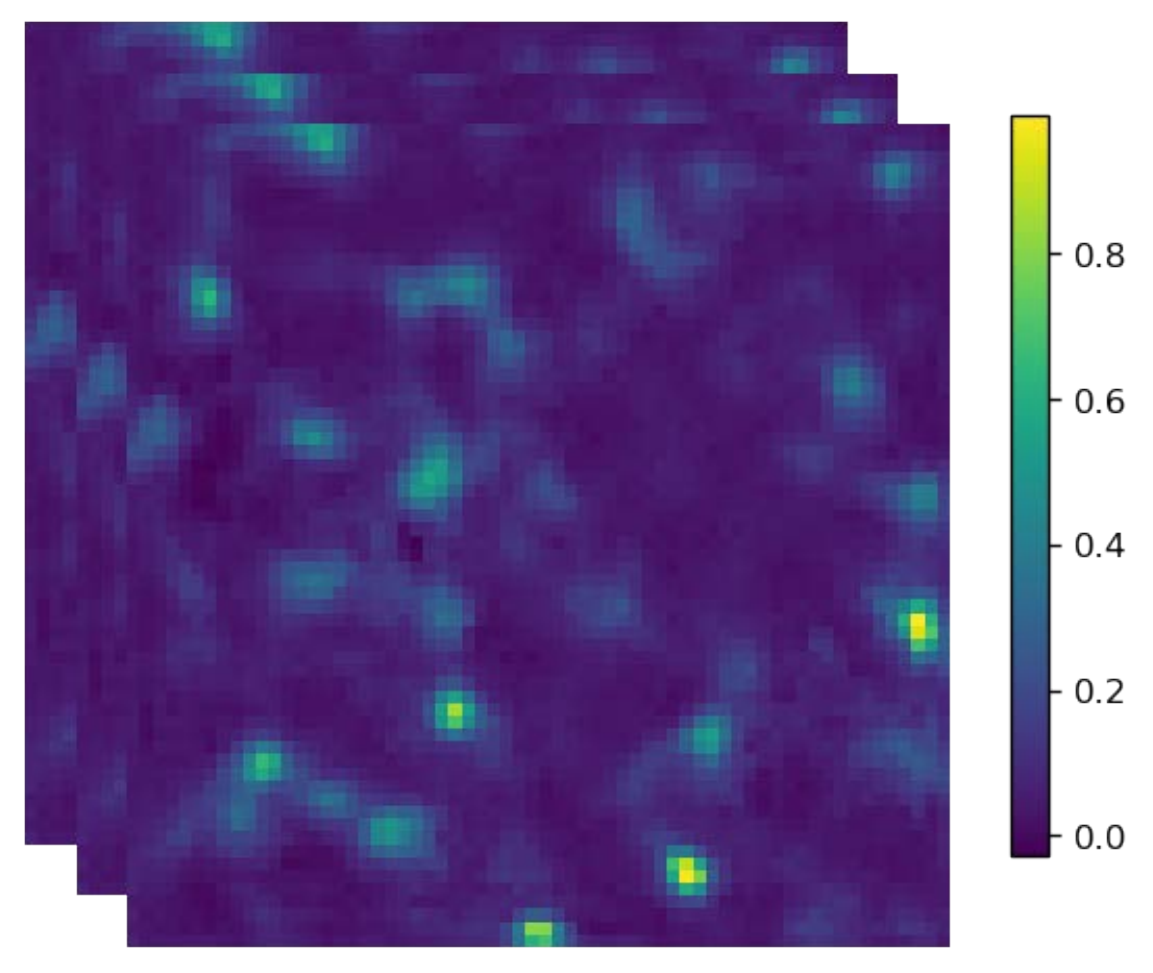}
    \includegraphics[width=0.25\textwidth]{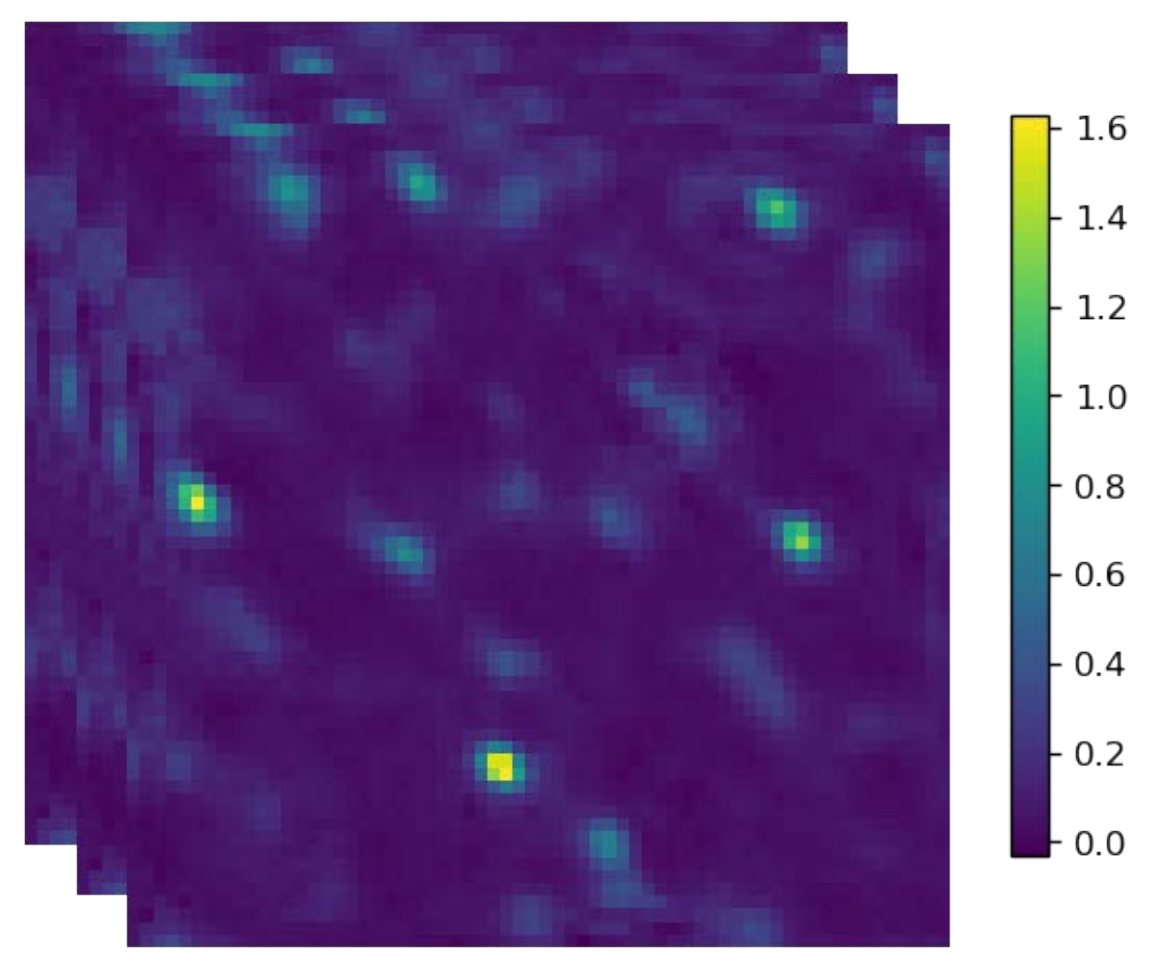}
    \includegraphics[width=0.25\textwidth]{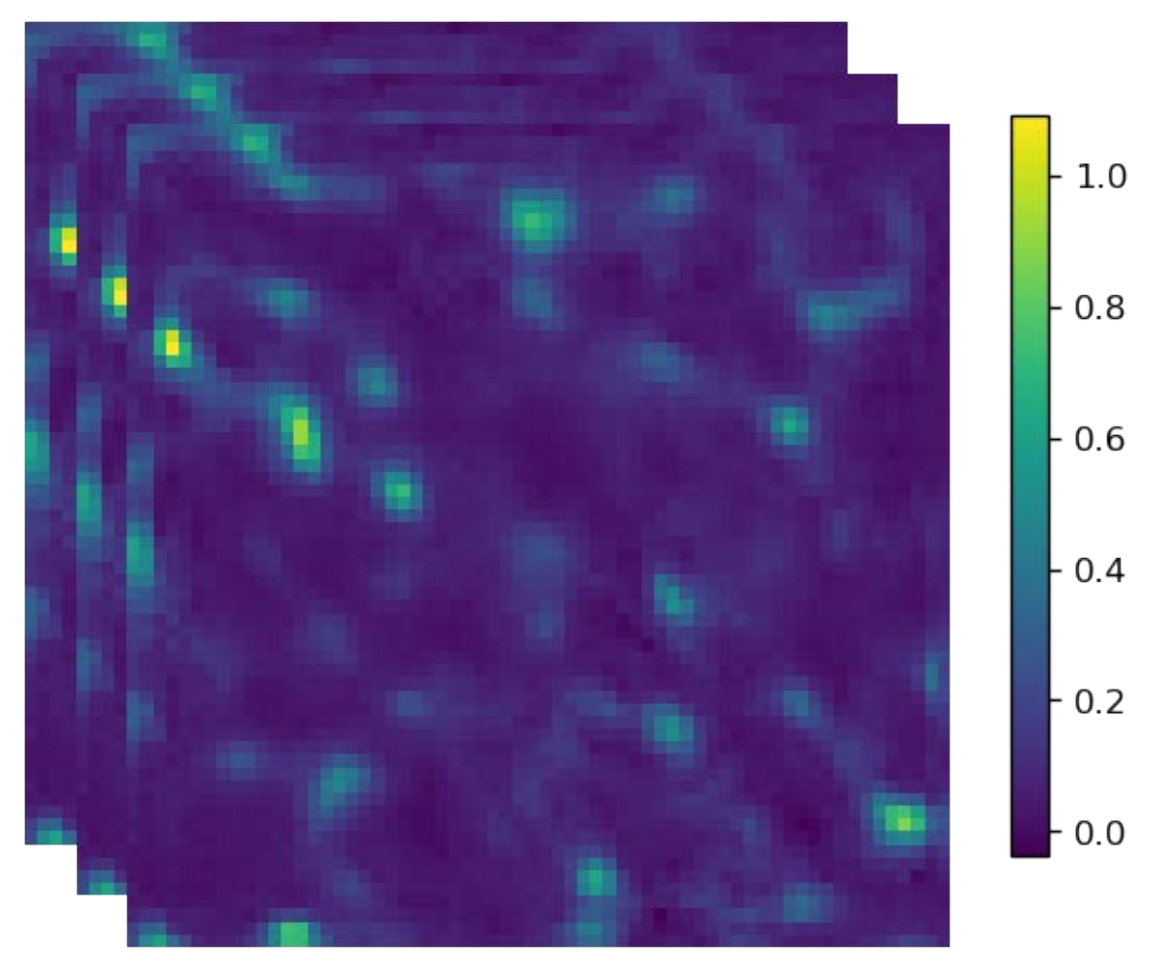}\label{fig:mapGunet}}\\
    \subfigure[]
    {\includegraphics[width=0.25\textwidth]{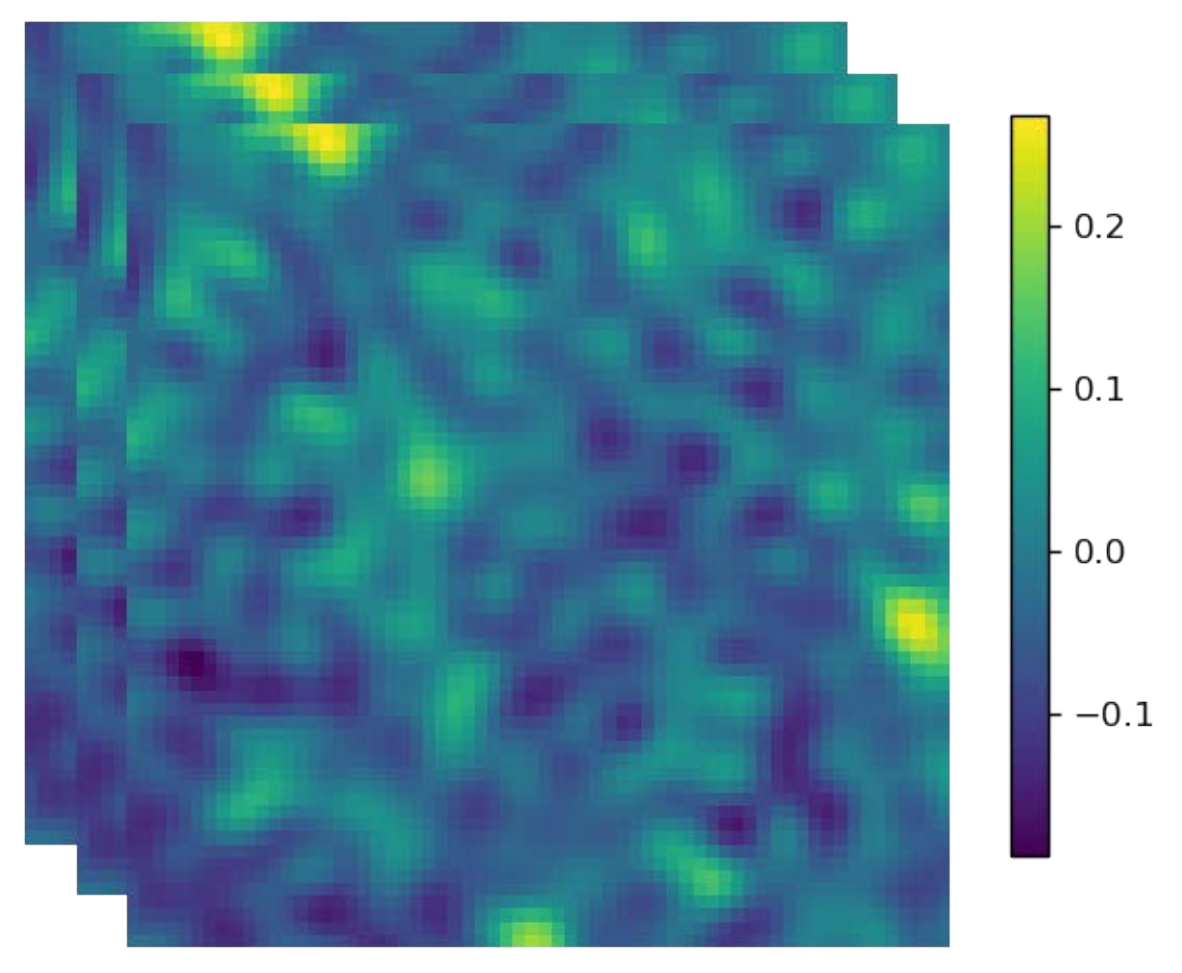}
    \includegraphics[width=0.25\textwidth]{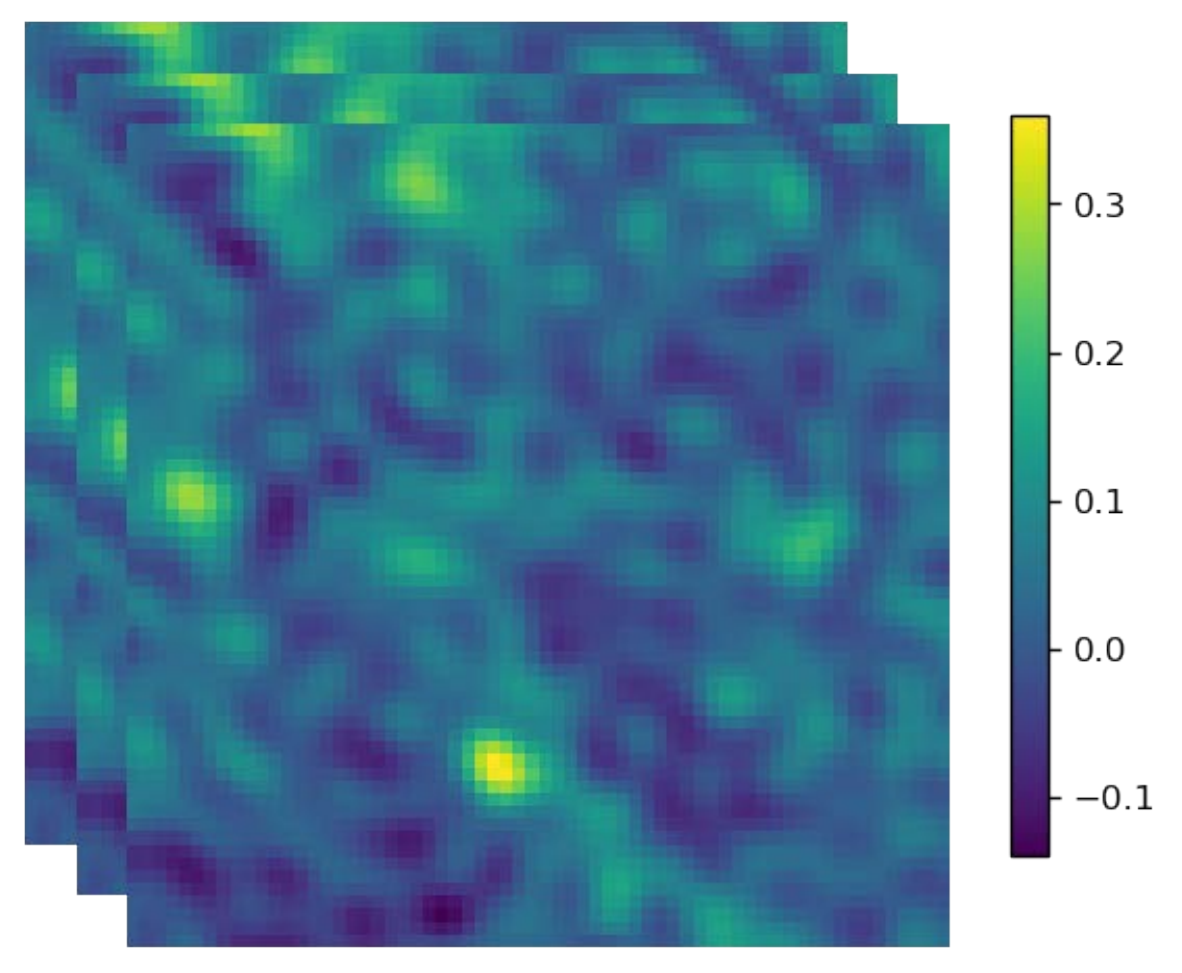}
    \includegraphics[width=0.25\textwidth]{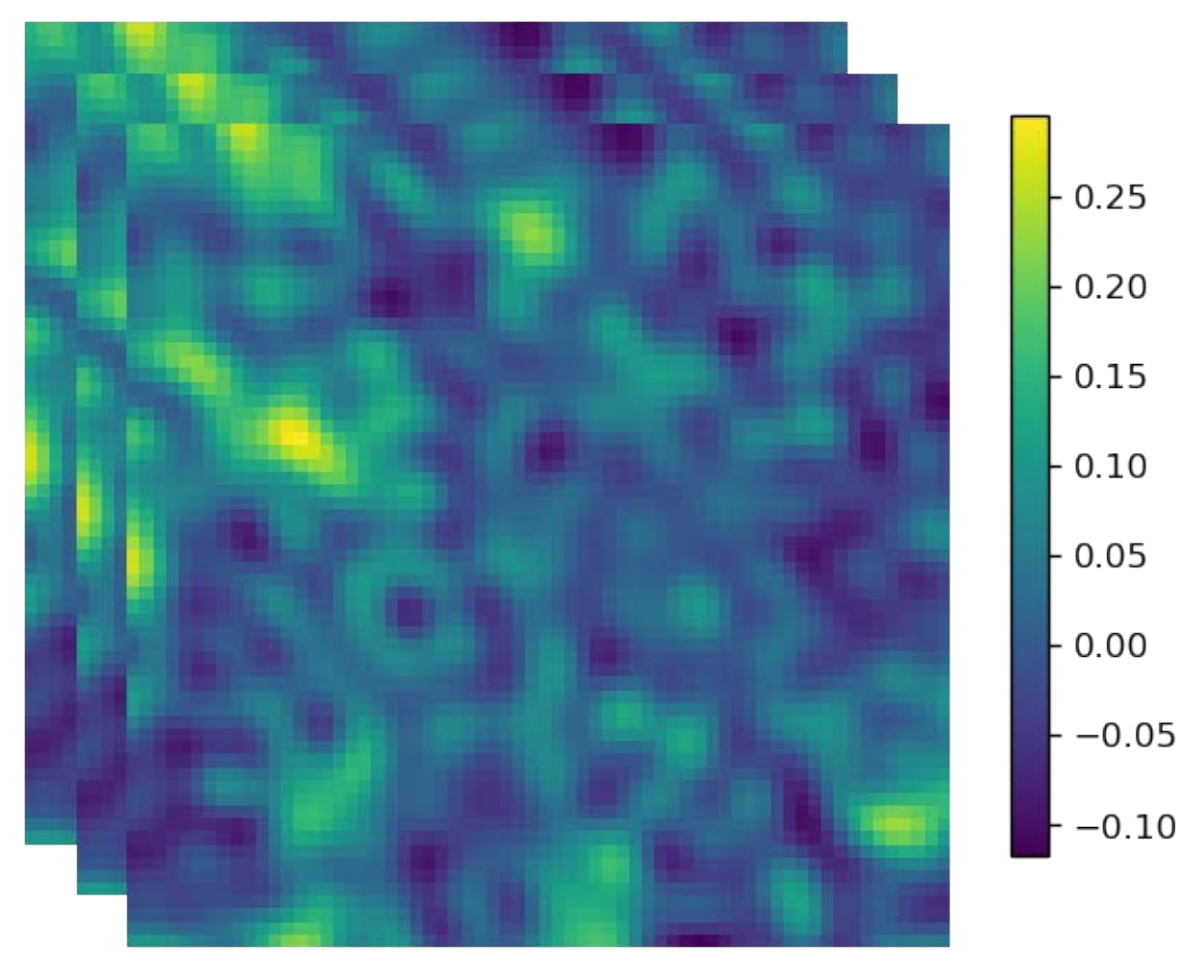}\label{fig:mapCpca}}\\
    \subfigure[]
    {\includegraphics[width=0.25\textwidth]{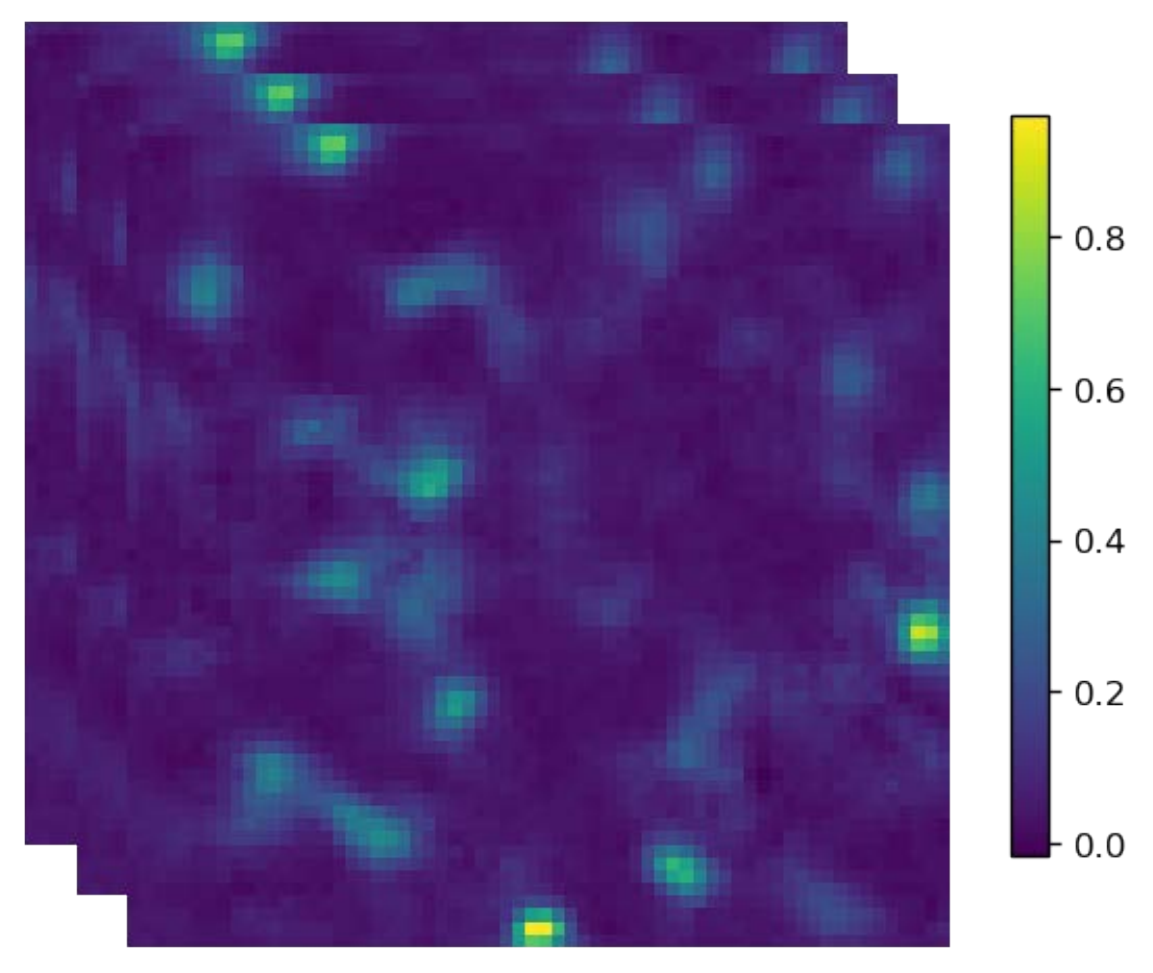}
    \includegraphics[width=0.25\textwidth]{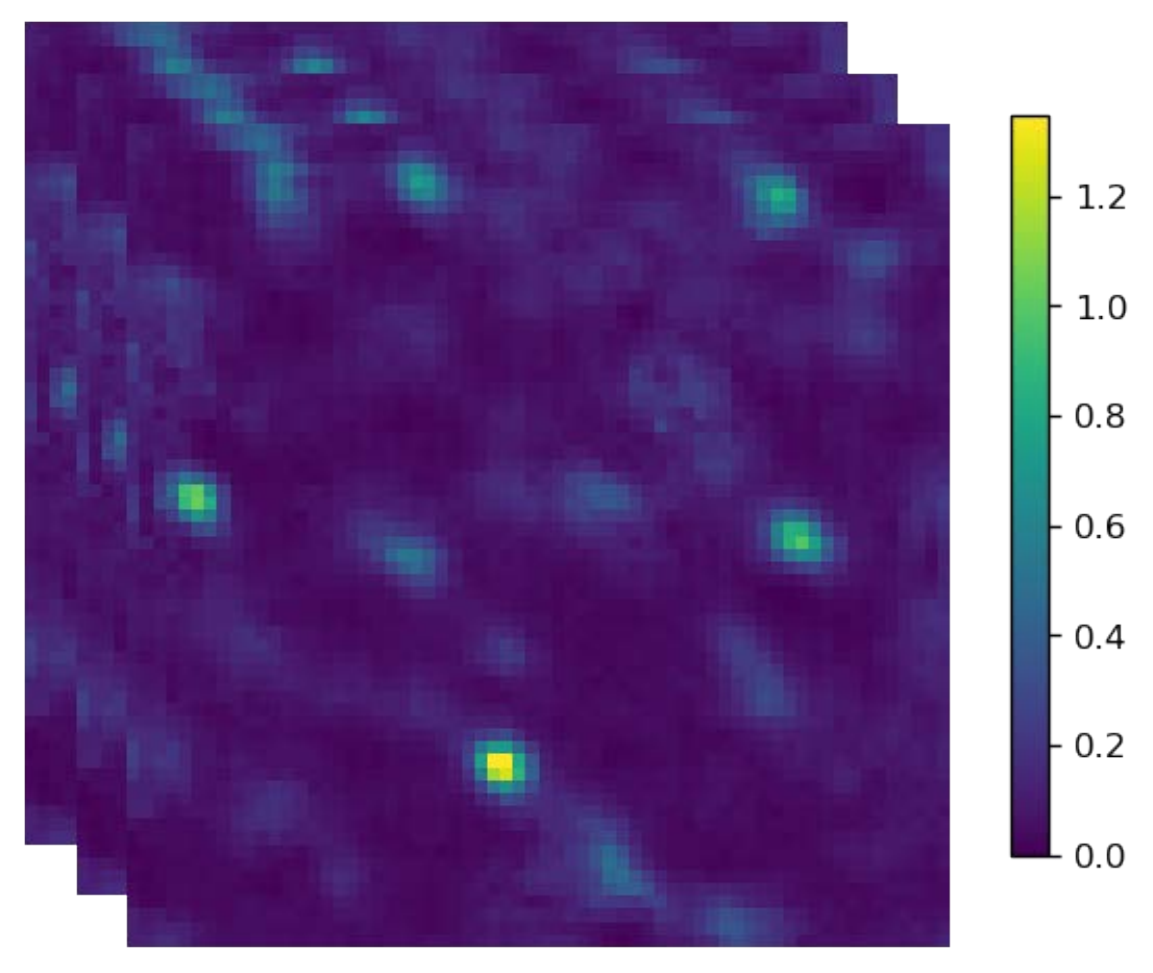}
    \includegraphics[width=0.25\textwidth]{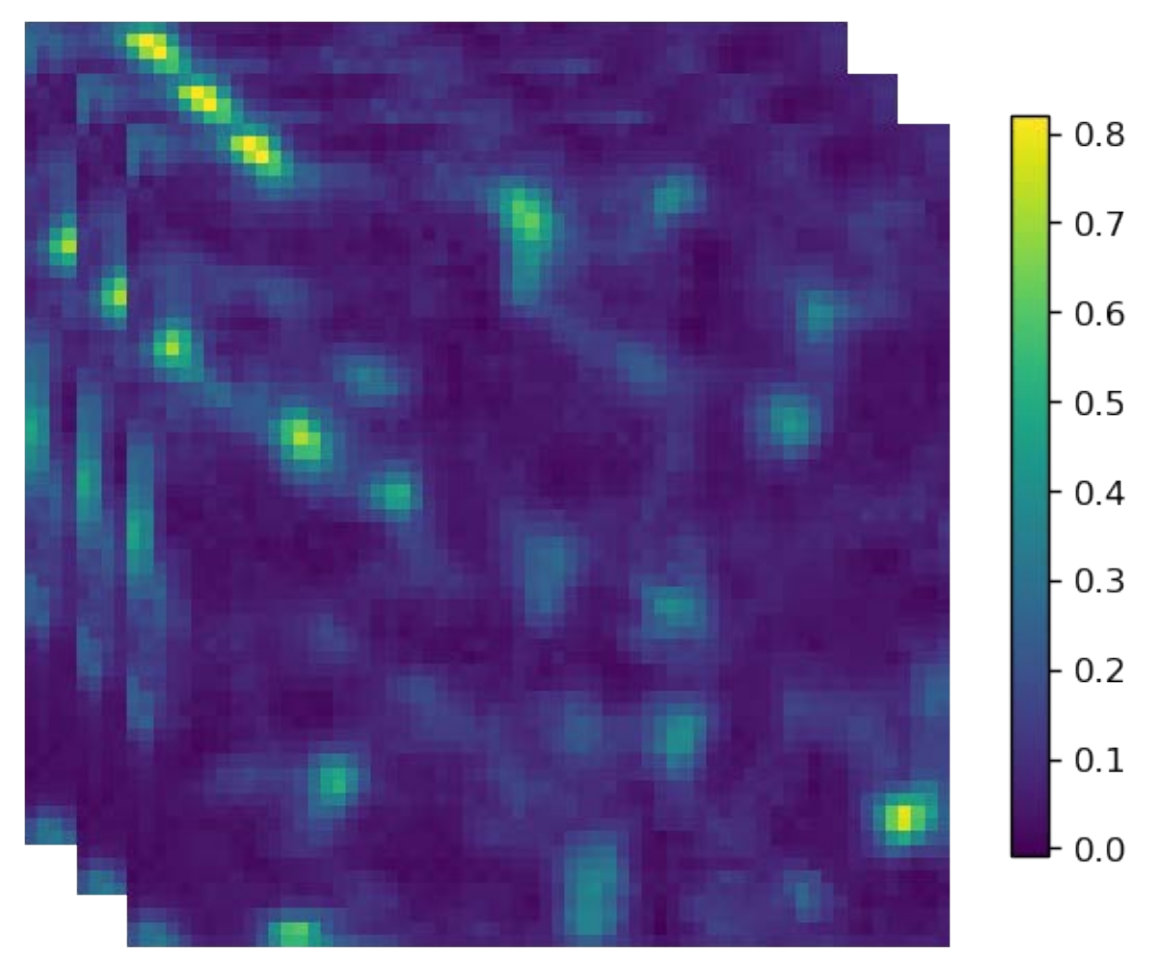}\label{fig:mapCunet}}
   \caption{Data cubes of the same sky patches as shown in Figure~\ref{fig:patch}, but with foreground subtracted. From top to bottom are initial HI map (a), Gaussian beam convolved map cleaned with PCA (b), Gaussian beam convolved map cleaned with PCA+U-Net (c), Cosine beam convolved map cleaned with PCA (d), and Cosine beam convolved map cleaned with PCA+U-Net (e). The unit of the cube is mK.}
  \label{fig:mapfgrm}
\end{figure*}

\begin{figure*}
  \centering
  \subfigure[]
    {\includegraphics[width=0.495\textwidth]{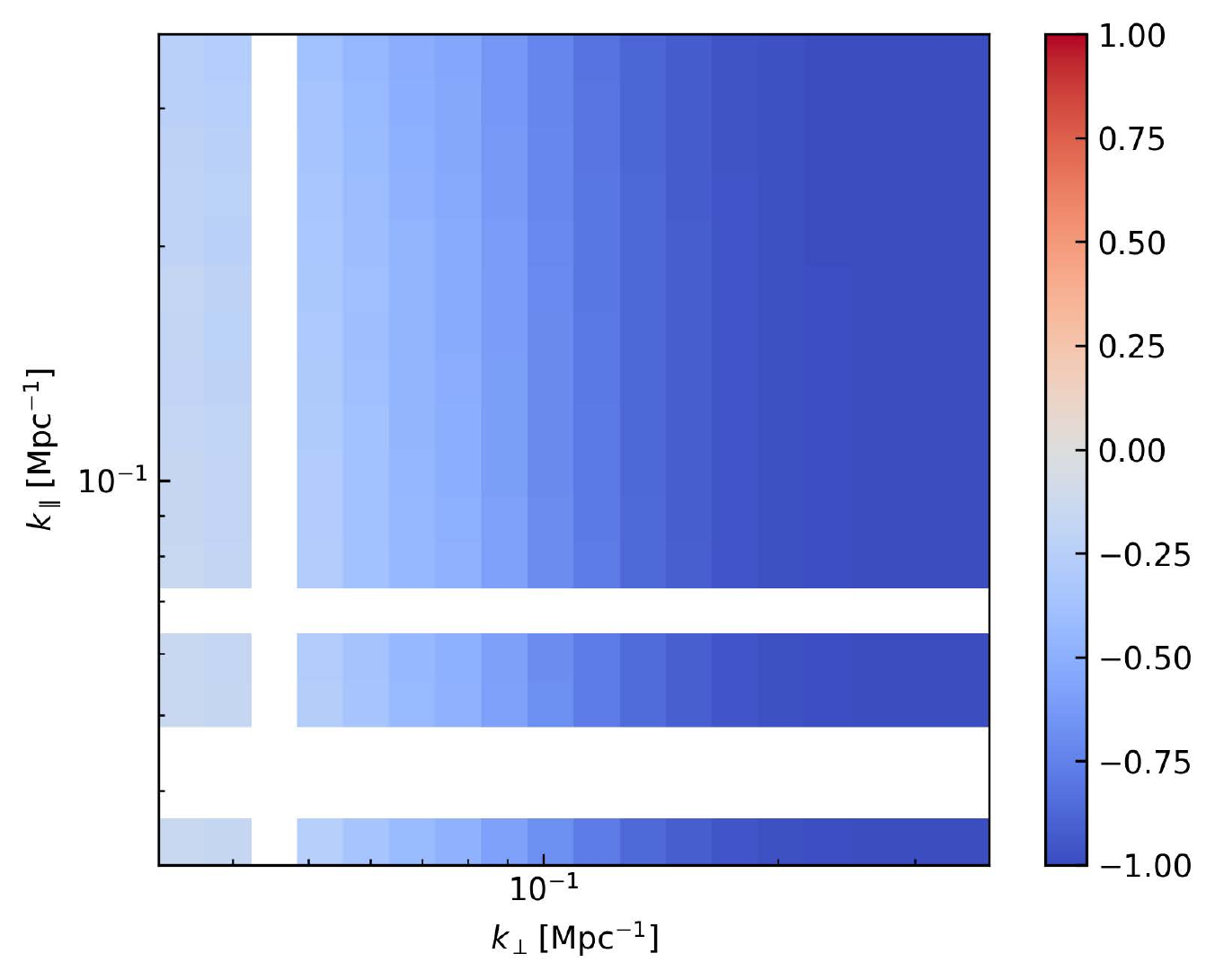}\label{fig:result2dpca0}}
  \subfigure[]
    {\includegraphics[width=0.495\textwidth]{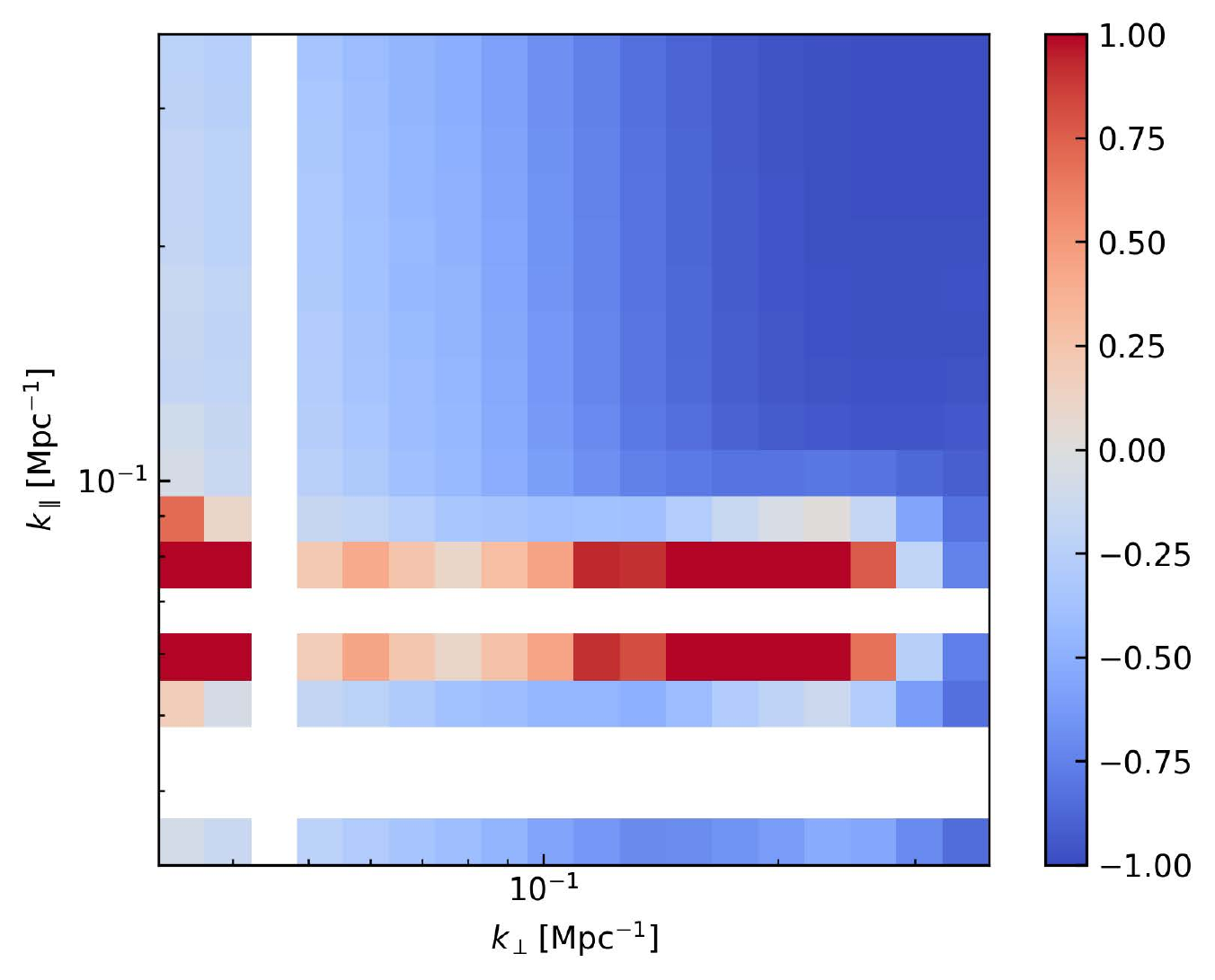}\label{fig:result2dpca1}}
    \caption{The 2D auto-correlation power spectrum ratio of the PCA cleaned map and the initial HI model.
  Panel (a) shows the result from the Gaussian convolved map and Panel (b) shows the result from the Cosine beam
  convolved map.}\label{fig:result2dpca}
\end{figure*}

\begin{figure*}
  \centering
  \subfigure[]
  {\includegraphics[width=0.495\textwidth]{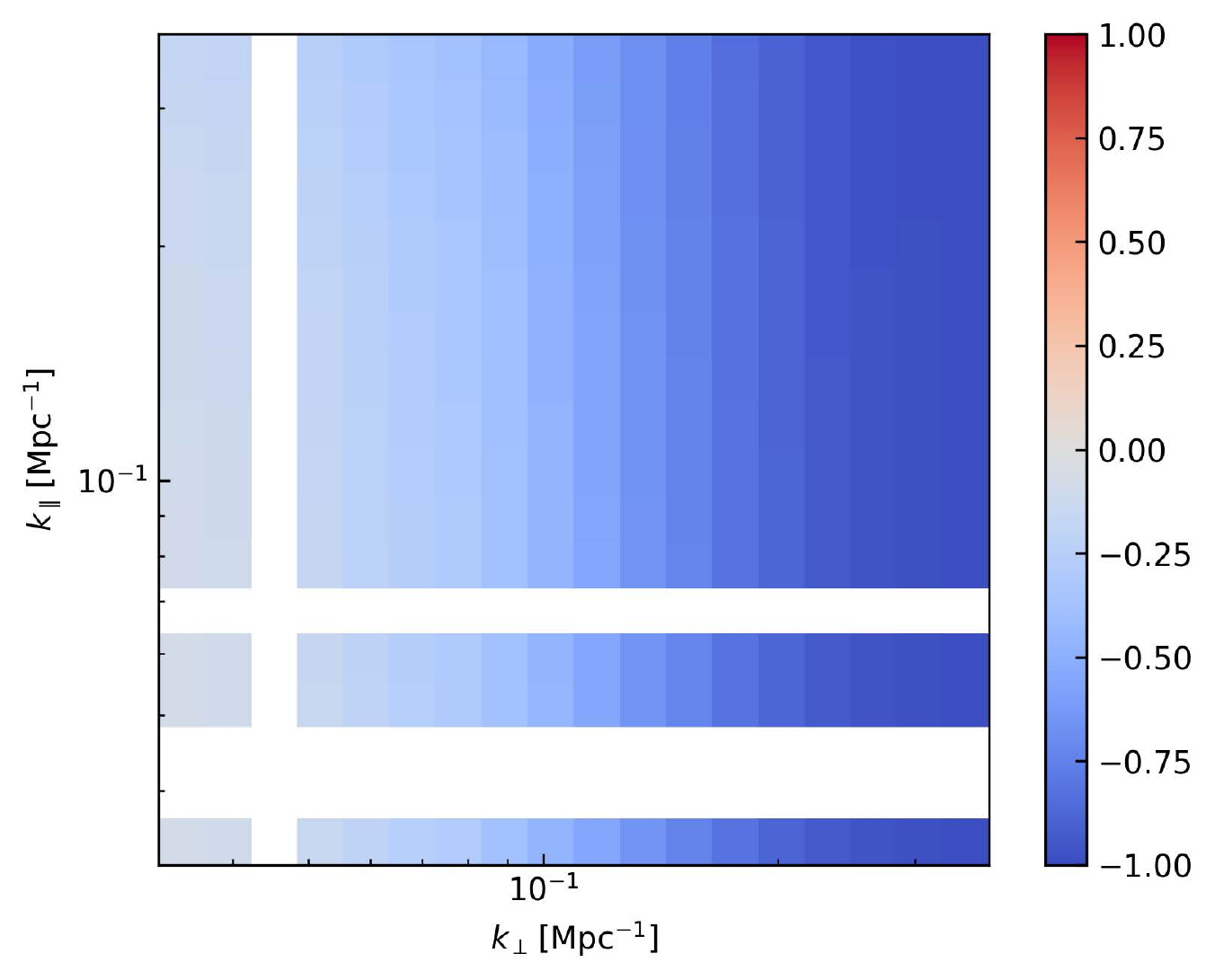}\label{fig:result2dpca_cross0}}
  \subfigure[]
  {\includegraphics[width=0.495\textwidth]{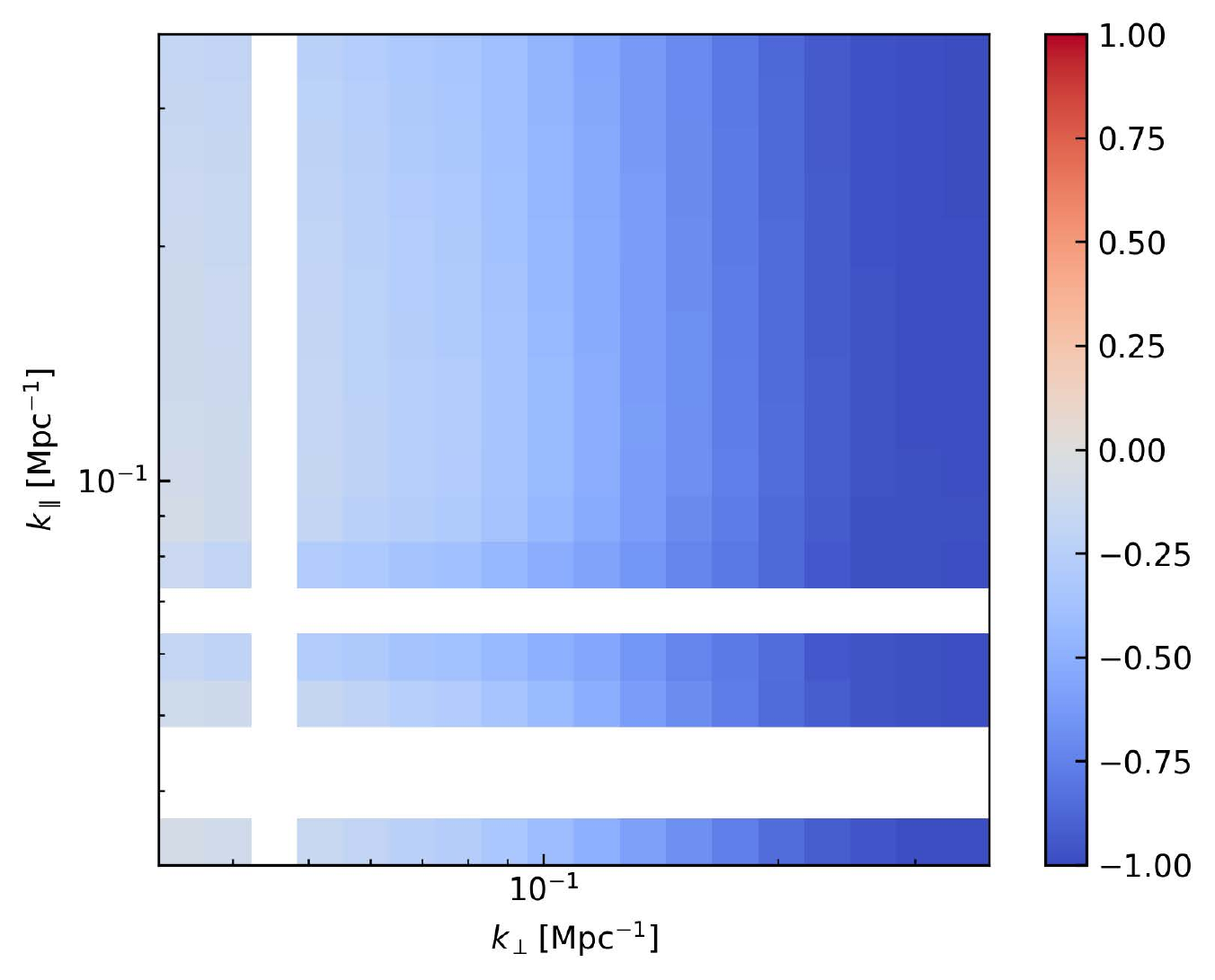}\label{fig:result2dpca_cross1}}
  \caption{The 2D cross-correlation power spectrum ratio of the PCA cleaned map and the initial HI model.
  Panel (a) shows the result from the Gaussian convolved map and Panel (b) shows the result from the Cosine beam
  convolved map.}\label{fig:result2dpca_cross}
\end{figure*}

Three different small patches of the foreground cleaned maps are shown in
different columns of Figure~\ref{fig:mapfgrm}. The different sky patches of the initial HI signal map
are shown in Figure~\ref{fig:HImap}. Figures~\ref{fig:mapGpca} and \ref{fig:mapGunet} show the
sky patches of the Gaussian beam convolved maps cleaned with PCA and PCA+U-Net, respectively.
Both the PCA and PCA+U-Net cleaned maps have good consistency with the initial HI map.
It shows that both PCA and PCA+U-Net foreground subtraction approaches perform well, and that the systematic effect caused by the simple Gaussian beam model can be
effectively eliminated with both methods.

Figures~\ref{fig:mapCpca} and \ref{fig:mapCunet} show patches of Cosine beam convolved maps
cleaned with PCA and PCA+U-Net, respectively. As we can see in Figure~\ref{fig:mapCpca},
the PCA cleaned maps have extra structures. Such structures can be eliminated
with the additional U-Net process, as shown in Figure~\ref{fig:mapCunet}.

We can verify the foreground subtraction efficiency by comparing the angular power spectrum to
the initial HI model. The sky patches are re-filled to their original positions after foreground subtraction,
forming the cleaned whole sky map.
The angular power spectrum is shown in Figure~\ref{fig:result_gc}.
The initial HI power spectrum model is shown with the blue line.
The orange line shows the result of the simulation map cleaned
with the first $3$ PCA modes. The green line shows the result with the combination
of PCA and U-Net foreground subtractions.

\subsection{PCA foreground subtraction}

We compare the results for the Gaussian beam convolved map and the Cosine beam convolved map.
With the Gaussian beam convolved map, the HI signal can be recovered by simply subtracting
a few of the PCA modes, e.g., $3$ modes subtraction in this analysis.
The variation on large scales is due to the effect of the signal loss during blind
foreground subtraction, and the reduction on small scales is due to the beam smoothing.
However, if the map is convolved with the Cosine beam, the HI signal cannot be recovered through
PCA foreground subtraction.
The reason is that the Cosine beam model induces much more systematic effects than the Gaussian beam.
As discussed in Section~\ref{sec:Beam_Model}, with the Gaussian beam model, only the
effect of beam size variation with frequency is induced. The Cosine beam model includes the beam sidelobes and
the frequency-dependent beam size ripple. The intensity of the bright sources outside of the field of view
can be observed through the beam sidelobe, which is non-smooth components. Meanwhile, the beam size ripple directly complicates the foreground spectrum shape.
Such effects can be further investigated through the 2D power spectrum.

With the data cubes of $192$ sky patches, we estimate the 2D auto-correlation power spectrum and plot the ratio between the foreground cleaned map and the initial HI model,
\begin{equation}\label{eq:2dpsratio}
R(k_\parallel, k_\perp)_{\rm auto} = \frac{\bar{P}_{\rm cln}(k_\parallel, k_\perp)}{P_{\rm HI}(k_\parallel, k_\perp)} - 1,
\end{equation}
where $\bar{P}_{\rm cln}(k_\parallel, k_\perp)=\frac{1}{192}\sum_{i=1}^{192}P^i_{\rm cln}(k_\parallel, k_\perp)$
and $P^i_{\rm cln}(k_\parallel, k_\perp)$ is the 2D power spectrum of one foreground-cleaned sky patch.
The results for PCA cleaned map are shown in Figure~\ref{fig:result2dpca}.
The left panel shows the result of the Gaussian beam convolved map.
The values of $R(k_\parallel,k_\perp)_{\rm auto}$ are
generally consistent with $0$ at large scales (i.e. $k_\perp \lesssim 0.06$ Mpc$^{-1}$),
which indicates that the HI signal can be recovered via PCA at large scales.
The negative $R(k_\parallel,k_\perp)_{\rm auto}$ values indicate the signal reduction,
especially on large $k_\perp$. Such signal reduction is mainly due to the beam smoothing effect.
We cannot observe significant signal reduction at small $k_\parallel$.
As shown in the literature \citep{Switzer:2015ria},
HI signal at such scales is always removed during the PCA foreground subtraction.
However, because we only remove the first $3$ PCA modes, such signal loss effect
is still negligible.

The result of the Cosine beam convolved map is shown in the right panel of
Figure~\ref{fig:result2dpca}.
There is significant extra power at $k_\parallel\sim 0.7~{\rm Mpc}^{-1}$.
Such extra power is mainly due to the beam size ripple induced by the Cosine beam model.
Our results are consistent with the simulation analysis in the literature \citep{Matshawule:2020fjz}.
In order to remove such a systematic effect, it needs aggressive mode subtraction,
which can cause non-negligible HI signal loss~\citep{Switzer:2015ria}.

{Note that the white strips in Figure~\ref{fig:result2dpca} do not have any physical meaning, which are due to the logarithmic binning being not uniform in the linear $k$-space. The same is also to Figures \ref{fig:result2dpca_cross}--\ref{fig:result2dunet_cross}}



In order to check the concordance between the foreground removed HI map and the initial
HI map, we estimate the cross-correlation power spectrum, $P_{\rm cln, HI}(k_\parallel, k_\perp)$.
The concordance can be evaluated
with the power spectrum ratio between the foreground cleaned map and the initial HI map,
\begin{equation}\label{eq:2dpsratio_cross}
R(k_\parallel, k_\perp)_{\rm cross} = \frac{\bar{P}_{\rm cln, HI}(k_\parallel, k_\perp)}{P_{\rm HI}(k_\parallel, k_\perp)} - 1,
\end{equation}
where $\bar{P}_{\rm cln, HI}(k_\parallel, k_\perp)$ represents the averaged cross-correlation power spectrum
across different sky patches. The results for PCA cleaned map are shown in
Figure~\ref{fig:result2dpca_cross}. The results of the Gaussian and Cosine beam convolved maps
are shown in the left and right panels, respectively.
Obviously, the extra power shown in $R(k_\parallel, k_\perp)_{\rm auto}$ of the Cosine
beam convolved case is eliminated through the cross-correlation power spectrum estimator.
This is because the power spectrum error due to the systematic effect can be
eliminated through cross-correlation.
The values of $R(k_\parallel, k_\perp)_{\rm cross}$ in both Gaussian and Cosine beam
convolved cases show the similar trend. The negative $R(k_\parallel, k_\perp)_{\rm cross}$ values
at large $k_\perp$ indicate the reduction of the correlation efficiency,
which is mainly due to the beam smoothing effect.
However, although the systematic effect can be removed via the cross-correlation power spectrum,
it still enlarges the power spectrum estimation error and biases the auto-correlation power spectrum.
The elimination of the systematic effect is crucial to the auto-correlation power spectrum detection.

\subsection{Additional U-Net foreground subtraction}

\begin{figure*}
    \centering
    \subfigure[]
    {\includegraphics[width=0.495\textwidth]{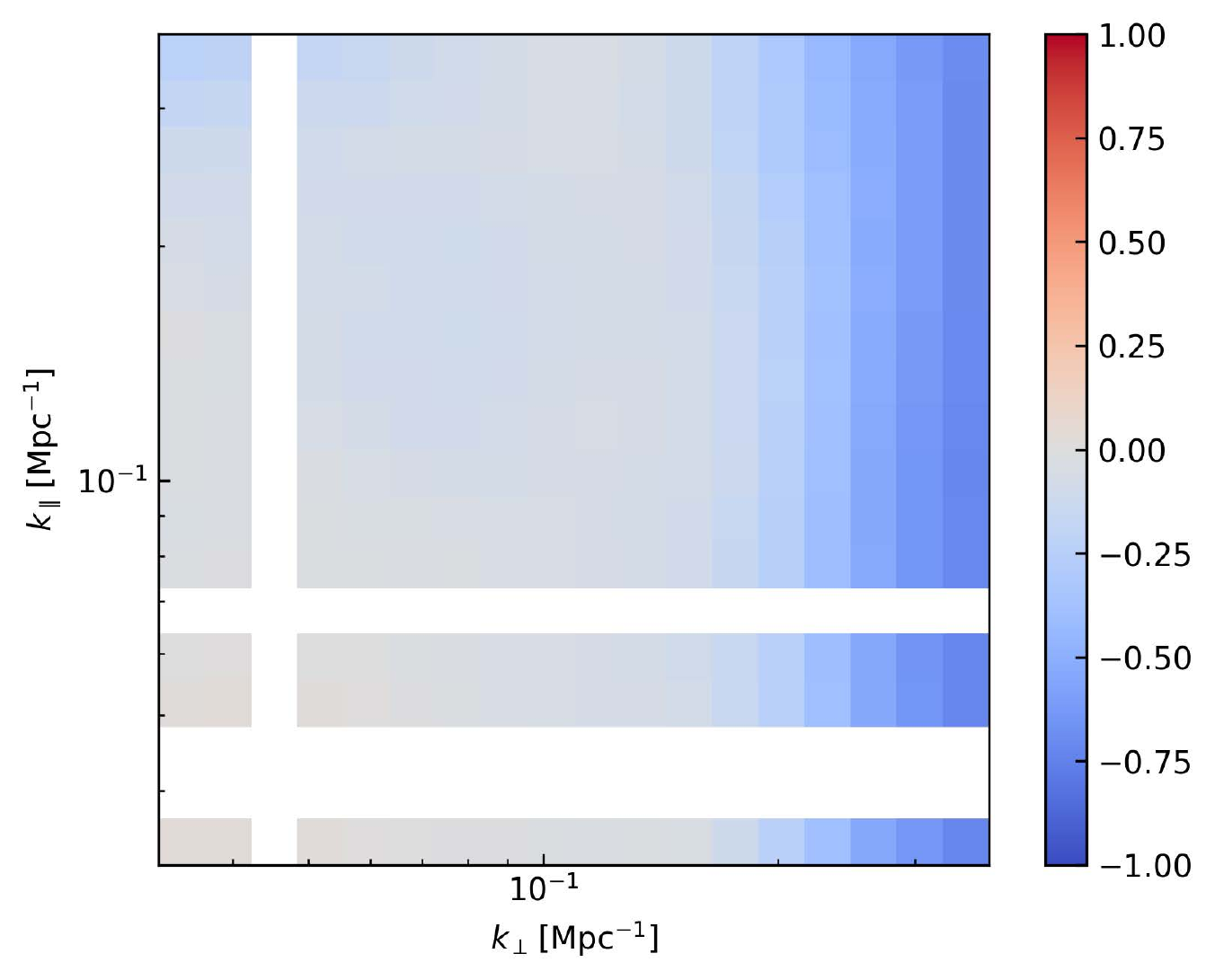}\label{fig:result2dunet0}}
    \subfigure[]
    {\includegraphics[width=0.495\textwidth]{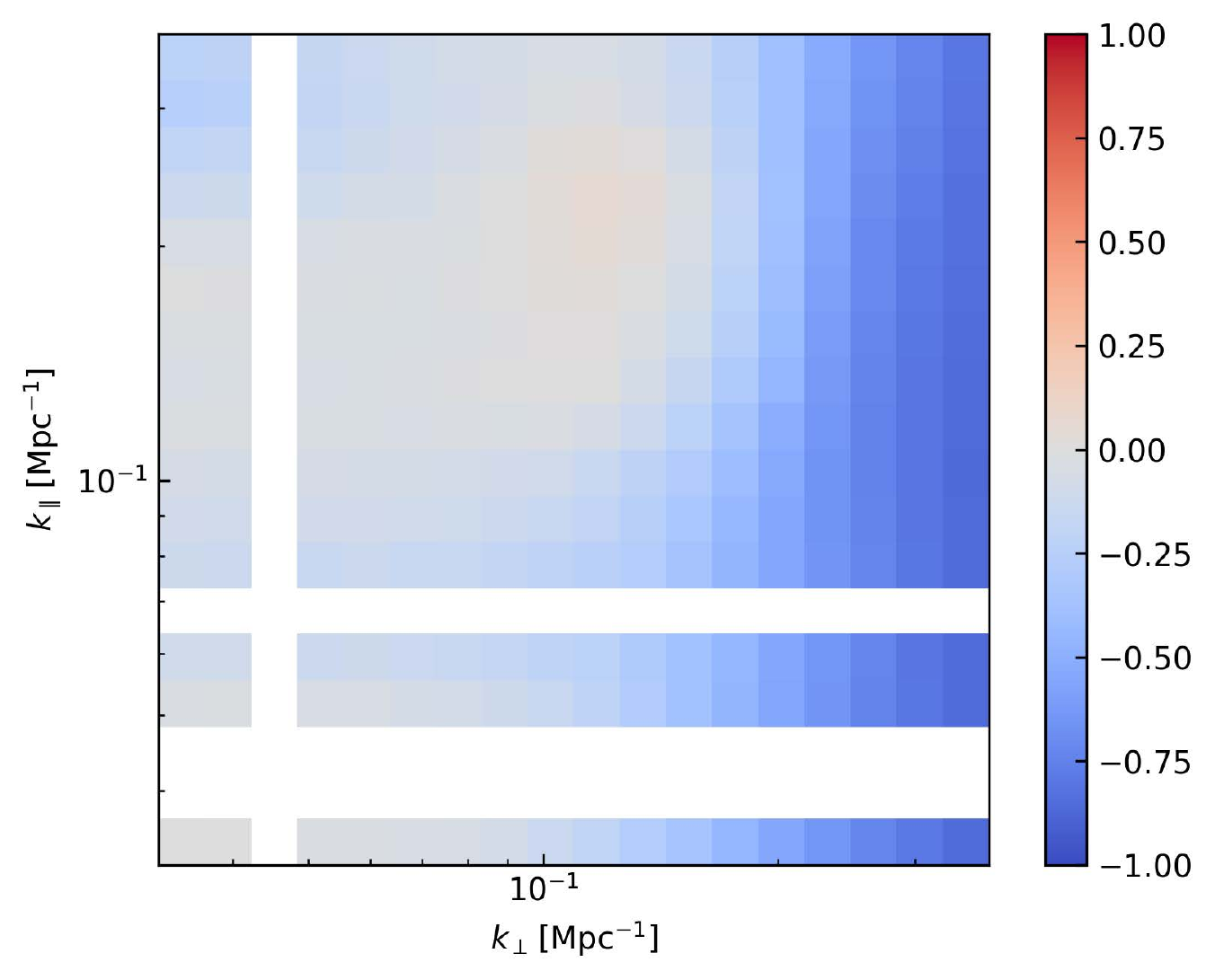}\label{fig:result2dunet1}}
    \caption{The 2D auto-correlation power spectrum ratio of the U-Net cleaned map and the initial HI model.
  Panel (a) shows the result from the Gaussian convolved map cross the initial HI map and Panel (b) shows the result from the Cosine beam convolved map cross the initial HI map.}\label{fig:result2dunet}
\end{figure*}

\begin{figure*}
    \centering
    \subfigure[]
    {\includegraphics[width=0.495\textwidth]{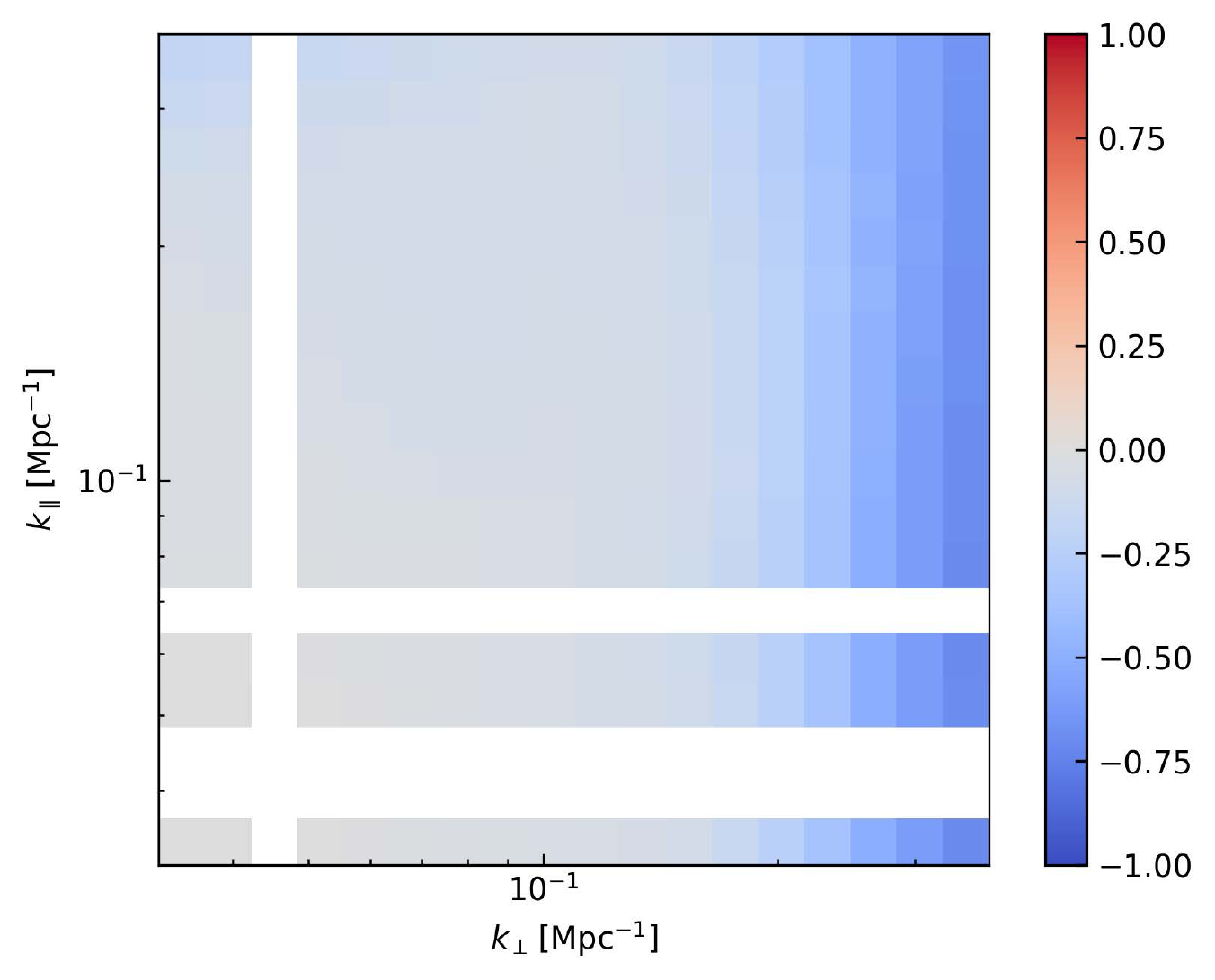}\label{fig:result2dunet_cross0}}
    \subfigure[]
    {\includegraphics[width=0.495\textwidth]{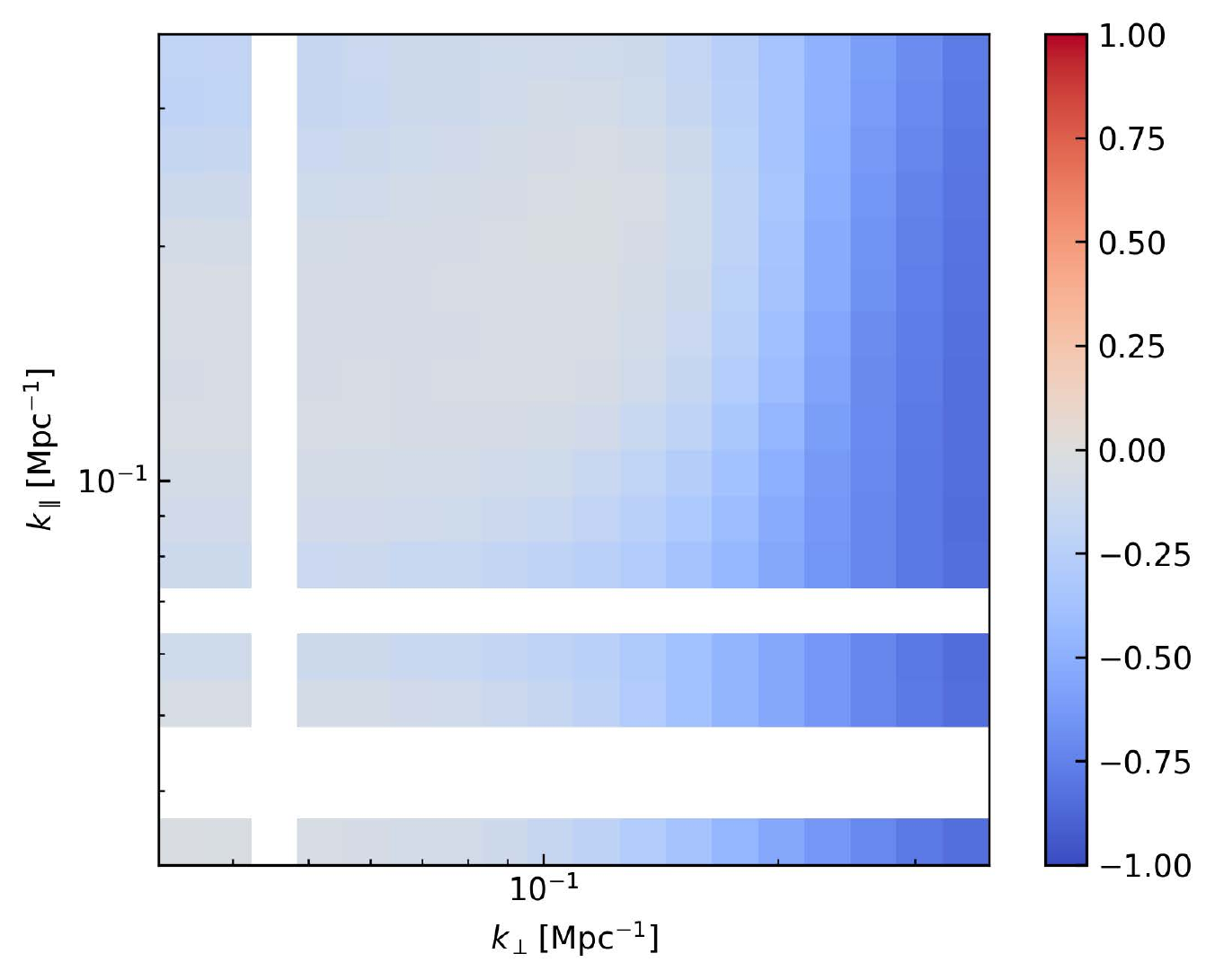}\label{fig:result2dunet_cross1}}
    \caption{The 2D cross-correlation power spectrum ratio of the U-Net cleaned map and the initial HI model.
  Panel (a) shows the result from the Gaussian convolved map cross the initial HI map and Panel (b) shows the result from the Cosine beam convolved map cross the initial HI map.}\label{fig:result2dunet_cross}
\end{figure*}

Instead of applying aggressive additional mode subtraction, we feed the residual maps
to the U-Net architecture. The angular power spectra of the U-Net cleaned maps
are shown with green lines in Figure~\ref{fig:result_gc}.
The U-Net foreground subtraction corrects angular power spectrum in both the
Gaussian and Cosine beam convolved cases. But in the Gaussian beam
convolved case, the correction from U-Net is not evident.
The significant improvement can be seen from the results in the Cosine beam convolved map,
as shown in Figure~\ref{fig:result_c}.
With the additional U-Net foreground subtraction, the angular power spectrum deviation
at large scales is significantly eliminated. The recovered angular power spectrum
is consistent with the initial HI power spectrum. Next, we can further investigate it with
the 2D power spectrum.

The $R(k_\parallel, k_\perp)_{\rm auto}$ values, as specified by Equation~(\ref{eq:2dpsratio}),
in the U-Net foreground subtracted sky patches are shown in Figures~\ref{fig:result2dunet}.
The results for the Gaussian and Cosine beam convolved cases are shown in the left and right
panels, respectively. These two results are generally consistent with each other,
which indicates that the U-Net foreground subtraction can remove the
systematic effect induced by either the simple Gaussian beam model or the complicated Cosine beam model.

The $R(k_\parallel, k_\perp)_{\rm cross}$ values, as specified by Equation~(\ref{eq:2dpsratio_cross}),
in the U-Net foreground subtracted sky patches are shown in Figure~\ref{fig:result2dunet_cross},
with the Gaussian beam convolved case in the left panel and the Cosine beam convolved
case in the right panel, respectively.
The $R(k_\parallel, k_\perp)_{\rm cross}$ values are generally $\sim 0$,
indicating that the recovered HI fluctuation signal is consistent with the initial HI signal.

\subsection{PCA v.s. U-Net}

In Figure~\ref{fig:result_g}, PCA is effective in reducing the foreground with Gaussian beam, but slightly worse than the U-Net network. As can be seen in Figure~\ref{fig:result_c}, the results of PCA are much worse than those of the U-Net when dealing with the Cosine beam convolutional sky map. This indicates that PCA fails in processing the sky map with convolved complex bundles, while the U-Net network can learn large scale information after the convolved bundles of the sky map. In contrast, the U-Net network can still recover the angular power spectrum of the original HI signal on large scales.

In Figures~\ref{fig:result2dpca}--\ref{fig:result2dunet_cross}, we plotted average auto-correlation and cross-correlation power spectra in the lateral~($k_{\perp}$) and radial~($k_\parallel$) directions. {We define a measure of concordance:
\begin{equation}\label{equ:acc}
    \Gamma= {\rm mean}\left\{R(k_\parallel, k_\perp)_{\rm X}\right\} + 1,
\end{equation}
where ${\rm mean}\{\cdot\cdot\cdot\}$ represents the mean value across the $k_\parallel$--$k_\perp$ bins
and $R(k_\parallel, k_\perp)_{\rm X}$
is the spectrum ratio
defined in Equations~(\ref{eq:2dpsratio}) and (\ref{eq:2dpsratio_cross}) with the subscript X denoting
the auto or cross power spectrum. }

We first discuss the case of auto-correlation power spectrum.
We obtain the concordances with the original HI map using both the PCA and U-Net methods, i.e., $\Gamma^{\rm gau}_{\rm pca}=60.73\%$, $\Gamma^{\rm cos}_{\rm pca}=28.46\%$, $\Gamma^{\rm gau}_{\rm unet}=77.38\%$, and $\Gamma^{\rm cos}_{\rm unet}=69.66\%$, respectively. So, it is clearly seen that the 3D U-Net method is much better than the PCA method in recovering the HI signal. We find that, in the case of the Gaussian beam, the concordance of using the U-Net method is better than that of PCA by $27.4\%$, and in the case of the Cosine beam,  the concordance of using the U-Net method is better than that of PCA by $144.8\%$.
The results of $R(k_\parallel, k_\perp)_{\rm cross} \sim 0$ at large scales for both the
PCA and U-Net foreground subtractions indicate that the recovered HI fluctuation
is consistent with the initial HI signal.

It should be noticed that, although the sky map is partitioned into 192 small sky patches,
the HI fluctuation on large scales can still be properly recovered. This is due to the fact that
either the foreground contamination or the systematic effect considered here are
highly correlated across frequencies. The results indicate that,
even with small sky patch observation, the systematic effect
induced by the telescope primary beam can be well eliminated using the U-Net architecture.
It is ideal for most of current stage HI IM experiments.

The U-Net architecture is one of the supervised deep leaning methods.
It means that the foreground subtraction with the U-Net architecture is
model dependent. The accuracy of the primary beam model affects the foreground subtraction efficiency.
However, the telescope primary beam, especially for telescope arrays, can be measured
via the `astro-holographic' observation \citep{Asad:2021mnras}.
Therefore, the U-Net based foreground subtraction sheds new light on eliminating the primary beam effect
for future HI IM experiments.

{PCA is currently one of the most effective blind foreground subtraction methods that has been widely discussed in different analyses with both simulated and actual observational data. Due to the systematic effect, most of the blind foreground subtraction methods, including the PCA method, can result in either significant foreground residual or substantial signal loss. In this work, we show that the additional foreground subtraction with the U-Net network can eliminate the primary beam effect and improve the efficiency of PCA foreground subtraction. It can potentially improve other blind foreground subtraction methods, such as ICA, as well as non-blind foreground subtraction methods.}


\section{Conclusion}\label{sec:con}
Deep learning, derived from artificial neural networks, can overcome the shortcomings of traditional feature extraction and matching, and has made significant breakthroughs in the field of image recognition. Deep learning has been very successful in astronomy by combining underlying features to form more abstract higher-level representations of attribute classes or features, trained with large amounts of specific data.

In the HI 21-cm IM survey, the foreground contamination on the cosmological signals is extremely severe, and the systematic effects caused by radio telescopes themselves further aggravate the difficulties in subtracting the foregrounds.
These foregrounds mainly originate from Galaxy synchrotron, Galaxy free-free, extragalactic free-free, and extragalactic point sources emissions, which are in total five orders of magnitude higher than the cosmological signals in the HI IM survey. Therefore, finding an effective way to subtract foregrounds is one of the crucial challenges for the HI IM observations.

{In this work, we used a similar network structure to \citet{Makinen:2020gvh}. Compared with \citet{Makinen:2020gvh}, we consider the frequency band of the MeerKAT telescope and, more importantly, we focus on the systematic effect induced by the radio telescope primary beam.} For the radio telescope primary beam effect, we consider the Gaussian beam model as a simple case and the Cosine beam model as a sophisticated case. We investigate whether the deep learning method, concretely the U-Net algorithm in this work, can play a crucial role in eliminating such primary beam induced systematic effect and improving the foreground subtraction efficiency.

The results of PCA foreground subtraction are different between using the Gaussian beam convolved map
and the Cosine beam convolved one.
When processing the Gaussian beam convolved map, the recovered angular power spectrum is
essentially consistent with the initial HI angular power spectrum.
Using the 2D power spectrum, we find that the values of $R(k_\parallel, k_\perp)_{\rm auto}$
[see Equation~(\ref{eq:2dpsratio})] are close to $0$ at large scales.
It indicates that PCA foreground subtraction can recover part of the HI signal.
While at small scales, the HI signal reduction is found, which is due to the beam smoothing effect.

However, when dealing with the Cosine beam covolved map, the angular power spectrum with $3$ PCA modes
subtraction deviates significantly from the initial HI angular power spectrum.
With the 2D power spectrum, we find that there is a significant extra power at
$k_\parallel\sim 0.7~{\rm Mpc}^{-1}$, which is relevant to the beam size ripple induced by
the Cosine beam model.
Therefore, PCA foreground subtraction fails in subtracting foregrounds in the Cosine beam convolved map.
It indicates that the PCA foreground subtraction cannot remove the complex systematic effect
induced by a sophisticated primary beam model, such as the Cosine beam model.

Besides, we evaluate the concordance of the results with the cross-correlation power spectrum
between the foreground removed HI map and the initial HI map.
The values of $R(k_\parallel, k_\perp)_{\rm cross}$ [see Equation~(\ref{eq:2dpsratio_cross})] in
both the Gaussian and Cosine beam convolved cases show the similar trend.
The negative $R(k_\parallel, k_\perp)_{\rm cross}$ values
at large $k_\perp$ indicate the reduction of the correlation efficiency,
which is mainly due to the beam smoothing effect.

Instead of directly using U-Net for foreground subtraction, we conservatively feed
the PCA-subtracted residuals into the U-Net network.
When dealing with the map convolved with the Cosine beam, the results are significantly improved.
In particular, the deviation of the large-scale angular power spectrum is significantly removed,
and the recovered angular power spectrum is consistent with the initial HI angular power spectrum.
By studying the 2D power spectrum, we find that the values of $R(k_\parallel, k_\perp)_{\rm auto}$
are consistent in general between using the Gaussian and Cosine beam convolved maps.
We quantify the results by defining a measure of concordance, as indicated by Equation~(\ref{equ:acc}).
We find that, in the case of Gaussian beam, the concordance with the original HI map using U-Net is better than that using PCA by $27.4\%$, and in the case of Cosine beam, the concordance using U-Net is better than that using PCA by $144.8\%$.
With the cross-correlation power spectrum, we find that $R(k_\parallel, k_\perp)_{\rm cross} \sim 0$,
which indicates that the recovered HI signal is consistent with the initial HI signal.
Therefore, the results show that the U-Net foreground subtraction method can
eliminate the systematic effects caused by the telescope primary beam.

In this work, we find that the U-Net based foreground subtraction method can significantly
improve the foreground subtraction efficiency, especially in the case with serious 
primary beam induced systematic effect.
{U-Net is based on the fully convolutional network and its architecture was modified and
extended to work with fewer training images and to yield more precise segmentation. 
Such supervised deep learning method relies on completeness of the simulated training set.
Because the detailed primary beam feature, as well as the systematic effect, 
can be different between different experiments, an accurate beam model is required if a different experiment is considered. 
Such beam model can be provided using either `astro-holographic' observation or EM simulation \citep{Asad:2021mnras}. Therefore, the U-Net based foreground subtraction 
method sheds new light on recovering the HI power spectrum for future HI IM experiments,
e.g. SKA and HERA \citep{2017PASP..129d5001D}.}

\section*{acknowledgments}

This work was supported by the National Natural Science Foundation of China (Grants Nos. 11975072, 11835009, 11875102, and 11690021), the Liaoning Revitalization Talents Program (Grant No. XLYC1905011), the Fundamental Research Funds for the Central Universities (Grant No. N2005030), the National 111 Project of China (Grant No. B16009), and the Science Research Grants from the China Manned Space Project (Grant No. CMS-CSST-2021-B01).



\bibliography{ref}{}

\begin{thebibliography}{}
\expandafter\ifx\csname natexlab\endcsname\relax\def\natexlab#1{#1}\fi
\providecommand{\url}[1]{\href{#1}{#1}}
\providecommand{\dodoi}[1]{doi:~\href{http://doi.org/#1}{\nolinkurl{#1}}}
\providecommand{\doeprint}[1]{\href{http://ascl.net/#1}{\nolinkurl{http://ascl.net/#1}}}
\providecommand{\doarXiv}[1]{\href{https://arxiv.org/abs/#1}{\nolinkurl{https://arxiv.org/abs/#1}}}

\bibitem[{Abdalla {et~al.}(2021)}]{Abdalla:2021nyj}
Abdalla, E., {et~al.} 2021, arXiv e-prints, arXiv:2107.01633.
\newblock \doarXiv{2107.01633}

\bibitem[{Aghanim {et~al.}(2020)}]{Planck:2018vyg}
Aghanim, N., {et~al.} 2020, Astron. Astrophys., 641, A6,
  \dodoi{10.1051/0004-6361/201833910}

\bibitem[{Alonso {et~al.}(2014)Alonso, Ferreira, \& Santos}]{Alonso:2014sna}
Alonso, D., Ferreira, P.~G., \& Santos, M.~G. 2014, Mon. Not. Roy. Astron.
  Soc., 444, 3183, \dodoi{10.1093/mnras/stu1666}

\bibitem[{Amiri {et~al.}(2022)}]{CHIME:2022kvg}
Amiri, M., {et~al.} 2022, arXiv e-prints, arXiv:2202.01242.
\newblock \doarXiv{2202.01242}

\bibitem[{Anderson {et~al.}(2018)}]{Anderson:2017ert}
Anderson, C.~J., {et~al.} 2018, Mon. Not. Roy. Astron. Soc., 476, 3382,
  \dodoi{10.1093/mnras/sty346}

\bibitem[{Ansari {et~al.}(2012)Ansari, Campagne, Colom, Goff, Magneville,
  Martin, Moniez, Rich, \& Yeche}]{Ansari:2011bv}
Ansari, R., Campagne, J.~E., Colom, P., {et~al.} 2012, Astron. Astrophys., 540,
  A129, \dodoi{10.1051/0004-6361/201117837}

\bibitem[{Asad {et~al.}(2021)Asad, Girard, de Villiers, Ansah-Narh, Iheanetu,
  Smirnov, Santos, Lehmensiek, Jonas, de Villiers, Thorat, Hugo, Makhathini,
  Jozsa, \& Sirothia}]{Asad:2021mnras}
Asad, K. M.~B., Girard, J.~N., de Villiers, M., {et~al.} 2021, Monthly Notices
  of the Royal Astronomical Society, 502, 2970–2983,
  \dodoi{10.1093/mnras/stab104}

\bibitem[{Bacon {et~al.}(2020)}]{SKA:2018ckk}
Bacon, D.~J., {et~al.} 2020, Publ. Astron. Soc. Austral., 37, e007,
  \dodoi{10.1017/pasa.2019.51}

\bibitem[{Bagla {et~al.}(2010)Bagla, Khandai, \& Datta}]{Bagla:2009jy}
Bagla, J.~S., Khandai, N., \& Datta, K.~K. 2010, Mon. Not. Roy. Astron. Soc.,
  407, 567, \dodoi{10.1111/j.1365-2966.2010.16933.x}

\bibitem[{Bandura {et~al.}(2014)}]{Bandura:2014gwa}
Bandura, K., {et~al.} 2014, Proc. SPIE Int. Soc. Opt. Eng., 9145, 22,
  \dodoi{10.1117/12.2054950}

\bibitem[{Battye {et~al.}(2013)Battye, Browne, Dickinson, Heron, Maffei, \&
  Pourtsidou}]{Battye:2012tg}
Battye, R.~A., Browne, I. W.~A., Dickinson, C., {et~al.} 2013, Mon. Not. Roy.
  Astron. Soc., 434, 1239, \dodoi{10.1093/mnras/stt1082}

\bibitem[{Battye {et~al.}(2004)Battye, Davies, \& Weller}]{Battye:2004re}
Battye, R.~A., Davies, R.~D., \& Weller, J. 2004, Mon. Not. Roy. Astron. Soc.,
  355, 1339, \dodoi{10.1111/j.1365-2966.2004.08416.x}

\bibitem[{Bonaldi \& Brown(2015)}]{Bonaldi:2014zma}
Bonaldi, A., \& Brown, M.~L. 2015, Mon. Not. Roy. Astron. Soc., 447, 1973,
  \dodoi{10.1093/mnras/stu2601}

\bibitem[{Bull {et~al.}(2015)Bull, Ferreira, Patel, \& Santos}]{Bull:2014rha}
Bull, P., Ferreira, P.~G., Patel, P., \& Santos, M.~G. 2015, Astrophys. J.,
  803, 21, \dodoi{10.1088/0004-637X/803/1/21}

\bibitem[{Chang {et~al.}(2010)Chang, Pen, Bandura, \& Peterson}]{Chang:2010jp}
Chang, T.-C., Pen, U.-L., Bandura, K., \& Peterson, J.~B. 2010, Nature, 466,
  463, \dodoi{10.1038/nature09187}

\bibitem[{Chang {et~al.}(2008)Chang, Pen, Peterson, \& McDonald}]{Chang:2007xk}
Chang, T.-C., Pen, U.-L., Peterson, J.~B., \& McDonald, P. 2008, Phys. Rev.
  Lett., 100, 091303, \dodoi{10.1103/PhysRevLett.100.091303}

\bibitem[{Chen(2012)}]{Chen:2012xu}
Chen, X. 2012, Int. J. Mod. Phys. Conf. Ser., 12, 256,
  \dodoi{10.1142/S2010194512006459}

\bibitem[{de~Oliveira-Costa {et~al.}(2008)de~Oliveira-Costa, Tegmark, Gaensler,
  Jonas, Landecker, \& Reich}]{deOliveira-Costa:2008cxd}
de~Oliveira-Costa, A., Tegmark, M., Gaensler, B.~M., {et~al.} 2008, Mon. Not.
  Roy. Astron. Soc., 388, 247, \dodoi{10.1111/j.1365-2966.2008.13376.x}

\bibitem[{{DeBoer} {et~al.}(2017){DeBoer}, {Parsons}, {Aguirre}, {Alexander},
  {Ali}, {Beardsley}, {Bernardi}, {Bowman}, {Bradley}, {Carilli}, {Cheng}, {de
  Lera Acedo}, {Dillon}, {Ewall-Wice}, {Fadana}, {Fagnoni}, {Fritz},
  {Furlanetto}, {Glendenning}, {Greig}, {Grobbelaar}, {Hazelton}, {Hewitt},
  {Hickish}, {Jacobs}, {Julius}, {Kariseb}, {Kohn}, {Lekalake}, {Liu}, {Loots},
  {MacMahon}, {Malan}, {Malgas}, {Maree}, {Martinot}, {Mathison}, {Matsetela},
  {Mesinger}, {Morales}, {Neben}, {Patra}, {Pieterse}, {Pober}, {Razavi-Ghods},
  {Ringuette}, {Robnett}, {Rosie}, {Sell}, {Smith}, {Syce}, {Tegmark},
  {Thyagarajan}, {Williams}, \& {Zheng}}]{2017PASP..129d5001D}
{DeBoer}, D.~R., {Parsons}, A.~R., {Aguirre}, J.~E., {et~al.} 2017, \pasp, 129,
  045001, \dodoi{10.1088/1538-3873/129/974/045001}

\bibitem[{Delabrouille {et~al.}(2013)}]{Delabrouille:2012ye}
Delabrouille, J., {et~al.} 2013, Astron. Astrophys., 553, A96,
  \dodoi{10.1051/0004-6361/201220019}

\bibitem[{Di~Matteo {et~al.}(2002)Di~Matteo, Perna, Abel, \&
  Rees}]{DiMatteo:2001gg}
Di~Matteo, T., Perna, R., Abel, T., \& Rees, M.~J. 2002, Astrophys. J., 564,
  576, \dodoi{10.1086/324293}

\bibitem[{{Fabian} \& {Klaus}(2019)}]{Isensee:2019AnAA}
{Fabian}, I., \& {Klaus}, M.-H. 2019, arXiv e-prints, arXiv:1908.02182.
\newblock \doarXiv{1908.02182}

\bibitem[{G\'orski {et~al.}(2005)G\'orski, Hivon, Banday, Wandelt, Hansen,
  Reinecke, \& Bartelman}]{Gorski:2004by}
G\'orski, K.~M., Hivon, E., Banday, A.~J., {et~al.} 2005, Astrophys. J., 622,
  759, \dodoi{10.1086/427976}

\bibitem[{Haslam {et~al.}(1982)Haslam, Salter, Stoffel, \&
  Wilson}]{Haslam:1982zz}
Haslam, C. G.~T., Salter, C.~J., Stoffel, H., \& Wilson, W.~E. 1982, Astron.
  Astrophys. Suppl. Ser., 47, 1

\bibitem[{Hothi {et~al.}(2020)}]{Hothi:2020dgq}
Hothi, I., {et~al.} 2020, Mon. Not. Roy. Astron. Soc., 500, 2264,
  \dodoi{10.1093/mnras/staa3446}

\bibitem[{Li {et~al.}(2020)}]{Li:2020ast}
Li, J., {et~al.} 2020, Sci. China Phys. Mech. Astron., 63, 129862,
  \dodoi{10.1007/s11433-020-1594-8}

\bibitem[{Li {et~al.}(2021)Li, Santos, Grainge, Harper, \& Wang}]{Li:2020bcr}
Li, Y., Santos, M.~G., Grainge, K., Harper, S., \& Wang, J. 2021, Mon. Not.
  Roy. Astron. Soc., 501, 4344, \dodoi{10.1093/mnras/staa3856}

\bibitem[{Lidz {et~al.}(2011)Lidz, Furlanetto, Oh, Aguirre, Chang, Dore, \&
  Pritchard}]{Lidz:2011dx}
Lidz, A., Furlanetto, S.~R., Oh, S.~P., {et~al.} 2011, Astrophys. J., 741, 70,
  \dodoi{10.1088/0004-637X/741/2/70}

\bibitem[{Liu {et~al.}(2014)Liu, Parsons, \& Trott}]{Liu:2014bba}
Liu, A., Parsons, A.~R., \& Trott, C.~M. 2014, Phys. Rev. D, 90, 023018,
  \dodoi{10.1103/PhysRevD.90.023018}

\bibitem[{Loeb \& Wyithe(2008)}]{Loeb:2008hg}
Loeb, A., \& Wyithe, S. 2008, Phys. Rev. Lett., 100, 161301,
  \dodoi{10.1103/PhysRevLett.100.161301}

\bibitem[{Makinen {et~al.}(2021)Makinen, Lancaster, Villaescusa-Navarro,
  Melchior, Ho, Perreault-Levasseur, \& Spergel}]{Makinen:2020gvh}
Makinen, T.~L., Lancaster, L., Villaescusa-Navarro, F., {et~al.} 2021, JCAP,
  04, 081, \dodoi{10.1088/1475-7516/2021/04/081}

\bibitem[{Mao {et~al.}(2008)Mao, Tegmark, McQuinn, Zaldarriaga, \&
  Zahn}]{Mao:2008ug}
Mao, Y., Tegmark, M., McQuinn, M., Zaldarriaga, M., \& Zahn, O. 2008, Phys.
  Rev. D, 78, 023529, \dodoi{10.1103/PhysRevD.78.023529}

\bibitem[{Masui {et~al.}(2013)}]{Masui:2012zc}
Masui, K.~W., {et~al.} 2013, Astrophys. J. Lett., 763, L20,
  \dodoi{10.1088/2041-8205/763/1/L20}

\bibitem[{Matshawule {et~al.}(2021)Matshawule, Spinelli, Santos, \&
  Ngobese}]{Matshawule:2020fjz}
Matshawule, S.~D., Spinelli, M., Santos, M.~G., \& Ngobese, S. 2021, Mon. Not.
  Roy. Astron. Soc., 506, 5075, \dodoi{10.1093/mnras/stab1688}

\bibitem[{McQuinn {et~al.}(2006)McQuinn, Zahn, Zaldarriaga, Hernquist, \&
  Furlanetto}]{McQuinn:2005hk}
McQuinn, M., Zahn, O., Zaldarriaga, M., Hernquist, L., \& Furlanetto, S.~R.
  2006, Astrophys. J., 653, 815, \dodoi{10.1086/505167}

\bibitem[{Mertens {et~al.}(2018)Mertens, Ghosh, \& Koopmans}]{Mertens:2017gxw}
Mertens, F.~G., Ghosh, A., \& Koopmans, L. V.~E. 2018, Mon. Not. Roy. Astron.
  Soc., 478, 3640, \dodoi{10.1093/mnras/sty1207}

\bibitem[{Nan {et~al.}(2011)Nan, Li, Jin, Wang, Zhu, Zhu, Zhang, Yue, \&
  Qian}]{Nan:2011um}
Nan, R., Li, D., Jin, C., {et~al.} 2011, Int. J. Mod. Phys. D, 20, 989,
  \dodoi{10.1142/S0218271811019335}

\bibitem[{Newburgh {et~al.}(2016)}]{Newburgh:2016mwi}
Newburgh, L.~B., {et~al.} 2016, Proc. SPIE Int. Soc. Opt. Eng., 9906, 99065X,
  \dodoi{10.1117/12.2234286}

\bibitem[{Olivari {et~al.}(2018)Olivari, Dickinson, Battye, Ma, Costa,
  Remazeilles, \& Harper}]{Olivari:2017bfv}
Olivari, L.~C., Dickinson, C., Battye, R.~A., {et~al.} 2018, Mon. Not. Roy.
  Astron. Soc., 473, 4242, \dodoi{10.1093/mnras/stx2621}

\bibitem[{Peel {et~al.}(2019)Peel, Wuensche, Abdalla, Ant{\'o}n, Barosi,
  Browne, Caldas, Dickinson, Fornazier, Monstein, Strauss, Tancredi, \&
  Villela}]{Pee:l2019BaryonAO}
Peel, M., Wuensche, C.~A., Abdalla, E., {et~al.} 2019, Journal of Astronomical
  Instrumentation

\bibitem[{Pritchard \& Loeb(2008)}]{Pritchard:2008da}
Pritchard, J.~R., \& Loeb, A. 2008, Phys. Rev. D, 78, 103511,
  \dodoi{10.1103/PhysRevD.78.103511}

\bibitem[{Reddi {et~al.}(2019)Reddi, Kale, \& Kumar}]{Sashank:2019abs}
Reddi, S.~J., Kale, S., \& Kumar, S. 2019, CoRR, abs/1904.09237

\bibitem[{{Ronneberger} {et~al.}(2015){Ronneberger}, {Fischer}, \&
  {Brox}}]{Ronneberger:2015UNetCN}
{Ronneberger}, O., {Fischer}, P., \& {Brox}, T. 2015, arXiv e-prints,
  arXiv:1505.04597.
\newblock \doarXiv{1505.04597}

\bibitem[{Santos {et~al.}(2005)Santos, Cooray, \& Knox}]{Santos:2004ju}
Santos, M.~G., Cooray, A., \& Knox, L. 2005, Astrophys. J., 625, 575,
  \dodoi{10.1086/429857}

\bibitem[{Santos {et~al.}(2015)}]{Santos:2015gra}
Santos, M.~G., {et~al.} 2015, PoS, AASKA14, 019, \dodoi{10.22323/1.215.0019}

\bibitem[{Santos {et~al.}(2017)}]{MeerKLASS:2017vgf}
Santos, M.~G., {et~al.} 2017, in {MeerKAT Science}: {On the Pathway to the
  SKA}.
\newblock \doarXiv{1709.06099}

\bibitem[{Seo {et~al.}(2010)Seo, Dodelson, Marriner, Mcginnis, Stebbins,
  Stoughton, \& Vallinotto}]{Seo:2009fq}
Seo, H.-J., Dodelson, S., Marriner, J., {et~al.} 2010, Astrophys. J., 721, 164,
  \dodoi{10.1088/0004-637X/721/1/164}

\bibitem[{Soares {et~al.}(2022)Soares, Watkinson, Cunnington, \&
  Pourtsidou}]{Soares:2021ohm}
Soares, P.~S., Watkinson, C.~A., Cunnington, S., \& Pourtsidou, A. 2022, Mon.
  Not. Roy. Astron. Soc., 510, 5872, \dodoi{10.1093/mnras/stab2594}

\bibitem[{{Spinelli} {et~al.}(2022){Spinelli}, {Carucci}, {Cunnington},
  {Harper}, {Irfan}, {Fonseca}, {Pourtsidou}, \& {Wolz}}]{2022MNRAS.509.2048S}
{Spinelli}, M., {Carucci}, I.~P., {Cunnington}, S., {et~al.} 2022, \mnras, 509,
  2048, \dodoi{10.1093/mnras/stab3064}

\bibitem[{Switzer {et~al.}(2015)Switzer, Chang, Masui, Pen, \&
  Voytek}]{Switzer:2015ria}
Switzer, E.~R., Chang, T.-C., Masui, K.~W., Pen, U.-L., \& Voytek, T.~C. 2015,
  Astrophys. J., 815, 51, \dodoi{10.1088/0004-637X/815/1/51}

\bibitem[{Switzer {et~al.}(2013)}]{Switzer:2013ewa}
Switzer, E.~R., {et~al.} 2013, Mon. Not. Roy. Astron. Soc., 434, L46,
  \dodoi{10.1093/mnrasl/slt074}

\bibitem[{Villanueva-Domingo \&
  Villaescusa-Navarro(2021)}]{Villanueva-Domingo:2020wpt}
Villanueva-Domingo, P., \& Villaescusa-Navarro, F. 2021, Astrophys. J., 907,
  44, \dodoi{10.3847/1538-4357/abd245}

\bibitem[{Wang {et~al.}(2021)}]{Wang:2020lkn}
Wang, J., {et~al.} 2021, Mon. Not. Roy. Astron. Soc., 505, 3698,
  \dodoi{10.1093/mnras/stab1365}

\bibitem[{Wolz {et~al.}(2017)}]{Wolz:2015lwa}
Wolz, L., {et~al.} 2017, Mon. Not. Roy. Astron. Soc., 464, 4938,
  \dodoi{10.1093/mnras/stw2556}

\bibitem[{Wolz {et~al.}(2022)}]{Wolz:2021ofa}
---. 2022, Mon. Not. Roy. Astron. Soc., 510, 3495,
  \dodoi{10.1093/mnras/stab3621}

\bibitem[{Wu {et~al.}(2021)}]{Wu:2020jwm}
Wu, F., {et~al.} 2021, Mon. Not. Roy. Astron. Soc., 506, 3455,
  \dodoi{10.1093/mnras/stab1802}

\bibitem[{Wuensche {et~al.}(2021)}]{Wuensche:2021dcx}
Wuensche, C.~A., {et~al.} 2021, arXiv e-prints, arXiv:2107.01634.
\newblock \doarXiv{2107.01634}

\bibitem[{Wyithe \& Loeb(2008)}]{Wyithe:2007gz}
Wyithe, S., \& Loeb, A. 2008, Mon. Not. Roy. Astron. Soc., 383, 606,
  \dodoi{10.1111/j.1365-2966.2007.12568.x}

\bibitem[{Wyithe {et~al.}(2008)Wyithe, Loeb, \& Geil}]{Wyithe:2007rq}
Wyithe, S., Loeb, A., \& Geil, P. 2008, Mon. Not. Roy. Astron. Soc., 383, 1195,
  \dodoi{10.1111/j.1365-2966.2007.12631.x}

\bibitem[{Çiçek {et~al.}(2016)Çiçek, Abdulkadir, Lienkamp, Brox, \&
  Ronneberger}]{Cicek:2016abs}
Çiçek, {\"O}., Abdulkadir, A., Lienkamp, S.~S., Brox, T., \& Ronneberger, O.
  2016, arXiv e-prints, arXiv:1606.06650.
\newblock \doarXiv{1606.06650}

\end{thebibliography}
\bibliographystyle{aasjournal}


\end{document}